# SPECTRAL AND TEMPORAL CHARACTERIZATION OF NANOSECOND AND FEMTOSECOND LASER PRODUCED PLASMA FROM METALLIC TARGETS

A THESIS

*Submitted by*

## SMIJESH N.

*For the award of the degree of*

DOCTOR OF PHILOSOPHY

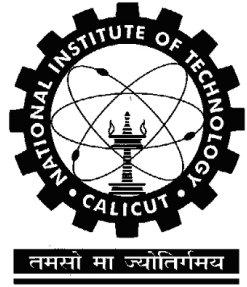

DEPARTMENT OF PHYSICS
NATIONAL INSTITUTE OF TECHNOLOGY CALICUT
NIT CAMPUS P. O., KOZHIKODE
KERALA, INDIA 673601

JULY 2014

*To my family and teachers*

# ACKNOWLEDGEMENTS

Keeping a step forward in my academic life and my personal life too, it will be unfair if I don't remember those people who helped me to succeed in taking this big step in my life. Prayers, support, encouragement and criticism from people had always been an asset to help me form myself.

A fraction of my academic achievements is dedicated to Dr. K. Chandrashekaran and Dr. Reji Philip; my PhD supervisors. Dr. K. Chandrashekaran did his best to help me complete the formalities and requirements related to my PhD right from the beginning. The time spent at NIT was made more helpful and promising by this person who gave his whole hearted support throughout the PhD tenure. Dr. Reji Philip, my supervisor at the Raman Research Institute, Bangalore, where I did my entire set of experiments, was a good person to be with. I had been with him since the time of my M. Tech project, where I developed my interest in the field of plasma and decided to do my PhD in the field. Whole hearted support from Dr. Chandrashekaran and Dr. Reji Philip made the initial steps of my PhD smooth. While Dr. Reji helped me by providing me his lab and other facilities required, the other side of my PhD work was supported and backed up by Dr. Chandrashekaran. Dr. Reji and his family were a big mental support to my work during the stay at Bangalore. They are thanked and are always placed close to my heart. Unforgettable moments with Reji ; a very good stress buster, especially during the lunch breaks and tea time are always cherished as they always helped me to relax out of the strenuous work during hectic times.

I would also like to thank Dr. M. N. Bandyopadhyay, Director, NITC, for providing all the facilities required to carry out my research, especially the course work related to the curriculum. I am also thankful to Dr. V. K. Govindan, Dean (Academic) and former Dean Dr. K. Prabhakaran Nair for their extended support throughout my research work.

Dr. Raghu. C, Dr. M. K Ravi Varma and Dr. Vari Sivaji Reddy, Department of Physics, NITC; Dr. Parameswaran P, Department of Chemistry, NITC and Dr. Soney Varghese from SNST, NITC; formed the Doctoral committee for my research programme. Their fruitful discussions, suggestions and criticisms were live throughout the time, which helped me improve a lot. I thank them sincerely for their constructive encouragement. I am grateful to all the teaching and non-teaching staff of NIT Calicut, for their ever loving support.

Dr. Achamma Kurian, who directed me into research and constantly encouraged me to join a PhD is one of the reasons for this moment in my life and I remember her with gratitude and love. I am always indebted to her and her support to me throughout these years. Dr. S. S. Harilal is yet another person whom I would like to mention with respect and love, who helped me with methods of writing a good paper and with some experimental insights.

Heartfelt thanks to Mr. Jinto Thomas; my silent mentor, a good experimentalist and over all a good person to be with; who helped me with his experimental skills to get out of dilemmas at some hard times during the course of my experiments. His beauty of work in the lab and way of executing things were always admired and it inspired me to take up such experiments for my PhD work. I always remember him with love and respect as he held me up while entering the experimental works in the field. I also thank him for providing me the ICCD which helped me to give a good final form to my thesis.

Suchand Sandeep, my elder brother in life and lab, well wisher and a good friend always, kept me on track during my PhD days. He really wanted to see me complete this huge step successfully in my life and hence was a solid support in all ways. I am happy that I could fulfil his wish satisfactorily and I cherish my sweet times with him in the lab during the initial days of lab work at the Raman Research Institute.

*Acknowledgements*

Initial period of my work which started from the scratch was carried out with the help of Aparna, who was a true supporter in experiments and a very good friend in my life. I cherish the weekends and the workaholic times we spent happily together at this moment. Later on, Jagdish joined me for a while. His theoretical support and doubts strengthened my work and he was one of those who accompanied my work during late nights. Apart from a lab mate and a good supporter, he became my good friend whom I don't want to miss for this entire lifetime. Jijil and Pranitha who came for their Master's project work, became my very good experimental partners later on. I whole heartedly thank them for their support, help and affection. I am also happy to have Suchand Sangeeth as a friend who spent some nice times with me during the PhD tenure.

Friends who always kept me live without frustrations are an integral part of my whole work. Words become utterly helpless to describe these people who are unique in their own ways. Their ways, presence and friendship can only be felt. Satheesh, Pramod and Anin are awesome people who are assets to my life. The tea-time, spare times and group talks are among some memories that I cherish a lot. My room-mate Aneesh and his friends need special mention as they were the only people who were with me, once I was back after strenuous work. They were extremely helpful and I am thankful to them for being in my life. Rekha, Sony chettan, Irshadettan, Sibu, Rajesh, Jayasree, and Haripadmam also equally helped me to keep my presence of mind throughout the PhD period. I also thank Jileesh, Suraj, Sunil and Ashok who joined my life and extended their strong support as they always did. I also like to specially mention Sreekanth, Nithin, Elizabeth and Anitta who were continuous supporters and good companions. They kept me live and sporty throughout, especially during the final times of writing a thesis. The routine of badminton practice at NITC gifted me a hand full of very good friends. Darshan, Jithin, Praveen Sir and Madhu sir were good people to be with and they always gave me positive energy that helped me work happily at NITC. Special regards to Abhilash sir who was a source of inspiration. The energy, support and energy they gave me in my life are priceless and unforgettable. I am happy to have them close to my heart always.

*Acknowledgements*

Labmates help in creating an environment favourable for work. Lab members at NITC gave me a very huge support and backup. They were my family, friends and more for the entire time of the PhD. Days that I stayed away from my family were made easier by them. I personally thank Siji, Sudheesh, Divyechi and Vijisha for their support and rocking memories. I thank Suhail, Arun, Aneesh J, Issac sir, Shiju, Praveen, Yoosuf, Lakshmikanth and Nishaina for their cooperation at NITC. I thank Augustin, Benoy, Aravindan, Ann Mary, Safakath, Athira, Nivya, Shafi and Priya at the Raman Research Institute for their memorable company. Harini, Shiva and Manju at RRI were nice to me and helped me in the official works at the institute and I thank them at this moment. I also thank Sarojini chechi and Abraham chettan at NITC for their valuable support which helped me a lot.

Family is an important part of life which is a source of inspiration, support and love. My mother's love, father's affection, care from sisters and support from in-laws filled my mind with peace and love throughout the period. My niece kept me bound tightly to the family during this period so that I felt the pool of love and my mind could relax to work more efficiently each day. One good friend for life - Neena and her family are also remembered with love at this moment for whatever love, care and support they gave me. I love them all and thank them whole heartedly for supporting me in my ventures.

Grace of god has not left me blank throughout these days at any point of my life. I thank one and all who are responsible for my achievements till now and who were with me to complete this venture successfully. I remember each one of them in my heart and thank them all whole heartedly.

Smijesh N.

# DECLARATION

*I hereby declare that this submission is my own work and that, to the best of my knowledge and belief, it contains no material previously published or written by another person nor material which has been accepted for the award of any other degree or diploma of the university or other institute of higher learning, except where due acknowledgment has been made in the text.*

NIT Calicut **Smijesh N.**
28 July 2014 Reg. No. P110002PH

# CERTIFICATE

*This is to certify that the thesis entitled **"Spectral and temporal characterization of nanosecond and femtosecond laser produced plasma from metallic targets"** submitted by **Mr. Smijesh N** to the National Institute of Technology Calicut for the award of the degree of Doctor of Philosophy is a bona fide record of the research work carried out by him under our supervision and guidance. The content of the thesis, in full or parts have not been submitted to any other Institute or University for the award of any other degree or diploma.*

*Dr. Reji Philip*
Associate Professor
Ultrafast and Nonlinear Optics Lab
Light and Matter Physics Group
Raman Research Institute,
Bangalore, India

*Dr. Chandrasekharan. K*
Professor
Laser and Nonlinear Optics Laboratory
Department of Physics
National Institute of Technology Calicut
Kerala, India

*NIT Calicut*
*28$^{th}$ July 2014*

*Signature of Head of the Department*

# CONTENTS

# ABSTRACT

Advent of high power lasers revolutionized several fields of physics related to light-matter interaction. One among them was the field of ultrafast processes, of which Laser Produced Plasma (LPP) forms an important part. This leap in ultrafast science leads to the emergence of fundamental studies of intense field physics.

LPP from nanosecond (ns, short-pulse) irradiations have been a field of study for the last few decades whereas the femtosecond (fs, ultrafast) LPP has attracted the research community only recently. Several studies are going on around the globe owing to its wide range of applications like pulsed laser deposition, nanoparticle generation, nano-imprinting, generations of light sources, plasma-based accelerators, etc. It is found that nanosecond and femtosecond LPP are entirely different in terms of the nature of plume generation and expansion, resulting in various effects that are theoretically predicted. The work presented in this thesis aims at the generation and characterization of both ns and fs LPPs, facilitating fundamental studies related to plasma parameters like the calculation of temperature, number density, identification of the plasma species and their properties etc., helping us to realize the applications of LPP.

Chapter 1 gives an idea of the motivation behind this work and a brief description of the same, followed by a formal description of plasma, its fundamental aspects, and parameters defining plasma behavior. Various methods for generating plasma in the laboratory, radiations emitted from the plasma, and major diagnostic methods used for characterizing laser produced plasma, along with their relative merits and demerits, are given in detail. A review on laser produced plasmas also is included.

Instrumentation required to perform optical emission spectroscopy and optical time of flight measurements and imaging of the plasma plume, generated by nanosecond/ femtosecond laser pulses, is discussed in Chapter 2. The laser system, pulse characterization techniques, synchronization circuits, and instruments used for detecting the emission spectra are explained in detail.

*Abstract*

The experimental techniques employed in the thesis for carrying out the experiments are detailed in Chapter 3. Execution of various techniques like optical emission spectroscopy, optical time of flight measurements and imaging of the hydrodynamics are explained in detail.

Experimental characterization and comparison of the temporal features of femtosecond and nanosecond laser produced plasmas generated from a solid Nickel target, expanding into a nitrogen background, is presented in Chapter 4. Dynamics of ions and fast and slow neutrals present in the laser produced nickel plasma is studied by the optical time of flight technique, using emissions from neutrals and ions, respectively. Velocities of these species are estimated from the time of flight data. Furthermore, the variation of temperature, number density and velocity of different species in the plume with respect to variations in ambient pressure, for both nanosecond and femtosecond irradiation, is also investigated.

A detailed study of femtosecond and nanosecond laser produced Zinc plasmas conducted using optical emission spectroscopy and optical time of flight techniques is presented in Chapter 5. Optical emission spectroscopy allows the estimation of temperature and number density of the plasma, while optical time of flight measurements throw light into the dynamics of ions and neutrals in the plasma. Imaging done using an intensified charge coupled device (ICCD) helps observe features of plasma expansion into the ambient gas with high time resolution.

In Chapter 6, line emission dynamics of neutrals and ions in the ns LPP nitrogen ambient is measured and compared. Emission from neutrals can be measured to larger axial distances compared to ions in the expanding plasma. The times of arrival of ionic and neutral species at larger fluences confirm the occurrence of fast atomic species via recombination of fast ions with electrons. Moreover, fast atomic species are absent at larger axial distances where ionic emissions are not detected. The estimated peak velocities of fast neutrals and ions are found to increase with an increase in the laser energy. Furthermore, an acceleration of fast neutrals and ions is observed in the plasma at close distances to the target. This is attributed to the space charge effect and is discussed.

Optical pump-probe technique is the best choice for detecting fast processes that demand a high time resolution, where fast detectors (except the streak camera) fail. A novel asynchronous laser plasma pump-probe scheme is employed to study the dynamics and transient nature of ns LPP, using a very low energy 100 fs, 80 MHz pulse train as the probe beam, allowing to probe the plasma at every 12.54 ns interval after its generation. It was found that the plasma acts as gain medium, imparting amplification to the probe beam. Details of this novel observation are given in Chapter 7.

Finally, Chapter 8 gives a brief conclusion of all the experiments done, their significance of obtained results to laser plasma literatures and future scope.

# SYMBOLS AND ABBREVIATIONS

| | |
|---|---|
| $r$ | : Distance |
| $n_e$ | : Number density of electrons |
| $n_i$ | : Number density of ions |
| $T_e$ | : Temperature of electrons |
| $B$ | : Magnetic field |
| $T_i$ | : Temperature of ions |
| $T_a$ | : Temperature of atoms |
| $T_g$ | : Temperature of gas molecules |
| $v, u$ | : Velocity of species |
| $v_s$ | : Velocity of scattered species |
| $K, K_B$ | : Boltzmann constant |
| $m$ | : Mass |
| $m_e$ | : Mass of electron |
| $m_i$ | : Mass of ions |
| $n$ | : Number density of plasma |
| A | : Normalization constant |
| e | : Charge of electron |
| $\lambda_D$ | : Debye length |
| $\phi(r)$ | : Electric potential |
| L | : Physical length |
| $N_D$ | : Plasma Parameter |
| $\omega$ | : Frequency of Plasma oscillations |
| $\omega_{pe}$ | : Electron Plasma frequency |
| $\omega_{pi}$ | : Ion plasma frequency |
| $\omega_p$ | : Total plasma frequency |
| $\omega_c$ | : Cyclotron frequency |
| $f_{pe}$ | : Frequency of oscillations in the plasma |

*Symbols and Abbreviations*

| | |
|---|---|
| ε | : Dielectric constant |
| $\varepsilon_0$ | : Permittivity of free space |
| $\varepsilon_l$ | : Linear dielectric permittivity |
| τ | : Mean time between collisions |
| $q_j$ | : Charge of species |
| P | : Fluid pressure |
| Γ | : Adiabatic constant |
| V | : Volume |
| E | : Electric field |
| H | : Magnetizing field |
| J | : Current density |
| σ | : Surface charge density |
| $\lambda_{mfp}$ | : Mean free path |
| λ | : Wavelength |
| $ä_k$ | : Acceleration of species |
| h | : Plank's constant |
| υ | : Frequency of radiation |
| $\chi_i$ | : Ionization potential |
| Z | : Atomic number |
| $I_i$ | : Intensities of the $i^{th}$ line. |
| $g_i$ | : Statistical Weights of emission |
| $A_i$ | : Transition Probabilities |
| $\Lambda$ | : Wavelength |
| $E_i$ | : Energies of upper level |
| $\Delta\lambda_{1/2}$ | : FWHM |
| w | : Electron impact parameter |
| A | : Ion broadening parameter |
| $N_e$ | : Electron number density |
| $\Delta E$ | : Difference in energy |
| z | : Distance from the target |
| $E_0$ | : Constant proportional to the laser energy density |
| $\rho_b$ | : Density of the background gas |

*Symbols and Abbreviations*

| | |
|---|---|
| $t$ | : Time |
| $z_0$ | : Distance moved by the species for the time of the order of the average life time of excited states. |
| $z_f$ | : Stopping distance of the plume |
| $\beta$ | : Slowing coefficient |
| $v_0$ | : Initial velocity of the ejected species |
| $I_{min}$ | : Ablation threshold |
| $\rho$ | : Density of the material |
| $L_v$ | : Latent heat of vaporization |
| $\kappa$ | : Thermal diffusivity |
| $\Delta t$ | : Laser pulse width |
| $\alpha_{ib}$ | : Inverse Bremsstrahlung |
| $t_{ei}$ | : Electron-ion energy transfer time |
| $t_h$ | : Heat conduction time |
| $t_p$ | : Pulse duration of the laser |
| 2D | : 2-Dimensional |
| AC | : Alternating Current |
| AOM | : Acousto-Optic Modulator |
| CCD | : Charged Coupled Device |
| CDS | : Crooke's Dark Space |
| CF | : Conflat Flange |
| CTE | : Complete Thermodynamic Equilibrium |
| CW | : Continuous Wave |
| DC | : Direct Current |
| DP | : Double Pulse |
| EUV | : Extreme Ultraviolet |
| FDS | : Faraday's Dark Space |
| FWHM | : Full Width Half Maximum |
| HIB | : Heavy Ion Beam |
| ICCD | : Intensified Charged Coupled Device |
| IR | : Infrared |
| KDP | : Potassium Dihydrogen Phosphate |
| LA | : Laser Ablation |
| LBO | : Lithium Triobate |
| LIBS | : Laser Induced Breakdown Spectroscopy |
| LPP | : Laser Produced Plasma |
| LTE | : Local Thermodynamic Equilibrium |
| Nd: YAG | : Neodymium – doped Yttrium Aluminum Garnet |
| Nd: $YVO_4$ | : Neodymium – doped Yttrium Orthovanadate |
| Ni | : Nickel |
| NIR | : Near Infra Red |
| NIST | : National Institute of Standards and Technology |
| OE | : Optical Emission |
| OES | : Optical Emission Spectroscopy |
| OTOF | : Optical Time of Flight |
| PLD | : Pulsed Laser Deposition |
| PK1 | : Peak 1 (fast peak) |
| PK2 | : Peak 2 (slow peak) |
| PK3 | : Peak 3 |
| PMT | : Photomultiplier Tube |
| REB | : Relativistic Electron Beam |
| RF | : Radio Frequency |
| SDG | : Synchronization and Delay Generator |
| SP | : Single Pulse |
| TEM | : Transverse Electromagnetic Mode |
| Ti | : Titanium |
| TSA | : Titanium Sapphire Amplifier |
| TTL | : Transistor Transistor Logic |
| UV | : Ultraviolet |
| Zn | : Zinc |

# LIST OF FIGURES

# LIST OF TABLES

# CHAPTER 1
# INTRODUCTION

*The motivation for doing this work and a brief description of the same are presented in this chapter, followed by a formal description of plasma, its fundamental aspects, and parameters defining plasma behavior. Various methods for generating plasma in the laboratory, radiations emitted from the plasma, and major diagnostic methods used for characterizing laser produced plasma, along with their relative merits and demerits, are given in detail. A review on laser produced plasmas also is included.*

## 1.1 MOTIVATION AND OVERVIEW

Plasma is the most common state of matter in the universe. Earth is one of the planets in this universe where the plasma state does not occur naturally, but about 99% of the universe happens to be in the plasma state. Plasma is a very good radiation source which can emit in the VISIBLE, IR, UV or X-ray regimes depending on its nature. Production of plasma in a laboratory can be achieved in many ways such as ac and dc discharges, particle accelerated plasmas, z –pinch, theta – pinch etc. Laser produced plasma (LPP) is another important method for generating plasma. In LPP, an intense laser beam focused onto a small area of the target ablates its surface within a short time, producing the plasma. The major applications of LPP include Pulsed Laser Deposition (PLD) [1], higher harmonic generation [2-5], generation of UV, EUV and X-rays [6-10], and nanoparticle and nanocluster generation [11-14]. Laser parameters (such as wavelength, fluence, pulse width, irradiation spot size) can be tuned and the plasma parameters can be optimized to implement the above mentioned applications successfully.

Nanosecond (ns) LPP has been extensively studied to understand the underlying fundamental physics of plasma generation such as laser-target energy coupling, laser-plasma interaction, and its expansion to the surroundings. Study of variations in the properties of the LPP with respect to changes in laser parameters and ambient conditions (nature and pressure of ambient gas) has been a hot topic of research for the last two decades. It was found that the size and distribution of nanoparticles generated by femtosecond (fs) laser ablation can be tuned and optimized under certain conditions, and this has become an active area of research in the last ten years. The generation and expansion of fs LPP is different from those of ns LPP in many ways, and are these are detailed in the respective chapters which follow. With a keen interest to understand the fundamental processes behind the generation and expansion of laser produced metal plasmas, this work mainly focuses on the effects of pulse width and pressure on the dynamics of both fs and ns LPPs.

## 1.2 BASIC CONSIDERATIONS OF PLASMAS

Plasma, more often referred to as the fourth state of matter, is essentially distinct from the other states, and has thermal energy more than the ionization energy. The fundamental concepts for explaining the physics of plasmas build up on varied fields such as mechanics, electrodynamics, statistical mechanics, fluid mechanics, and kinetic theory of gases. Unlike other states of matter, plasma responds readily to electric and magnetic fields since it is composed of charged species such as electrons and ions. A paper by I. Langmuir titled 'Oscillations in ionized gas', in the Proceedings of the National Academy of Sciences (August 1928), reports the observation of low amplitude voltage oscillations of certain frequencies in a strongly ionized gas at low pressures. Langmuir and Tonks (1929) then explored collective behavior and long-range force in a charged system containing free electrons, ions and neutral species through a series of experiments. They defined plasma as a "collection of charged particle sufficiently dense such that space charge effects can result in a strongly coherent behavior". In other words, plasma is a quasi-neutral gas of charged and neutral particles which exhibits collective behavior. A fundamental characteristic behavior of plasma is its ability to shield out electric potentials that are applied to it, known as Debye Shielding [15-18].

While the macroscopic force is transferred via collisions in a system of neutral gas, the situation is totally different in plasma. Charges in the plasma move around creating local concentrations of positive and negative charges, giving rise to electric fields whose motion generates electric currents, and hence magnetic fields. The Coulomb field between two slightly charged regions of plasma separated by a distance $r$ varies as $1/r^2$ and for a given solid angle the volume of a local region in the plasma can affect another volume as $r^3$. Thus, the elements of plasma are found to exert force to larger distances. Movement of species in the plasma not only depends on the local conditions but also on the state of the plasma in remote regions, due to the long range Coulomb forces which are much larger than the ordinary collisional forces in ordinary gas. This property is referred to as the 'collective' behavior of the plasma.

Plasma properties depend strongly on bulk and average properties, and the parameters used to define plasma characteristics include the degree of ionization, the plasma temperature, electron number density (more specifically the electron temperature and number density under local thermodynamic equilibrium), and magnetic field in the region of plasma. The fundamental parameters that are used to characterize plasma are the electron number density ($n_e$), temperature ($T_e$) and steady state magnetic field ($B$). Other parameters which are useful in plasma characterization such as the Debye length, Larmour radius, plasma frequency, thermal velocity, and the cyclotron frequency can be derived from the fundamental plasma parameters [15-18].

### 1.2.1 Electron temperature

Plasma consists of various species like the electrons, ions and neutrals (generally known as the plasma species) that are not heated equally, and hence their mean kinetic energies are different. The kinetic energy of electrons is much higher than that of the ions, and the kinetic energy of the ions is higher compared to that of the neutrals. Since the kinetic energies corresponding to various species are different, they posses different temperatures, i.e. $T_e$, $T_i$, and $T_a$ for the electrons, ions and neutrals respectively, which are in the order $T_e>T_i>T_a$. In plasma, the temperature of the electrons can be several magnitudes higher than that of the neutrals and ions. This can be either due to the strong heating of the electrons by the plasma sources or because of the efficient energy transfer in collisions between the electrons (two-body collisional energy transfer is much more efficient for collision between particles of similar masses) [15-18]. The electron temperature $T_e$ is usually measured in Kelvin (K) in or electron volts (eV).

In plasma, species are in random motion and if electrons, ions, and neutrals are in thermal equilibrium, they can be described by the Maxwellian distribution which is given (in one dimension) by,

$$f(u)du = A\, exp\left[\frac{-mu^2}{2KT_e}\right] \qquad (1.1)$$

where $f(u)du$ is the number of particles per m³ with a velocity between u and u + du, $1/2\, mu^2$ is the kinetic energy, and $K$ is the Boltzmann constant. Density of particles per m³ is given by,

$$n = \int_{-\infty}^{+\infty} f(u)du \qquad (1.2)$$

Where A, the normalization constant is given by,

$$A = n\left(\frac{m}{2\pi mKT_e}\right)^{1/2} \qquad (1.3)$$

The width of the distribution is characterized by the temperature of the plasma.

### 1.2.2  Debye shielding

One of the fundamental characteristics of plasma is the shielding out of electric potentials that are produced upon the introduction of electric charges, by pushing similar charges away and pulling the opposite charges towards the point of perturbation. This creates a cloud of positive/negative charge around the perturbed point which maintains the macroscopic electrical neutrality of the plasma. Unlike cold plasma, shielding may not be perfect in the case of thermal plasma where the particles at the edge of the cloud (where the electric field is weak) have enough thermal energy to escape from the electrostatic potential. At the edge of the cloud where the potential energy is approximately equal to $KT_e$ of the particle, shielding is imperfect and potentials of the order of $\frac{KT_e}{e}$ can leak to the plasma to produce a finite electric field there. Calculation of the approximate thickness of such a charge cloud allows to measure the effective shielding distance using Poisson's equation in one dimension. This measure, known as the Debye length ($\lambda_D$), is given by,

$$\lambda_D = \left(\frac{\epsilon_0 K_B T_e}{n_e e^2}\right)^{1/2} \tag{1.4}$$

where $\lambda_D$ is the length over which the Coulomb field is no longer felt in the plasma. The extent of this screening can be calculated using Poisson's equations with the source terms being the test particle and its associated cloud, whose contribution is determined using the Boltzmann relation for the particles that cause screening. This is a self-consistent calculation for the potential because the shielding cloud is affected by its self-potential.

For $L \ll \lambda_D$; the potential $\phi(r)$ is identical to the potential of a test particle in vacuum. For $L \gg \lambda_D$; the test particle is completely screened by its surrounding shielding cloud. $\lambda_D$ is the radius about which the shielding can be felt in a plasma and this explanation makes sense only if $\frac{4}{3}\pi n_e \lambda_D^3 \gg 1$ (this is a condition for the plasma to be collisionless). For the shielding to be relevant, Debye Length ($\lambda_D$) must be small compared to the overall dimensions of the plasma. If this is not satisfied, then no point in the plasma could be outside the shielding cloud. The effect of local concentration of charges or electric potentials present in plasma is shielded out within a short distance maintaining the quasi neutrality. Plasma is quasi neutral, i. e. neutral enough that one can consider $n_i \simeq n_e$ which is equal to the plasma density ($n$), but not so neutral that all the electromagnetic forces vanish.

### 1.2.3 The plasma parameter

It is clear that the concept of Debye Length is valid only if there are enough number of particles in the charge cloud. This is because, if there are only one or two particles in the sheath region (outermost part of the cloud), then the concept of Debye shielding will be statistically invalid. We can calculate the number of particles ($N_D$, which is called the plasma parameter) in a "Debye Sphere" (Debye Sphere is an imaginary sphere whose radius is the Debye Length ($\lambda_D$) mentioned above) as:

$$N_D = \frac{4}{3} n_e \pi \lambda_D^3 = \frac{1.38 \times 10^6 T_e^{3/2}}{n_e^{1/2}} \qquad (1.5)$$

where $T_e$ is expressed in $^oK$. This in turn explains the condition for the collective behavior as $N_D \gg 1$.

### 1.2.4 Plasma oscillations

Perturbation applied to a local point in the plasma moves electrons in the plasma, which are pulled back to the equilibrium position due to columbic attraction, thereby maintaining the quasi neutrality. These perturbations induce slight deviations from quasi neutrality which set the electrons into collective oscillation. The electron plasma frequency of these non-damped Langmuir oscillations is given by,

$$\omega_{pe} = \left(\frac{n_e e^2}{\epsilon_0 m_e}\right)^{1/2} \qquad (1.6)$$

and the total plasma frequency is given by,

$$\omega_p^2 = \omega_{pe}^2 + \omega_{pi}^2 \qquad (1.7)$$

where $\omega_p$, $\omega_{pe}$ and $\omega_{pi}$ are the total, electron plasma, and ion plasma frequencies respectively. This can be approximated as $\omega_p \approx \omega_{pe}$ because of the large ion-electron mass ratio. By substituting standard values in the equation, the frequency of oscillations in the plasma can be calculated as,

$$f_{pe}(Hz) = {\omega_{pe}}/{2\pi} = 8.98 \times n_e^{1/2} \qquad (1.8)$$

Generally, unmagnetized plasmas support plasma oscillations and electromagnetic waves. Once the electrons in the plasma are displaced from their mean positions in a uniform background of ions, field builds up in the opposite direction of the movement of electrons causing them to oscillate around their equilibrium position. These collective motions which occur when the equilibrium condition is disturbed cannot be sustained in the plasma, and hence they break up into a natural frequency of oscillations known as the plasma frequency. The generated plasma frequency $\omega_p$, which is related to the oscillation of electron density can be written as:

$$\omega_p = \sqrt{\frac{n_e e^2}{m_e \varepsilon_0}} \approx 50\, n_e^{1/2} \qquad (1.9)$$

where $n_e$ is expressed in $m^{-3}$. It is clear that $\omega_p$ depends only on the number density in the plasma. Hence, plasma with higher number density will have space charge oscillation of larger frequency, and vice-versa.

## 1.3 CRITERIA FOR PLASMAS TO EXIST

For an ionized gas to be called plasma, certain conditions must be satisfied out of which two conditions are already discussed above. The third condition is related to collisions in the system. Weakly ionized gas in a jet exhaust, for example, does not qualify as plasma because the charged particles collide so frequently with the neutral atoms that their motion is controlled by ordinary hydrodynamic forces rather than by electromagnetic forces. If ω is the frequency of typical plasma oscillations and $\tau$ is the mean time between collisions with neutral atoms, then the condition $\omega\tau > 1$ must be satisfied for the ionized gas to behave like plasma. In addition to the condition $L \gg \lambda_D$, "collective behavior" demands the conditions $N_D \gg 1$ and $\omega\tau > 1$ also to be satisfied, for the plasma to exist.

## 1.4 BASIC PLASMA PHENOMENA

### 1.4.1 Fluid description

In fluid description, plasma is considered to be composed of two or more interpenetrating fluids (which correspond to the electrons, ions and neutrals respectively) which can interact via self generated electric and magnetic fields due to their thermal motion. The fluid picture of plasma is most appropriate when the plasma is somewhat collisional (where the electrons and ions separately relax to a local thermodynamic equilibrium on a short time compared to the time in which substantial changes in the plasma occur and in regions small compared to the physical length of the plasma). The self consistent equation of motion for such a system is given by,

$$m_j n_j \left(\frac{dv_j}{dt} + v_j . \nabla v_j\right) = -q_j[E + (v_j \times B)] - \nabla . P_j - m_j n_j \vartheta_{jk}(v_j - v_k) \quad (1.10)$$

where $m_j$ and $q_j$ are respectively the mass and charge of the species j; $P_j$, $n_j$ and $v_j$ are the fluid pressure, density and velocity, respectively, and $\vartheta_{jk}$ is the average frequency of collisions with species *k*. Pressure $P_j = \gamma_j n_j k T_j$, where *k* is the Boltzmann constant, $T_j$ is the temperature of the fluid in Kelvin, and $\gamma_j$ is the adiabatic constant (equal to 3 for one dimensional, 2 for two dimensional and 5/3 for three-dimensional plasmas respectively). Plasma temperature ultimately depends on $\gamma$ for each species. The dynamics follows conservation of net flux over a volume V, described by the continuity equation,

$$\frac{\partial n_j}{\partial t} + \nabla . (n_j v_j) = 0 \quad (1.11)$$

Along with the Maxwell's equations,

$$\epsilon_0 \nabla . E = \sigma \quad (1.12)$$

$$\nabla \times E = \dot{B} \tag{1.13}$$

$$\nabla \cdot B = 0 \tag{1.14}$$

$$\nabla \times H = J + \epsilon_0 \dot{E} \tag{1.15}$$

Where $\sigma = n_i q_i + n_e q_e$ and $J = n_i q_i v_i + n_e q_e v_e$. Movement of charged particles creates electric and magnetic fields and the particles thus move in the generated field. Hence all the above equations must be solved simultaneously to get a self-consistent solution for the dynamics of plasma using the fluid theory.

Fluid dynamics is applicable to plasma in a strong magnetic field perpendicular to the fluid velocity which allows large collision frequency complementing the Maxwellian distribution of velocity of species. But this theory will not be sensitive for any deviation from Maxwellian distribution, i.e., when collision is less, so that the temperature cannot be described accurately. In general, a deviation from Maxwellian distribution is always possible and the solution for such a situation is obtained using a kinetic description.

### 1.4.2 Kinetic description

We can see that the collision time in some cases is so long that we can ignore the collisions in the plasma. For such plasmas, referred to as collisionless plasmas, the fluid theory does not give an appropriate theoretical picture. In such cases, the double adiabatic theory is used which can help to overcome the limitations of the fluid description to an extent [15-18]. The best way to define such a system with deviations from Maxwellian distribution is to use the distribution function $f_j$ (*r, v, t*) for each species (which is a function of seven independent scalar variables) governed by the Boltzmann equation,

$$\frac{\partial f}{\partial t} + v \cdot \nabla f + \frac{q}{m}(E + v \times B) \cdot \frac{\partial f}{\partial t} = \left[\frac{\partial f}{\partial t}\right]_c \tag{1.16}$$

where $\left[\frac{\partial f}{\partial t}\right]_c$ represents collisions. "Collisionless" plasmas are hot enough so that this term can be set to zero, in which case eqn. (1.16) is called the Vlasov equation [15-18].

Though plasmas can often be treated as fluids, plasma physics differs from hydrodynamics because of the non-Maxwellian distributions that require the kinetic treatment described above. It differs from electromagnetism as well, because the details of the dielectric tensor $\epsilon$ are treated from the particulate point of view. Indeed, the complex motions of charged particles in electric and magnetic fields support a rich variety of wave phenomena that do not occur in ordinary fluids or dielectrics. An important fact to be noted is that many plasmas are collisionless to a very good approximation, especially those encountered in astrophysics and space plasma physics.

## 1.5  TYPES OF PLASMA

Based on properties such as temperature, degree of ionization etc., plasmas can be classified into different types as follows.

### 1.5.1  Cold plasmas

In a low pressure gas discharge the degree of ionization is ~ $10^{-4}$ so that the gas consists mostly of neutral species. Collisions between electrons and gas molecules are not so frequent and a non-thermal equilibrium exists between the energies of electrons and gas molecules. Therefore the gas molecules are at room temperature even though the electronic temperature is very high. This type of plasmas where $T_e > T_i > T_g$, ($T_e$, $T_i$ and $T_g$ are the temperatures of electron, ion and gas molecules respectively) are called cold plasmas.

### 1.5.2  Hot plasma

In high pressure gas discharges collisions between electrons and gas molecules occur so frequently that the temperatures of electrons and gas molecules are approximately equal, and they are in thermal equilibrium. Hence the system approaches Local thermodynamic equilibrium (LTE), and such plasmas are known as hot plasmas. Hot plasmas are also called thermal plasmas.

### 1.5.3 Collisional plasma and collisionless plasma

Collisions between charged particles in a plasma differ fundamentally from those between molecules in a neutral gas because of the long range Coulomb forces present in the system. Binary collision processes (i.e. number of collisions is relatively very low) can be defined only for weakly coupled plasmas. These binary collision processes are modified by collective effects when the many-particle process of Debye shielding happens in the system in a crucial manner. For very large values of the plasma parameter ($N_D$), we can speak of binary collisions and a collision frequency $\nu_{ss\prime}$, which measures the rate at which particles of species $s$ are scattered by those of species $s'$,

$$\nu_s \cong \sum_{s\prime} \nu_{ss\prime} \qquad (1.17)$$

and for an electron, this can be written as

$$\nu_e \cong \left(\frac{m_i}{m_e}\right)^{1/2} \nu_i \qquad (1.18)$$

The collision frequency $\nu$ is a measure of the frequency with which the particle trajectory undergoes major angular change due to Coulomb interactions with the surrounding particles. Collision frequency is sometimes referred to as the "90° scattering rate" since it is the inverse of the typical time needed for enough number of collisions to occur so that the particle trajectory is deviated through 90°.

It is important to define the mean-free-path which measures the typical distance that a particle travels between collisions (*i.e.*, 90°scattering events), which is given by

$$\lambda_{mfp} = \frac{v_t}{\nu} \tag{1.19}$$

If $\lambda_{mfp} \ll L$; ,where L is the observation length-scale, then the plasma is referred to as collision-dominated or collisional plasma. The opposite limit of large mean-free-path corresponds to collisionless plasma.

The typical magnitude of the collision frequency is $\nu \sim \frac{\ln N_D}{N_D} \omega_p$; where $\omega_p$ is the plasma oscillation frequency. The condition $\nu \ll \omega_p$ represents weakly coupled plasma, which follows that the collisions do not interfere seriously with the plasma oscillations in such systems. If $\nu \gg \omega_p$, then it represents a strongly coupled plasma where collisions effectively prevent plasma oscillations in the system. This is in accordance with the basic picture of strongly coupled plasma which can be modeled as a system dominated by coulomb interactions that does not exhibit conventional plasma dynamics. Collision frequency can be written in terms of the number density, the plasma parameter, and the temperature of the plasma, as:

$$\nu \approx \frac{e^4 \ln N_D}{4\pi\varepsilon_0^2 m^{1/2}} \frac{n_e}{T_e^{3/2}} \tag{1.20}$$

Thus it is clear that diffuse and high temperature plasmas tend to be collisionless whereas dense and low temperature plasmas are more likely to be collisional in nature [19].

## 1.6 METHODS OF PLASMA PRODUCTION

With few exceptions like lightning or auroras, most of the plasmas on the earth are man-made. There are various methods by which plasma can be produced in a laboratory, and these include arc discharge, glow discharge, Radio Frequency (RF) discharge, heating with particle beams, heating by lasers, ohmic heating, and Pinch devices.

### 1.6.1 Arc discharge

Electrical discharge that happens at high pressure is referred to as arc discharge, where a high voltage is applied between two electrodes kept in a tube filled with gas, at a pressure of hundreds of Torr, generating plasma. The theory of ionization for this kind of system was explained by Townsend who assumed the following: ionizing electron starts off between two collisions with zero velocity in the direction of the field, it loses all its gained energy between two collisions, and the probability of ionization by electron-atom collision is unity as soon as the kinetic energy of the electron at the time of impact is equal or greater than the ionization energy of the atom. Some variants of the arc discharge are brush discharge, spark discharge and corona discharge.

### 1.6.2 Glow discharge

Electrical discharge produced in gases at low pressures is referred to as glow discharge. A fixed potential is applied to the tube filled with gas at atmospheric pressure and then the tube is gradually evacuated. When this is done, initially at atmospheric pressure no discharge is observed. Irregular streaks with cracking noise appear in the tube at pressures around 130 mbar, which result from the ionization caused by stray electrons that have acquired sufficient energy, thereby starting the process of gas breakdown. Excitation of gas atoms or molecules by the accelerating electrons creates a luminous column or a positive column in the tube which is observed around 10 mbar. Cathode glow appears at 3-4 mbar pressure, which is the result of electrons produced by secondary emissions by the ions hitting the cathode which ionize the surrounding gas. On further acceleration, they gain sufficient energy giving rise to what is known as positive column. At around 1 Torr pressure, positive luminous column shortens, Faraday's Dark Space (FDS) extends and the cathode glow detaches itself from the cathode creating Crooke's Dark Space (CDS). Further, at around a pressure of 0.1 Torr, CDS increases in length and the positive column breaks into striations. CDS fills the entire tube and the tube walls start glowing at still lower pressures, and thereafter, at very low pressures, the whole tube starts conducting. There is not much difference between glow discharge by DC, and AC up to a frequency of $10^5$ to $10^6$ Hertz. Beyond $10^6$ Hertz the regime of RF discharges starts which have entirely different characteristics.

### 1.6.3 RF discharge

There are two types of discharges that come under this category, namely, E-Type and H-Type discharges. To create plasma using an E-Type discharge, an oscillating electric field is applied across two electrodes, whereas in the case of plasma production using an H-Type discharge, an oscillating electric field is produced by the oscillating magnetic field due to a solenoidal current round the gas tube. In the case of plasma produced by an RF discharge the losses associated with the electrodes are minimal, and since the electrodes in these systems are located outside the tube, highly reducing the contamination, RF discharges are used to produce high purity spectroscopic sources. Since the relatively massive ions are practically immobile on the RF time scale, a high concentration of ions with very low energy (velocity) spread is found in the central region of the RF discharge, making it a useful ion source in accelerators like cyclotrons.

### 1.6.4 Using particle beams

This is yet another method to produce plasmas which use either Relativistic Electron Beams (REB) or Heavy Ion Beams (HIB). For plasmas produced using REB, a beam of high energy (MeV) is produced using a high voltage discharge in a vacuum diode, using a Marx Bank. Beam focusing is done using electrostatic or magnetic lenses. REB produces good quality plasmas, but focussability problems due to Columbic repulsion are a major drawback. HIB can also be used for producing plasma provided vacuum conditions are met. Plasmas produced using REB or HIB have temperatures and densities similar to those produced by LPP.

### 1.6.5 Using intense laser beams

This class of plasma is generally referred to as Laser Produced Plasma (LPP), and can be usually produced using laser pulses of nanoseconds or lower pulse duration. It is required to focus the laser beam using a lens to obtain a high intensity at a particular spot on a target. Such a focused beam can have intensity varying between $10^{12}$ W/cm$^2$ to $10^{18}$ W/cm$^2$. The typical parameters, i.e., number density and temperature, will be around $10^{21}$ cm$^{-3}$ and 100eV to few KeV, respectively.

### 1.6.6 Ohmic Heating

In ohmic heating, an AC signal is passed through the primary of a transformer and a load R is connected to the secondary. In such a configuration the load gets heated which can initiate the production of plasma. This method of plasma production is used in a Tokamak.

## 1.7 RADIATIONS FROM PLASMAS

Radiations are emitted by plasmas due to the various processes occurring in the plasma. Emissions observed from various experiments are discussed below.

### 1.7.1 Black Body Emission

Plasmas in Complete Thermal Equilibrium (CTE) emit black body radiation. The plasma is opaque to radiation and the emission happens from the surface of the plasma. The wavelength spectrum peaks at $\lambda_{max}(\text{Å}) = \frac{2500}{T_e(eV)}$ and the frequency peaks at $\omega_{max} = 3.9 \times 10^{11} T\ (in\ K) rad/s$. The total emission peak power is found to be proportional to $T^4$.

### 1.7.2 Bremsstrahlung Radiation

"Bremsstrahlung" is derived from the German words 'Bremse' (brake) and 'Strahlung' (radiation), meaning 'braking radiation'. Radiation is emitted when a Coulomb field or another charged particle deflects a charged particle. Since the electron is free before and after the emission, the radiation is also referred to as free-free radiation. In plasmas, Bremsstrahlung radiation generally occurs due to the deceleration of electrons by the field of ions. Plasma is transparent to the bremsstrahlung radiation and hence the radiation is emitted by the whole body. Peak of the wavelength spectrum is given by $\lambda_{max}(\text{Å}) = \frac{6200}{T_e(eV)}$, and the spectrum of Bremsstrahlung radiation is found to be monotonously decreasing with frequency. The power emitted in this kind of radiation is found to be $\frac{\sum_k q_k^2 \ddot{a}_k^2}{6\pi\varepsilon_0 c^3}$, where $q_k$ is the charge undergoing an acceleration $\ddot{a}_k$. The total power of the emitted radiation is proportional to $T_e^{1/2}$.

### 1.7.3 Recombination Radiation

Recombination radiation is emitted when a free electron combines into the bound state of an ion. If $\varepsilon_0$ is the energy of the free electron and $\chi_i$ is the ionization potential of the i$^{th}$ level of ion, then the energy of the emitted photon is $h\nu = \varepsilon_0 + \chi_{i,z}$, where $\varepsilon_0 = \frac{1}{2}mv^2$. Since $\varepsilon_0$ is continuously varying the recombination radiation spectrum also is continuous. As the electron is free before recombination and bound afterwards, the radiation is also referred to as free-bound radiation. For low atomic number and high $T_e$ Bremsstrahlung radiation is predominant over recombination radiation. But for high atomic number and low $T_e$, recombination radiation predominates Bremsstrahlung radiation.

### 1.7.4 Line Radiation

Emissions due to the transition of an electron between two bound states are referred to as line radiation. Total power available in the line emission is proportional to $Z^6$. Those transitions to the ground state of the ion are the strongest and they are referred to as resonance transitions. They are also called bound-bound radiations. The selection rules are as follows:

$\Delta L = \pm 1$ (parity must change)

$\Delta j = 0, \pm 1$ ( $j_1 + j_2 \geq 1$)

ΔS=0 (multiplicity should be the same).

Transitions between states of different multiplicity also can take place, and these are referred to as inter combination transitions.

### 1.7.5 Cyclotron radiation

In a strong magnetic field (B), electrons will spiral around the magnetic lines of force (B) with a frequency $\omega_C = \frac{Be}{m}$. When $\omega_C > \omega_P$, radiation will be emitted by the plasma. This process is also referred to as Magnetic Bremsstrahlung and the frequency is approximately equal to $1.76 \times 10^{11}$B T rad/s. Ions also emit these radiations, but since ions are heavier than electrons, the radiation will be feeble. The radiation is circularly polarized in the direction of the magnetic field, linearly polarized in the plane perpendicular to the field, and elliptically polarized in other directions. Unlike in other radiations discussed earlier, cyclotron radiation is anisotropic in nature.

### 1.8 PLASMA DIAGNOSTICS METHODS

Number density ($n_e$) and Temperature ($T_e$) are the two basic parameters required to characterize a plasma. Methods employed to estimate these parameters are referred to as plasma diagnostic techniques, which can be performed either using the radiation emitted by the plasma or by using external radiation as a probe. One cannot simultaneously measure $n_e$ and $T_e$ using these techniques, and the presence of magnetic fields can complicate the measurements. We consider only non-magnetized plasmas such as laser produced plasma in the following discussion.

### 1.8.1 Plasma Spectroscopy

Spectroscopic techniques provide a variety of methods to diagnose a wide range of plasmas. Absorption spectroscopy, emission spectroscopy, spectroscopy of scattered radiation, time-resolved spectroscopy, and time of flight spectroscopy are some of the commonly used techniques for plasma diagnostics. Plasma spectroscopy is easy to implement in any plasma system, may it be high or low temperature/density plasma. It can also be employed in magnetically confined plasmas like the tokamaks. The most commonly used spectroscopic technique is emission spectroscopy which measures the electromagnetic radiation (which is regarded as the environment of the radiating species) emitted by the plasma (related to the plasma parameters or characteristic parameters of the radiating atoms). It is an easy and straightforward approach to diagnose the plasma but since the emitted radiation is measured, the information obtained will be integrated over the line of sight of the measurement. Spectroscopic techniques are particularly useful to establish the relationship between emitted radiations and plasma parameters such as the number density and temperature [20].

### 1.8.2 Langmuir probe

Langmuir probe, probably the simplest among electric probe diagnostics to measure plasma, consists of a wire inserted into the plasma that measures the current at various applied voltages. The technique is an intrusive one and the wire must be carefully designed so that it does not interfere with the plasma nor the plasma is destroyed by it. The interpretation of the measured current-voltage (I-V) curves is very difficult and has spawned a large literature of theoretical papers. These electric probes measure the local plasma parameters using a stationary or slow time varying electric (and/or magnetic) field to emit or collect charged particles from the plasma. One gets a respective current according to the bias voltage provided to the probe, from which the plasma parameters (like $T_e$ and $n_e$) can be deduced. The electric field between the plasma and the metallic probe helps the collection of charges which is dependent on the probe size $r_p$ and the thickness (or spatial extension) of the plasma sheath attached to the collecting surface, which is related to $\lambda_D$. The electron and ion Larmour radii introduce additional lengths in unmagnetized plasma whereas the mean free path between collisions plays an important role in collisional and weakly ionized plasmas. Based on the type of plasma and the nature of the measuring systems there will be orders of magnitude differences in the measurements done, and this disparity has lead to the lack of a unified explanation. For instance, interference of the probe with tokamak plasmas can lead to perturbations which serve as source of impurities to the plasma, which is a major drawback of the technique. These measurement techniques constitute an active field of research and are particularly well suited for low density cold plasmas, low pressure electric discharges, and ionospheric and space plasmas. The most widespread use of Langmuir probes at present is in the semiconductor industry, where radiofrequency (RF) sources are used to produce plasmas for etching and deposition. These partially ionized plasmas require special techniques in probe construction and theory [21-23].

### 1.8.3 Interferometry

Interferometry is another important diagnostic tool which has an advantage that it does not require local thermodynamic equilibrium (LTE) to be satisfied for the measurements. The principles of basic laser interferometry are employed and the analysis of fringe structures allows the study of plasma parameters [24-25]. In an interferometric set up, a laser pulse/beam is initially split into two and then allowed to recombine later in space and time, to form the fringes. The expanding plasma is kept along one of the arms of the interferometer whereas the probe beam is left either parallel to the target or orthogonal to the plume expansion direction. The underlying principle used in this method is that the refractive index profile of the expanding laser produced plasma is directly proportional to the number density of free electrons in the plasma. The fringe amplitude of the combined probe and reference beams determines the variation in refractive index, which is in turn related to the change in phase shift caused in the probe beam by the plasma. A two dimensional map of the electron density is obtained after performing interferometric analysis of the plasma. Various configurations used for this purpose are the Michelson Interferometer, Mach-Zehender interferometer, Nomarski interferometer etc. Even though the technique is highly useful for analysing plasmas, there are several drawbacks for this method too. For example, fringe visibility, which is an important factor for interferometric techniques, is either lost or becomes diminished when the plasma density approaches a fraction of a percent of the critical density. Similarly, refraction and opacity effects can significantly limit the usefulness of interferometric diagnostics. This leads to improper measurements with less accuracy which makes these kinds of techniques difficult to implement, especially in the case of highly dense plasma. The attenuation of the probe beam also may lead to adverse effects in measurements. The resolution of interferometric methods depends on the duration of the probe pulse as well as the gating time of the detector.

### 1.8.4 Thomson scattering

Thomson scattering is a technique that has proved its superiority in the measurement of electron temperature ($T_e$) and number density ($n_e$) in fusion plasma devices like tokamaks [26].The method is direct, active, accurate and unambiguous, and is widely used for localised and simultaneous measurements of the above parameters. The technique, which is used to measure the number density of fusion plasma, provides good spatial resolution. As long as the power of the probe beam which produces electron oscillation is kept low, the measurement will not disturb the plasma. In the Thomson scattering experiment light scattered by the plasma electrons is used for the measurements, and the plasma electron temperature is measured from the Doppler shifted scattered spectrum while the density is measured from the total scattered intensity. The scattered power is proportional to $\sigma_T n_e L$, where *L* is the length of the plasma traversed by the light beam. This power is too small in scattering experiments and stray light is a serious issue in such experiments. In practice, the fraction of incident electrons will not be more than $10^{-15}$, taking into consideration the quantum efficiency of the detector and other experimental conditions. Hence, while doing diagnostics using Thomson scattering one should be extremely careful with the experimental conditions and collection of the signals [27].

### 1.8.5 Shadowgraphy

Shadowgraphy can be considered as an alternative technique to derive the temperature profile, longitudinal expansion velocities etc. of the plasma species [28]. It is the plasma diagnostic technique which yields us the refractive index profile (shadowgraphy is dependent on the second derivative $\left(\frac{\partial^2 n}{\partial y^2}\right)$ of the refractive index, which gives us an idea about the contrast ratio of the technique). Refractive index is measured using the beam deflection angle measurement, and the shadowgraphy images are recorded using CCD cameras connected to the PC which are analyzed using image acquisition software. Expansion velocities are calculated from the original target surface position and the time taken for recording the shadowgram. The radiation hydrodynamic model along with the expansion velocities yields the temperature gradient of the plasma.

**1.9 REVIEW OF LASER PRODUCED PLASMAS**

LPP can be generated by focusing laser pulses onto the target using a lens once the laser intensity is greater than the ablation threshold of the material. Material ablation, plasma generation and expansion happen quickly after laser irradiation, in which the nature of interaction and ablation are dependent on various laser parameters and surrounding atmosphere. These studies have become one of the promising areas of research for understanding fundamental and application oriented laser produced plasma physics. A brief review of laser produced plasmas discussing its origin, historical developments and the present status are detailed below.

LPP using the 'giant' laser pulse was first reported in 1964, and plasma properties such as absorption of laser energy by the plasma and its magnetic properties were studied [29]. Soon after this, gas breakdown at optical frequencies [30-31] and scattering of electromagnetic radiation by LPPs and the corresponding Doppler shift were reported [32]. A self generated magnetic field associated with a laser produced spark was also reported in 1966 [33]. Thereafter the field of LPP became important for various applications including the production of vacuum ultraviolet radiation, production of X-rays, higher harmonic generation, particle acceleration etc.

Electron temperature and number density, the important parameters of the plasma, vary with time and space as the LPP is highly transient in nature. There are several methods available to investigate these parameters which include optical emission spectroscopy (OES), Optical time of flight (OTOF) measurements, Interferometry, Thomson scattering, Polarimetry, shadowgraphy, Langmuir probe etc. OES is one of the best nondestructive methods used for the investigation of temperature and number density but are not applicable at the earlier stages of plasma expansion due to the large continuum emission. Interferometry cannot be applied when the number density is too high or too low, and in this case, selection of the probe wavelength is also critical. Langmuir probe is a good and effective tool to understand the space averaged density but it is not possible to measure the density at the plasma core because the probe disturbs the plasma. Out of the above mentioned techniques Thomson scattering is the best method, but setting up the experiment and extracting signal from the noise is very difficult. Therefore for making accurate measurements, any of the above two methods are employed to understand LPP parameters. Imaging of the plasma using ICCD is also a good tool to understand the expansion dynamics of the LPP.

Laser parameters such as the fluence, irradiation spot size, and wavelength largely affect the properties of LPP. Variation of plasma properties with respect to the above mentioned parameters using nanosecond (ns) pulses to understand the fundamental physics of laser ablation was largely studied in the last two decades. It was shown that the emission intensity, temperature and number density of the plasma increases with laser wavelength and fluence whereas these parameters decreases with time and axial expansion direction. [34-40]. It is also reported that ambient conditions (both the nature and pressure of ambient gas) play an important role in the expansion dynamics of LPP, which is also an active area of ongoing research. The generation and expansion of LPP is different for various pulse durations, and this is also an active area of research for both fundamental (mostly related to dynamics of the plasma) and application oriented research (Pulsed Laser Deposition (PLD), nanoparticle and nano-cluster generation etc.). ns LPP has been widely studied but research related to femtosecond (fs) LPP is relatively scarce, and therefore the emission dynamics of an expanding fs LPP is another major area of ongoing research work.

Apart from the above, laser wake field acceleration, where particle acceleration is achieved via laser produced plasma, is an upcoming research field. Similarly, acceleration of trapped particles using Bessel beam plasmas is a forthcoming attraction of the field. Moreover, theoretical work and simulation of LPP under various conditions is an equally fast growing area of plasma physics.

## 1.10 CONCLUSION

Plasma, a distinct state of matter, is introduced and briefly described in this chapter, giving an insight into various processes that occur in the plasma and various techniques used to diagnose the plasma, whatever be its nature of origin. The underlying motivation, which is an important part of any research work, is explained in detail along with a brief review of the previous works done in the field. A brief mention of the research possible in the field is also given in the chapter.

# CHAPTER 2
# INSTRUMENTATION

*Instrumentation required to perform optical emission spectroscopy and optical time of flight measurements and imaging of the plasma plume, generated by nanosecond/femtosecond laser pulses, is discussed in this chapter. The laser system, pulse characterization techniques, synchronization circuits, and instruments used to detect the emission spectra are explained in detail.*

## 2.1 INTRODUCTION

Laser-produced plasmas (LPP) are formed by focusing a high intensity laser pulse onto a target (in any form of matter). Such plasmas expand rapidly, and the plasma parameters vary temporally and spatially, imparting them a transient nature. As soon as the laser interacts with the target, plasma expands isothermally within the duration of the incident laser pulse, which turns out to be adiabatic afterwards. High power laser systems are necessary to generate LPP, and to diagnose the transient plasma generated, detectors need to be synchronized with the laser pulse using fast electronic circuits. Details of instruments used such as the laser systems, vacuum chambers and detectors are described in this chapter.

Optical emission spectroscopy (OES), Optical Time of Flight (OTOF), and imaging using CCD cameras are the non-destructive measurement techniques employed in the present work to yield plasma parameters such as the temperature and number density. The plasma emission is collected using a lens system, which is fed into a monochromator and Charge Coupled Device (CCD) configuration for OES studies, and a monochromator, Photo-Multiplier Tube (PMT) and digital storage oscilloscope (DSO) combination for OTOF studies. The two dimensional expansion of the plasma plume is simultaneously imaged using an intensified CCD (ICCD) camera. The recorded signals are digitally saved for further analysis. For the generation and detection of the plasma, mutual synchronization of the instruments is necessary so that the timings of the different devices including CCD, PMT and ICCD are under control. For nanosecond (ns) and femtosecond (fs) laser plasma generation we have used a Q-Switched Nd: YAG Laser and a mode-locked Ti: Sapphire laser respectively.

## 2.2 LASER SYSTEM

The laser system is an integral part of the experimental setup, which delivers optical pulses as short flashes of light. The shortest optical pulse generated in a commercial laser system so far has a duration of about 5 fs ($5 \times 10^{-15}$ s), corresponding to a few optical cycles (few-cycle pulses) at the operating wavelength of 800 nm. Due to the short pulse durations and the potential for tight focusing, ultrafast optical pulses can be used to obtain extremely high intensities even at moderate pulse energies. Therefore, amplified fs pulses are essential for high-intensity physics, for studying phenomena such as multi-photon ionization and higher harmonic generation, and for generating shorter pulses with attosecond durations. A 100 fs Ti: Sapphire laser and a 5 ns ($5\times10^{-9}$ s) Nd: YAG laser have been used for the fs and ns LPP studies respectively, reported in the present work. Laser pulses of 100 fs width (FWHM) are initially generated in a mode-locked Ti: Sapphire oscillator, which are then amplified by a chirped pulse Ti: Sapphire amplifier (CPA) to the required pulse energies which are orders of magnitude higher in comparison.

## 2.2.1 The ultrafast (femtosecond) oscillator

The Ti: Sapphire laser falls into the category of tunable lasers, with emission in the red and near-infrared regions between 700 and 1000 nanometers. The femtosecond oscillator (*Tsunami, Spectra Physics*, USA) consists of a folded cavity with an acousto-optic modulator (AOM) and active feedback for generating femtosecond pulses by the technique of regenerative mode-locking. A one centimeter long Ti: Sapphire crystal rod pumped by a diode-pumped Nd:YVO$_4$ laser (*Millennia Pro S, Spectra Physics*) is the active medium in the system. Titanium ions (Ti$^{3+}$) in the Ti: Sapphire crystal is responsible for the laser action. Transitions corresponding to absorption occur over a broad range of wavelengths from 400 nm to 600 nm. Even though the fluorescence band starts from 600 nm, lasing action occurs only for wavelengths beyond 670 nm because the long wavelength side of the absorption band overlaps the short wavelength side of the fluorescence spectrum. Mode-locking of the ultrafast optical cavity is achieved through an acousto-optic Modulator (AOM) in a regenerative mode-locking configuration, which uses an RF signal to drive the AOM. The ultrafast oscillator produces pulses of 100 fs duration (approximately) at a repetition rate of 80 MHz. Figure 2.1 shows the ultrafast oscillator along with the Nd:YVO$_4$ pump laser.

**2.2.2   Pump laser for the ultrafast oscillator**

A diode-pumped solid state (DPSS) laser (*Millennia Pro-S*, *Spectra Physics, USA*) pumps the ultrafast Ti: Sapphire oscillator. The millennia Pro S- Series laser is a frequency doubled, all solid-state Nd: $YVO_4$ continuous wave laser capable of producing 5.2 W output power at 532 nm. The Nd: $YVO_4$ crystal is pumped by a diode laser (809 nm, 40 W) coupled to the crystal via an optical fiber module. The fiber coupling transforms the emitted astigmatic light of the diode laser into a round beam of exceptional brightness which is suitable for end-pumping the Ti: Sapphire oscillator. The T80 power supply module houses the 40 W diode laser bar and the attached fiber assembly. The power supply is air cooled and requires no water or external cooling connections. A compact recirculation chiller regulates the temperature of the Nd: $YVO_4$ crystal in the laser head. The fundamental emission from the Vanadate crystal is frequency-doubled using a Lithium Triobate (LBO) crystal to produce 532 nm giving an output with $TEM_{00}$ spatial intensity distribution and a nominal beam width of 2.3 mm. Since the Millennia Pro S is all solid- state, it is more reliable and easier to operate with lower optical noise and higher electrical to optical conversion efficiency, compared to an ion laser.

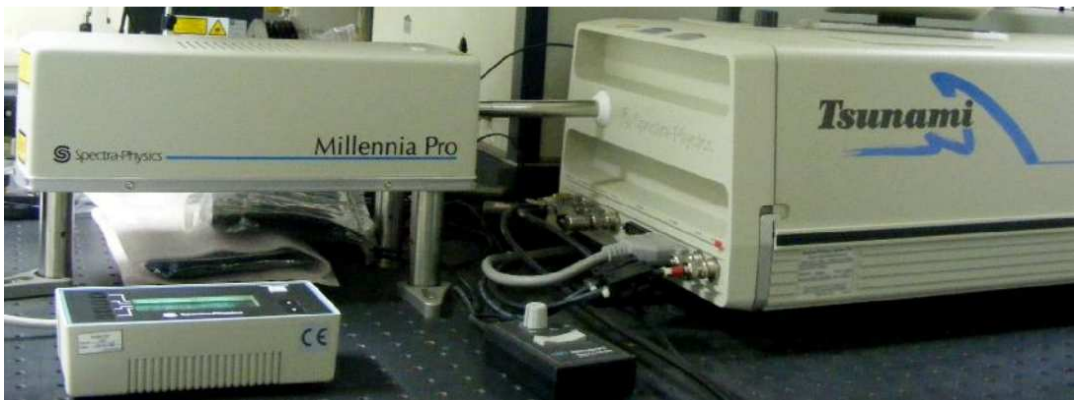

Figure 2.1. The ultrafast oscillator (*Tsunami*) and the *Millennia Pro* pump laser.

**2.2.3   Nd: YAG laser (Quanta Ray)**

Amplification by TSA requires a pump radiation which is provided by a frequency doubled, Q-switched Nd: YAG laser (Quanta Ray, Spectra Physics) operating at 10 Hz with a nominal pulsewidth of 7 ns. This flash lamp pumped laser is capable of producing pulses ~270 mJ energy at 532 nm (using a KDP crystal for frequency doubling). The fundamental 1064 nanometers output from the system is used for short pulse (i.e. ns) LPP generation in the present work. The maximum energy available from the instrument at 1064 nm is 150 mJ.

Output from the Nd: YAG laser and the ultrafast oscillator must be synchronized for the TSA to operate. Here, timings associated with the switching of Pockel cell in the CPA very critical. In order to ensure that only a single pulse enters the resonator from the 82 MHz train, the input Pockel cell must be switched at the same time, with respect to the mode locked pulse train. To achieve this, switching of the Pockel Cell is synchronized with the RF signal generated by the mode-locker of the ultrafast oscillator via a Synchronization and Delay Generator (SDG) unit. The Nd: YAG laser system can also work as an independent unit when needed.

**2.2.4   Chirped Pulse Amplifier (TSA-10)**

The energy of mode-locked single pulses produced by the ultrafast oscillator (*Tsunami*) is too low (~ 10 nano Joules), and it has to be amplified to the milli Joules level in order to obtain the high energy pulses required for intense laser field experiments. Direct amplification of ultrashort pulses is not possible as it will damage the amplifying medium itself (due to the high intensity of the 100 fs pulses), and therefore, the technique of Chirped Pulse Amplification (CPA) is used for amplification. In the CPA technique, the pulse is initially stretched in time by a large factor (typically 10,000) to reduce the peak power, amplified, and then compressed back to the original pulsewidth. Pulse stretching and compression are achieved by the use of diffraction gratings.

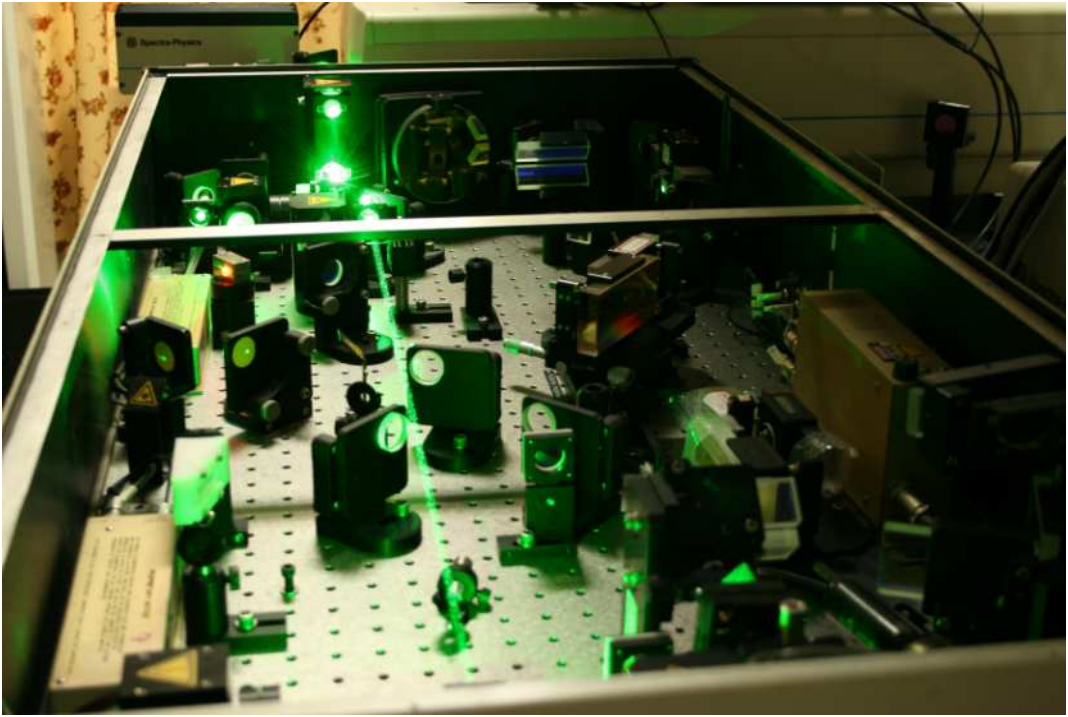

Figure 2.2. The Chirped Pulsed Amplifier (CPA) laser used for the experiments.

The Ti: Sapphire Amplifier (TSA 10, Spectra Physics) employed in the present work is shown in Fig 2.2. It is pumped by a pulsed Nd: YAG laser (Quanta Ray) operating at 10 Hz repetition rate. The 100 fs mode-locked pulse train at a repetition rate of 82 MHz from the ultrafast oscillator is sent to a pulse stretcher in the TSA-10 (which broadens the pulses to about 300 ps width), from where a single pulse is picked using an electro-optic modulator. This single pulse is seeded into a regenerative cavity which contains an optically pumped Ti: Sapphire crystal. The pulse oscillates in the cavity, gaining energy in each round trip as it traverses the crystal. After several round trips, the pulse gains sufficient energy and is reflected out of the cavity using another electro-optic switch. Typically an input pulse of few nano-joules energy can be amplified to over 2 mJ using a single Ti: Sapphire laser rod in the regenerative cavity. For further amplification by a factor of 4, the cavity dumped pulse is fed to a double-pass linear amplifier. The amplified pulse is then sent to a grating compressor, which compresses the pulse back to its original width (100 fs).

The TSA generates 100 fs pulses of 10 mJ pulse energy at a repetition rate of 10 Hz. Figure 2.3 shows the complete laser system used for plasma production in the laboratory.

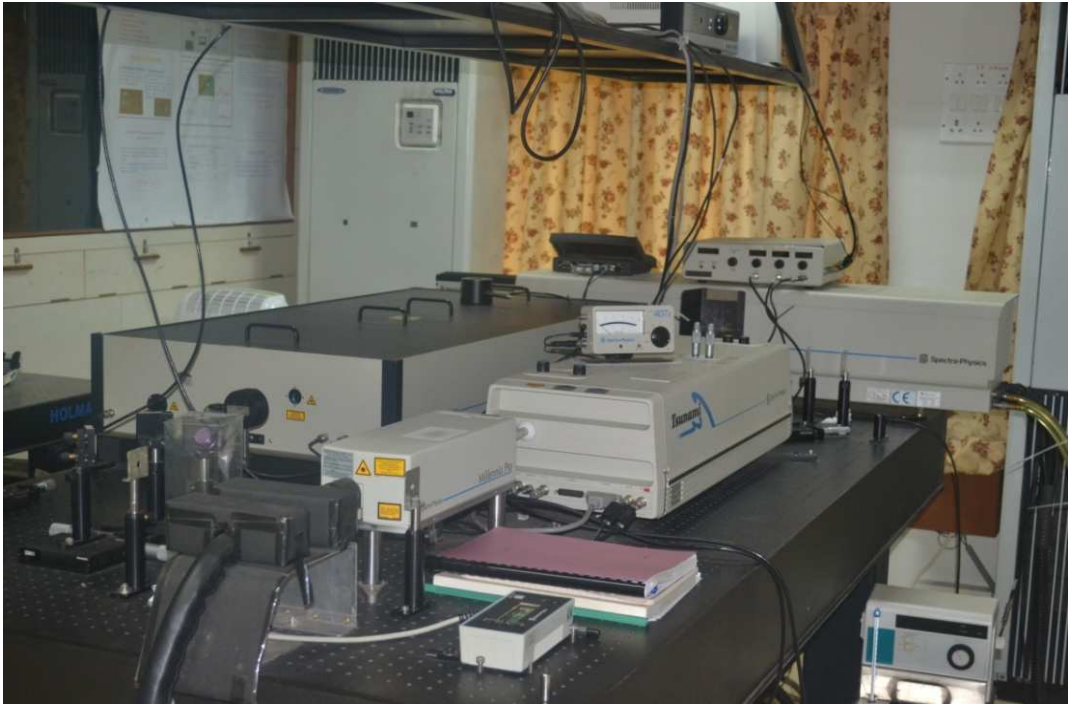

Figure 2.3. Ultrafast laser system used for plasma production. The system consists of a diode pumped Nd: $YVO_4$ laser, Ti: Sapphire femtosecond oscillator, Ti: Sapphire regenerative amplifier, and Q-Switched Nd: YAG laser.

## 2.3 VACUUM CHAMBER

The vacuum chamber used is 1 meter in diameter and 40 centimeter in height, and has a volume of approximately 300 liters. There are 16 ports in the main chamber and two ports in the small chamber. A high vacuum compatible target manipulator, fixed on top of the chamber (details given in section 2.6) is used for solid-target studies. Oil-free vacuum is used in our experiments because pumping using a rotary pump will leave a layer of oil on the target surfaces which will affect surface quality, especially on polished or coated surfaces. To obtain an oil-free Vacuum, the chamber is pumped using a turbo molecular pump backed by a dry pump. A Pfeiffer 2000 l/s turbo molecular pump (TPU 2101 PC) backed by a Pfeiffer 180 l/min (Unidry DBP 050-4) dry pump is used. The chamber pumping units are shown in Figure 2.4. The dry pump can evacuate the chamber up to $5\times10^{-2}$ Torr in about five to eight minutes, and a pirani gauge is used to measure the rough vacuum ranging from 750 Torr to $3.75\times10^{-4}$ Torr. Lower pressures from $7\times10^{-3}$ Torr to $1.5\times10^{-9}$ Torr are measured using a cold cathode gauge. The turbo molecular pump in conjunction with the dry pump can pump the chamber down to $1\times10^{-6}$ Torr in about twenty five minutes.

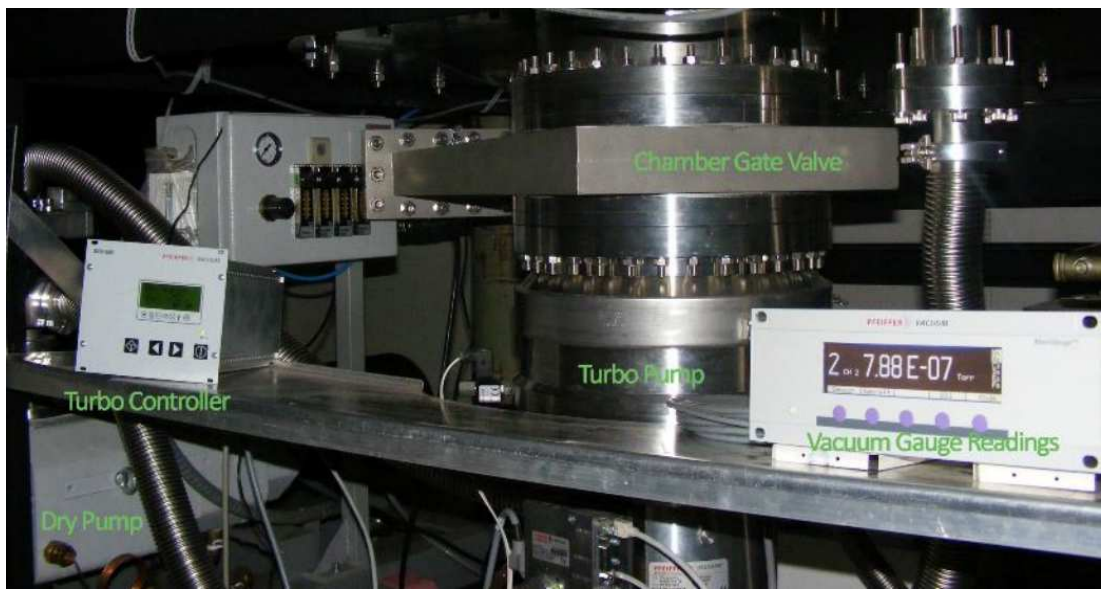

Figure 2.4. Chamber pumping unit [41].

**2.4 THE SOLID TARGET MANIPULATOR**

The target needs to be moved after every shot (or a limited number of shots at the same point) to produce plasma from a fresh sample surface. In the case of solid targets, a vacuum compatible translation stage is necessary for this purpose. We used an ordinary XY translation stage fixed outside the vacuum chamber, the translation of which is coupled via a Wilson seal to the vacuum compatible target holder placed inside the chamber. The target manipulator used can be moved in both the horizontal and vertical directions (50 mm in both directions) with a positional accuracy of 12.5 micrometers. A 100 CF port on top of the chamber is used to attach the solid target manipulator. The leak rate of the target manipulator is studied and found to be $2\times10^{-10}$ cc/s. The schematic of the target manipulator is shown in Figure 2.5.

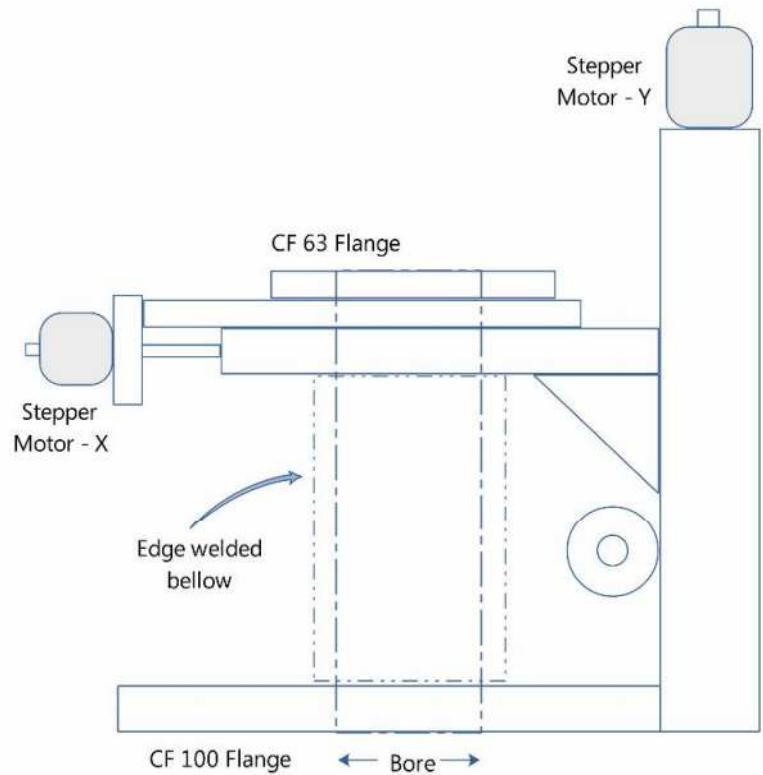

Figure 2.5. Schematic of the ultra high vacuum compatible solid target manipulator [41].

## 2.4.1 Target motion controller

The stepper motor of the target manipulator needs to be controlled and synchronized with the laser pulses for proper measurements. The motion can be controlled in the X or Y direction from one end to the other via an automated custom made electronic circuit, controlled by a program written in LabVIEW. Control signals from the PC to the target manipulator are sent via the LPT port (parallel port). Since the stepper motor requires relatively high currents to run, a buffer amplifier circuit is included between the computer and the motor. Moreover, to prevent PC damage, the back emf generated by the motor needs to be isolated using optocouplers. Details of this circuit are given elsewhere [41].

## 2.5 DETECTION SYSTEMS

Laser-induced plasma is a rich source of X-rays and ultraviolet, visible, and infrared radiations. Appropriate detectors are required to characterize the laser pulses and to measure the plasma radiation. A brief description of the detectors used in our measurements is given below.

### 2.5.1 Beam profiler

A beam profiler (*BeamLux II, Metrolux*, USA), which is the third generation laser beam diagnostic systems from MetroLux, was used to measure the spatial profile of the laser beam. It allows high resolution real-time monitoring and quantitative characterization of spatial light intensity distribution. It is well-suited for all pulsed and continuous wave light sources, and image sizes ranging from a few micrometers to a few millimeters can be investigated with an optimum resolution of 6.5 microns. It operates over a broad wavelength range ~ 320 -1100nm.

### 2.5.2 Laser energy and Power meters

Laser pulse energies and average powers were measured using (i) pyroelectric laser energy meter (RjP-735, *Laser Probe Inc.*), (ii) pyroelectric laser power meter and energy meter (*Laserstar, Ophir*), and (iii) Analog laser power meter with thermopile (*407 A, Spectra Physics*).

### 2.5.3 Spectrometer

Plasma emission collected using the imaging system (a combination of two lenses) is sent to a Czerny-Turner spectrometer (*iHR 320, Horiba Jobin Yvon*) where it is spatially dispersed with a resolution of about 0.06 nm at the exit slit, with a scan speed of 160nm/sec. The iHR 320 is an automated, triple grating spectrometer with a focal length of 320 mm (f/4.1 aperture). The on-axis triple grating turret, mounted on a drive, supports three gratings which can be individually selected via an *Origin* program written in the Visual BASIC platform. The spectral range is typically from 150nm to 1500nm. There is provision for fixing a compact CCD detector (*Synapse, Horiba Jobin Yvon*) as well as PMT (*R943-02, Hamamatsu*) to the iHR 320 spectrometer.

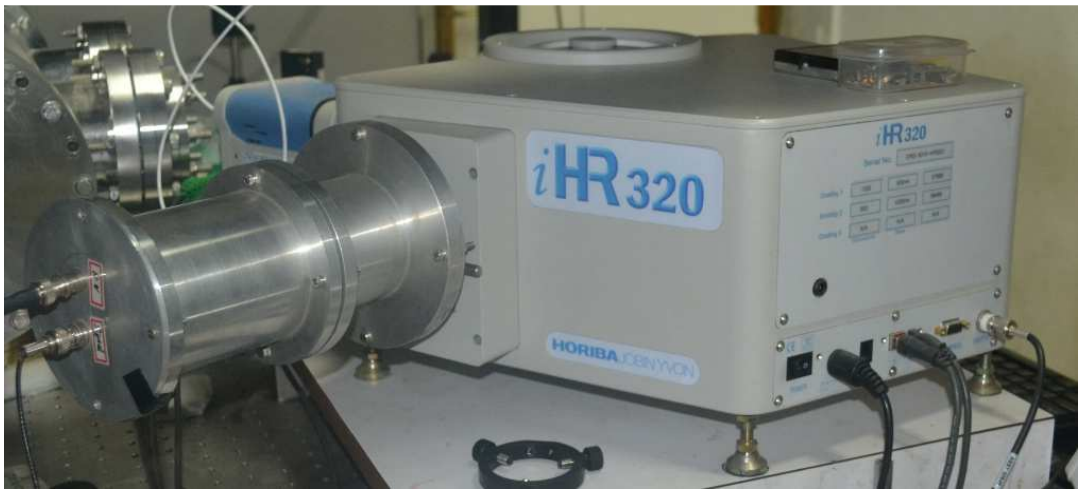

Figure 2.6. The iHR 320 spectrometer equipped with CCD and PMT detectors.

### 2.5.4 Digital Phosphor Oscilloscope

A fast digital phosphor oscilloscope (*DPO7354, Tektronix*) with four input channels is used for the acquisition of OTOF signals. It has an analog bandwidth of 3.5 GHz, and a real time sampling rate of 10 GS/s when all channels are active, which increases to 40GS/s on a single channel. Since the rise time of DPO7354 (110 picoseconds) is better than that of the PMT (2 ns) by an order of magnitude, the DPO7354 can capture and display OTOF signals accurately. The instrument works on a Windows Xp platform, and both ASCII data and screen images can be stored in digital format for further analysis.

**2.5.5   Intensified charge coupled device**

The intensified charge coupled device (ICCD) is an excellent imaging device capable of time gated data acquisition, commonly used for capturing images of LPP expansion for various time delays. From ICCD images taken at regular time delays, the nature of plume expansion can be directly investigated. The ICCD we used (*4-Picos, Stanford Research Systems Inc.*) has an 18 mm image intensifier with S-20 photocathode, and can start acquiring images 40 ns after the trigger delay from the trigger pulse. The minimum gate width is 200 picoseconds. The ICCD camera is controlled using the application software *4-spec line* and the communication is performed via RS 232 serial interface.

**2.6 CONCLUSION**

Brief descriptions of the instruments used for conducting the investigations reported in this thesis are given in this chapter. A detailed discussion of the LPP and OTOF experimental setups is given in the next chapter.

# CHAPTER 3
# EXPERIMENTAL TECHNIQUES FOR THE INVESTIGATION OF LASER PRODUCED PLASMA

*Laser-produced plasma (LPP) is transient in nature, and the parameters which unfold rapidly in time are crucially dependent on material properties and ambient gas pressure, in addition to laser parameters such as fluence, pulse width, wavelength, and beam spot size. Similar to other plasmas, key parameters like temperature and number density should be estimated to characterize the plasma. Plasma temperature can be estimated using techniques such as X-ray and visible spectroscopy, and number density can be estimated using the methods of emission spectroscopy, Langmuir probe, microwave and visible interferometry, and Thomson scattering. This chapter discusses the techniques used for characterizing laser produced plasmas, their pros and cons, and the other techniques used for the present experimental work.*

**3.1 INTRODUCTION**

After the invention of high power pulsed lasers, the study of intense laser-matter interaction has been a major area of research undertaken by different labs around the world. The advent of new technologies facilitated the production of sub-picosecond laser pulses, which can be amplified using the chirped pulse amplification technique to produce an intense source of monochromatic radiation suitable for producing plasmas. The large intensities associated with these lasers create a highly excited state of matter leading to nonlinear processes such as second and third harmonic generation [42-48], multiphoton ionization and over the barrier ionization [49], resulting in the formation of plasma. The plasma thus produced (Laser Produced Plasma (LPP)) is a rich radiation source emitting in the Visible, UV, and X-ray regions [7-8, 50]. Hence LPP is one of the currently relevant research areas in which lot of research is going on, that finds applications in varied fields such as surface cleaning [51], pulsed laser deposition [52-54], nanoparticle [11-14] and nanocluster production [55], particle acceleration [56-59] etc. The ablation yield, material removal, ionization and plasma plume expansion to the surroundings depend on parameters such as laser fluence [12, 60-61], pulse width [62-68], wavelength [38-40, 69-72], irradiation spot size [73] and material properties, along with the ambient pressure [74-89].

In the case of LPP generated by nanosecond (ns) lasers, the generated plasma will interact with the trailing part of the relatively long laser pulse and absorb energy from it. On the other hand, for LPP generated by femtosecond (fs) lasers, the laser pulse is too short to interact with the plasma since the electron-ion collision time and heat conduction time are in the order of picoseconds [65, 90]. The entire energy content of the pulse is deposited on the target resulting in efficient material ablation [62] and plasma formation. Unlike ns excitation, fs excitation minimizes the possibility of ejected particles interacting with the laser field preventing further ionization, resulting in the formation fast ions, atoms, nanoparticles and nanoclusters occurring at different time scales [12-14, 91-92]. This leads to a different emission dynamics for fs LPP compared to ns LPP. fs LPP is rich in excited neutrals at initial times whereas ns LPP are dominated by ions. This can be explained by the different ablation processes involved: while ns lasers tend to generate ions of lower charge states and produce neutral species via recombination, fs lasers generate ions of higher charge states emitting at much shorter wavelengths, followed by thermal vaporization of the target, producing mostly neutral species. It has been experimentally proven that the expansion is spherical for ns LPP both in vacuum and at atmospheric pressure, while the expansion of fs LPP has a narrower angular distribution. This narrower angular distribution is due to the strong self generated magnetic field created during the early expansion stage of fs LPP, which pinches the expanding plume and limits its expansion in a direction normal to the target [67]. Furthermore, higher plasma temperature and lower number density are characteristic of plasma generated by larger laser pulse durations at higher pressures [89, 93]. Moreover, continuum emission also increases for larger pulse duration.

Interaction of LPP with the ambient gas leads to an expansion involving several physical processes such as thermalization, attenuation, diffusion, electron-ion recombination, and shock wave formation, which is a fairly complex phenomenon. Plasma expands freely into vacuum or low background pressures, and the plume dynamics is characterized by the stronger interpenetration of the ambient gas with the plume, at higher ambient pressures. At still higher pressures, expansion dynamics of the laser produced plasma is fully determined by the properties and nature of the ambient gas. The increase in background pressures causes an increase in fluorescence emission from all the species due to enhanced collisions of the expansion front with the gas, and subsequent inter plume collisions. Shock wave formation and slowing of the plume due to spatial confinement of the plasma happen at higher pressures. The presence of ambient gas may lead to reactive scattering, thermalization of the plume, enhanced condensation etc. leading to cluster and nanoparticle formation. Thus the expansion and dynamics of the laser produced plasma is governed by the properties of the plasma and the surrounding gas. For example, ICCD measurements performed in a ns Aluminum LPP have shown the possibility of plume splitting at low pressures [94]. Plume expansion at lower pressures can be accurately described by a Monte Carlo Simulation, but this model is insufficient at moderate and high pressures [95-96]. At higher pressures a blast wave model fits the experimental results well in the earlier stages of plasma expansion [97-98] whereas a shock wave model predicts the plume length with reasonable accuracy. A three dimensional combined model for plasma plume formation and its expansion is available, which benefits from the advantages of both microscopic and continuous descriptions [99]. The two available descriptions for ablation by ultrashort pulses are the plasma-annealing model and the two-temperature model [100-102]. Dynamics of the ablated species can be theoretically modeled using a shock wave model at higher pressures [103], a blast wave model at moderate pressures, and molecular dynamics simulations [104] at extremely low pressures.

Even though material removal efficiency and laser plasma dynamics leading to the production of nanoparticles has been discussed in literature, a complete theoretical understanding of the expansion process is yet to be achieved. However, it is shown that at relatively lower laser fluences femtosecond laser produced plasmas consist of two different components, viz. atomic species and nanoparticles, in the earlier (~ 200 ns to 500 ns) and later (~ 100 µs) stages of plasma expansion, respectively. Fast atomic species can be observed once the fluence of the irradiating laser is increased well beyond the ablation threshold. Similarly, nanoparticle production can be maximized by optimizing the laser fluence. Furthermore, it is reported that ultrafast ablation of high-Z materials can lead to higher nanoparticle yield when compared to low Z materials [91]. The growth of nanoclusters from nanoparticles can be explained using diffusion limited aggregation at lower pressures [55, 105].

It has been shown experimentally that fs LPP plumes consist of fast ions, neutral atoms and nanoparticles, and that plume expansion depends on the ambient conditions and initial parameters. The plume emits continuum during the initial expansion stages due to the free-free and free-bound transitions, and acts as a black body radiator at later stages of expansion due to the formation of nanoparticles. Compared to ns excitation, fs LPP plume has a relatively complex structure, even though the processes leading to plume formation are simpler. Thus plasma diagnostics is essential for tailoring plasma properties for various applications. Different diagnostic techniques available for the characterization of LPP include optical emission spectroscopy, laser-induced fluorescence, Langmuir probing, mass spectrometry, photo-thermal beam deflection, Thomson scattering, interferometry etc. [106]. Conventional techniques employed for the observation and analysis of LPP include ion probe studies and time-resolved and time-gated emission spectroscopy. Temporal evolution of the transient plasma can be recorded using fast detectors such as photomultiplier tubes (PMT) and intensified charge coupled devices (ICCD) [107]. These studies are essential for understanding and modeling various processes in fundamental plasma physics and plasma hydrodynamics. Optical Emission Spectroscopy (OES) is a nondestructive technique used for the estimation of temperature and number density of LPP using radiative emission from the plasma. The present chapter gives a detailed description of experimental techniques employed for LPP diagnostics, such as the OES and Optical Time of Flight (OTOF) methods.

## 3.2 OPTICAL EMISSION SPECTROSCOPY (OES)

Spectroscopy is a branch of science which studies the interaction of electromagnetic radiation with matter. Optical emission spectroscopy (OES) is a useful non-destructive spectroscopic technique which can be used for the characterization of LPP. Laser power, and the sensitivity and wavelength range of the spectrograph and the detector, are important in OES. Because of its remarkable features, OES can be used to study a wide range of materials like metals, alloys, biological samples, pollutants, explosives etc.

The phenomenon of laser ablation plays a key role in the formation of LPP. If the energy deposited on the target is much greater than then the latent heat of vaporization of all its constituents, they get completely vaporized and removed causing ablation. The factors that affect ablation of the material mainly include the laser pulse width, spatial and temporal fluctuations, power fluctuations, mechanical, chemical and physical properties of the target, etc. In short the generation of LPP consists of three steps, namely, laser-matter interaction, removal of material followed by ablation, and finally plasma formation. Plasma will emit continuous as well as characteristic (discrete) radiation. Since there is a delay of several microseconds between continuum generation and characteristic radiation, the detector should be gated in time to obtain discrete spectra, to get a conclusive idea about the emitter. Detection and spectral measurement of the emission using a spectrograph provides information on the elemental composition of the material which has undergone ablation.

At lower irradiances ($<10^8$ W/cm$^2$) and longer pulse durations (several µs) differential absorption occurs and thermal processes are dominant. At higher irradiances ($\geq 10^8$ W/cm$^2$) and smaller pulse durations (nanoseconds) ablation takes place. Short and energetic laser pulses deposit greater energy on the substrate and the surface temperature rises above the vaporization temperature, inhibiting further vaporization till a critical temperature is reached. More uniform target heating leading to an explosive release of material with a composition close to the original sample composition occurs here. When a laser pulse of intensity greater than the breakdown threshold of the material is used, local heating leads to an intense evaporation of the material, forming plasma. When the laser pulse falls on the material, the initial part of the energy is utilized for the vaporization and ionization of the substance whereas rest of the energy is absorbed by the vapor, causing expansion in the plasma plume which emits continuum and atomic lines.

From the optical emission spectra, key parameters such as electron temperature and number density of the plasma can be calculated. Temperature of the plasma is calculated either by the ratio of line intensity method or via Boltzmann plot method [20]. Electron temperatures ($T_e$) in the current study are evaluated using intensity ratios of line emissions from similar species emitting two wavelengths which are decaying to the same lower energy state. Assuming local thermodynamic equilibrium (LTE) in the system, the temperature is estimated using the equation,

$$\frac{I_1}{I_2} = \frac{g_1}{g_2} \frac{A_1}{A_2} \frac{\lambda_2}{\lambda_1} \exp\left[\frac{-(E_1 - E_2)}{K_B T_e}\right] \quad (3.1)$$

where $\lambda_i$, $A_i$, $g_i$, $I_i$ and $E_i$ ($i = 1,2$) are the wavelength, transition probability, statistical weight, line intensity, and energy of the excited state respectively. The electron number density ($N_e$) was evaluated from Stark broadening of line emission at a particular wavelength using the equation

$$\Delta\lambda_{1/2} = 2w\left(\frac{N_e}{10^{16}}\right) + 3.5\,A\left(\frac{n_e}{10^{16}}\right)^{\frac{1}{4}}$$
$$\times \left(1 - \frac{3}{4}N_D^{-\frac{1}{3}}\right) w\left(\frac{n_e}{10^{16}}\right) \quad \text{Å} \quad (3.2)$$

where $\Delta\lambda_{1/2}$ is the full width at half maximum (FWHM) of emission, $w$ is the electron impact parameter, $N_D$ is the number of particles in the Debye Sphere, and $A$ is the ion broadening parameter. In addition to Stark broadening, broadening mechanisms such as Doppler broadening and Pressure broadening also will contribute to line broadening in LPPs. Instrumental broadening is another factor which can increase measured linewidths in OES. Doppler broadening arises due to different velocity components in different regions of the plume. Typical FWHM of Doppler width in LPP is always one or two orders of magnitude less than that of Stark broadening. Pressure broadening depends on the ground state number density and oscillator strength of the transition, and normally it can be neglected since it is very small. Instrumental broadening can be minimized by reducing slit widths in the spectrometer. For non-hydrogenic ions Stark broadening is dominated by electron impact, and therefore the ion correction factor is negligible. As a consequence the Stark broadening equation reduces to

$$\Delta\lambda_{1/2} = 2w\left(\frac{N_e}{10^{16}}\right) \text{ Å} \quad (3.3)$$

The theoretical value of $w$ can be obtained from literature [108-109]. A Lorentzian curve will fit to the emission profile for a given wavelength. Knowing the electron impact parameter corresponding to the estimated temperature, $N_e$ can be calculated.

Temperature calculations are performed by assuming the plasma to be in LTE. Under a valid LTE, different temperatures of the plasma such as ion temperature, electron temperature and neutral temperature can be considered to be approximately equal. The validity of LTE can be checked using the McWhirter criterion [110] for each case independently, which is given as

$$N_e \geq 1.6 \times 10^{14} T_e^{0.5} (\Delta E)^3 cm^{-3} \quad (3.4)$$

Where $\Delta E$ is the energy difference between upper and lower energy levels in eV, and $T_e$ is the electron temperature in eV. If the value of $N_e$ calculated using equation 3.3 is greater than that calculated using equation 3.4, then LTE approximation is valid. In a transient system like the plasma, LTE is valid only when collision time between particles is low compared to the time duration where significant changes occur in the plasma.

### 3.3 OPTICAL TIME OF FLIGHT (OTOF) MEASUREMENTS

Femtosecond and nanosecond LPP are different in the nature of their generation and expansion. In ns LPP, ablated species in the plasma will interact with the trailing edge of the irradiating laser pulse since material ablation happens in picoseconds. The important processes involved in the generation of LPP are (1) laser–matter interaction, (2) laser- plasma interaction, (3) plasma plume-background interaction, and (4) plasma expansion [111]. In fs LPP, laser-plasma interaction is absent as the pulsewidth is very short and ablation happens only after the laser pulse ceases.

Plasma plume expanding freely in vacuum shows a free adiabatic expansion. The ejected species will slow down with increase in ambient pressure and axial distance. For sufficiently high pressures, when the ablated mass is small compared to the mass of the background gas in motion, a shock wavefront is formed whose propagation is given by

$$z = \left(\frac{E_0}{\rho_b}\right)^{1/5} t^{2/5} + z_0 \qquad (3.5)$$

where $z$ is the distance from the target, $E_0$ is a constant proportional to the laser energy density, and $z_0$ is the distance moved by the species during a time of the order of the average life time of excited states. If the mass of the species ejected is higher than that of the background mass (i.e. at lower ambient pressure), the ejected species slows down and this can be treated under a classical drag force model whose expansion can be expressed as

$$z = z_f[1 - \exp(-\beta t)] + z_0 \qquad (3.6)$$

where $\beta$ is the slowing coefficient and $z_f$ is the stopping distance of the plume. That is, $z_f = \frac{v_0}{\beta}$, where $v_0$ is the initial velocity of the ejected species.

Time and space resolved line emission spectroscopic information is an effective tool to estimate the velocities of ions and neutrals moving in an expanding LPP. Line emission dynamics can be recorded using a fast PMT attached to a spectrometer. Recombination and ionization mechanisms present in expanding LPPs are studied in the present work. OTOF studies of ions and neutrals along the expansion direction (i.e. normal to the target surface) for a large range of ambient pressures also are done.

## 3.4 ICCD IMAGING

Plume expansion dynamics of LPP can be investigated by acquiring single-shot photographs of plume emission in the x-z plane using an intensified charge coupled device (ICCD) operating in the time gated mode (2 ns and 5 ns temporal resolutions for ns irradiation, and 10 ns for fs irradiation). From 2D imaging of a three dimensionally expanding plasma, processes like plume splitting and instabilities happening at different time scales during expansion can be recorded. Plume emission is captured by the ICCD equipped with a 1360 × 1024 array with a 2:1 magnification. All images recorded are found to be fairly symmetric about the surface normal. Measurements at various pressures for fs and ns irradiation allow one to distinguish angular distributions of expanding plumes. Immediately after plasma formation the plume intensity is high in the core, and it reduces as time elapses due to thermal leak and reduced collisions. Experimental results are discussed in detail in chapters 4 to 6.

## 3.5 EXPERIMENTAL

A typical Laser induced breakdown spectroscopy (LIBS) system consists of a high power pulsed laser, spectrometer, high sensitivity and time gated detector with fast response time, and an interface with a computer that can rapidly process and interpret the acquired data. Nd: YAG is the most commonly used laser for plasma generation because its wavelengths are available in the NIR (1064nm), visible (532nm), and UV (266nm) regions. Typical pulse durations are around 10 nanoseconds, and intensity at the focal point may exceed $1GW/cm^2$. Most commonly used spectrometer is of the Czerny-Turner type, which will be either a monochromator (scanning type) or a polychromator (non-scanning type). Polychromators are usually used in LIBS as they allow simultaneous data acquisition in a broad wavelength range. The typical response of grating spectrometers (with multiple gratings) is in the range of 170 nm (in deep UV region) to 1100 nm (in near IR) region, which matches with the response range of CCD detectors. Hence all elements whose emission falls in this wavelength range can be easily analyzed using the LIBS technique. Another important part of instrumentation for LIBS is the delay generator, which accurately gates the detector's response. Gating helps to yield spectra with best signal to background ratios.

We used two laser sources in our experiments. The first is a regeneratively amplified ultrafast Ti: Sapphire laser (*TSA-10, Spectra Physics*) operating at 796 nm, pumped by a Q-Switched Nd: YAG laser and seeded by a mode-locked Ti: Sapphire oscillator *(Tsunami, Spectra Physics)*. The second is a Q-Switched Nd: YAG laser (*Quanta Ray, Spectra Physics*) operating at 1064 nm. Maximum pulse energy of the former is 10 mJ with a pulsewidth of ~100 fs (which is monitored using an autocorrelator (*SSA-F*, *Positive Light*)) and that of the latter is 150 mJ at 7 ns. The lasers are set to run at 10 Hz for energy stability, but plasma measurements are taken in the single-shot mode using a fast mechanical shutter placed in the beam path which opens only when required. An electronic synchronization circuit is used to drive the shutter so that single pulses can be selected from the 10 Hz pulse train at will. Laser beam is focused on to the surface of high purity (>99.99%) 5 mm × 5 mm × 3 mm solid targets of zinc and nickel (ACI Alloys Inc.) to produce the plasma. In order to avoid pitting of the target surface, the sample is moved by about 1 mm using a stepper motor driven X-Y translator after each laser pulse so that the following laser pulse falls at a new point on the target surface. Plasma plume is studied with the help of an imaging system which can be moved horizontally to observe different axial positions of the plume, in order to get spatial information (see Figure 3.2). Line emissions corresponding to different species in the plasma are dispersed using a high resolution (~0.06 nm) spectrometer (*Horiba Jobin Yvon, iHR 320*) and is recorded using a CCD (*synapse*) attached to it. The spectral lines recorded are identified with the help of a standard NIST atomic spectra database. In order to understand the line emission dynamics, a fast photomultiplier tube (*R943-02*, *Hamamatsu*) with ~ 1ns rise time is employed. Time evolution data from the PMT was recorded on a fast oscilloscope *(DPO 7354, Tektronix)*. The experiment is repeated for various nitrogen pressures ranging from $10^{-6}$ Torr to $10^2$ Torr. The pressure inside the plasma chamber, which is controlled by a needle valve, is monitored using a pirani gauge (*Pfeiffer vacuum*) at rough vacuum level and cold cathode gauge at high vacuum level. Apart from this, ICCD (4 Picos, Stanford Computer Optics, USA) measurements under similar conditions of pressure and energy also are performed yielding hydrodynamics of the plume for selected gatewidths.

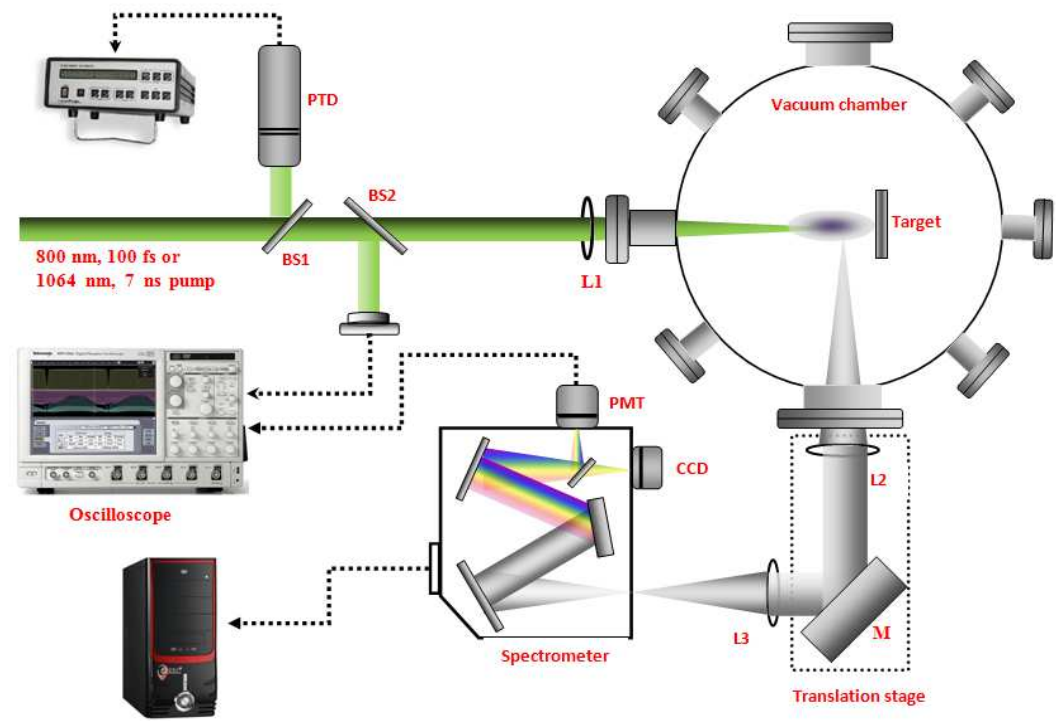

Fig 3.1. Schematic of the experimental set up used for optical emission spectroscopy (OES) and OTOF spectroscopy studies. Here a CCD is used for recording the spectra and a PMT is used for measuring OTOF data.

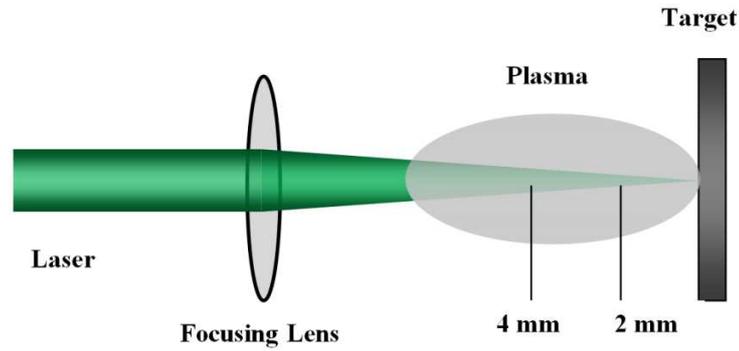

Figure 3.2: Geometry of the present LPP experiment. Focused laser pulses irradiate a solid metal target to form an expanding plasma plume. Spectral measurements of the plume are typically made at distances of 2 mm and 4 mm from the target surface.

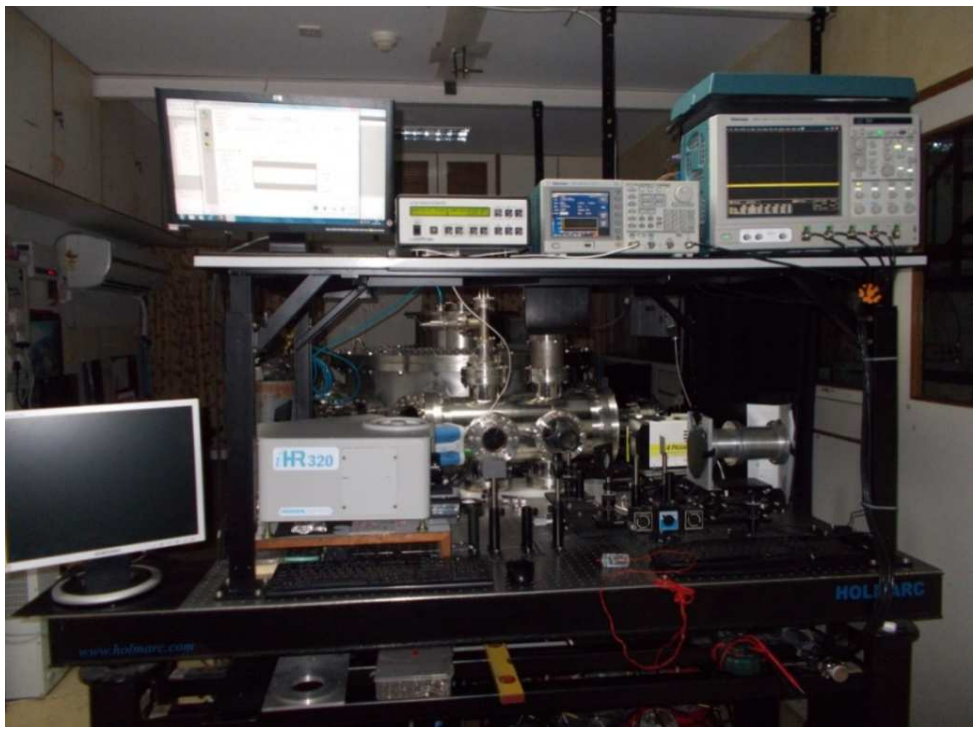

Figure 3.3. A view of the actual experimental setup used for OES and OTOF studies.

Figure 3.1 shows the schematic of the experimental layout, and a view of the actual experimental setup is shown in figure 3.3. The actual set up consists of the laser, vacuum chamber, target manipulator, a photo-thermal detector, oscilloscopes, monochromator etc. These instruments are synchronized with one another using a custom made electronic circuit; details of which are available in the Ph.D. thesis of C. S. Suchand Sandeep [41].

## 3.6 CONCLUSION

This chapter describes the necessity for doing optical emission spectroscopy and optical time of flight measurements for characterizing spatially expanding laser produced plasmas. Experimental details of optical emission spectroscopy and optical time of flight, along with the double-pulse experiment requirements, are presented, addressing the components used and their significance. The schematic and actual experimental set up are explained and presented in detail. Details regarding the synchronization and gating of the detectors are presented.

# CHAPTER 4
# INFLUENCE OF AMBIENT PRESSURE AND LASER PULSEWIDTH ON THE SPECTRAL AND TEMPORAL FEATURES OF ALASER PRODUCED NICKEL PLASMA

*Experimental characterization and comparison of the temporal features of femtosecond and nanosecond laser produced plasmas generated from a solid Nickel target, expanding into a nitrogen background, is presented in this chapter. Dynamicsof the ions and fast and slow neutrals present in the laser produced nickel plasma is studied by the optical time of flight technique,using emissions from neutral Ni I species at 361.9 nm and ionized Ni II speciesat 428.5 nm,respectively. The velocities of these species are estimated from the time of flight data. Furthermore, the variation of temperature, number density and velocity of different species in the plume with respect to variations in ambient pressure, for both nanosecond and femtosecond irradiation, is also investigated.*

**4.1 INTRODUCTION**

Experimental characterization and comparison of the temporal features of ns and fs LPPs generated from a solid nickel target, expanding into a nitrogen background, is presented in this chapter. When the ambient pressure is varied from $10^{-6}$ Torr to $10^2$ Torr, the plume intensity is found to increase rapidly as the pressure reaches 1 Torr. Electron temperatures ($T_e$) calculated from optical emission spectroscopy is found to be mostly pressure-independent for fs excitation, whereas an enhancement is seen in the range of $10^{-4}$ to $10^{-3}$ Torr for ns excitation. OTOF measurements of emission from neutral Ni (Ni I) at 361.9 nm ($3d^9(^2D)\ 4p \rightarrow 3d^9(^2D)\ 4s$ transition) reveals a single peak (the fast peak) in ultrafast excitation and two peaks (fast and slow peaks) in short-pulse excitation. The fast and slow peaks represent recombined neutrals and un-ionized neutrals respectively. The dynamics of emission from ionized Nickel (Ni II) also has been carried out using the 428.5 nm ($3p^6\ 3d^6\ (^3P) \rightarrow 3p^6 3d^9$) transition. Velocities of neutrals and ions are determined from OTOF measurements carried out at axial distances of 2 mm and 4 mm respectively from the target surface. A detailed account of these studies is given below.

100 fs pulses at 800 nm from a regeneratively amplified Ti: sapphire laser (*TSA-10, Positive Light*) with a maximum energy of ~ 10 mJ, and 7 ns pulses at 1064 nm from a Q-Switched Nd: YAG laser (*Quanta Ray, Spectra Physics*) with a maximum energy of ~ 150 mJ, were used for irradiating the target. In both cases the energies were adjusted so that the laser fluence was the same (~16 Jcm$^{-2}$, i.e. spot size is 200 µm and 560 µm for fs and ns excitations respectively) on the target surface, which is well above the ablation threshold. A high purity (better than 99.99%) nickel sample of dimensions 5cm×5cm×3mm (from *ACI Alloys Inc.*, San Jose) was the target. Optical emission spectra and temporal dynamics were measured by means of a spectrometer (*iHR320, Horiba JobinYvon*) equipped with a CCD detector (*Synapse, Horiba JobinYvon*) and PMT (*R943-02, Hamamatsu*). Plasma emission was collected and focused onto the spectrometer using two lenses and a mirror, which were fixed on a high resolution translation stage which could be moved to record emission from any part of the plume (Figure 1). Emission lines in the recorded spectra (with a resolution of 0.06 nm) were identified by comparingit with the standard NIST atomic spectral database[112]. The optical time of flight (OTOF) signals from the neutral Ni I line at 361.9 nm and ionic Ni II line at 428.5 nm were measured at various ambient pressures using the PMT and recorded on a fast digital storage oscilloscope (*DPO 7354, Tektronix*).Imaging of the plasma plume was performed using a gated ICCD (*4-Picos, Stanford Research Systems*), which revealed the temporal evolution of both ultrafast (fs) and short-pulse (ns) LPP plumes so thattheir expansion into the surrounding space could be characterized.

## 4.2 OPTICAL EMISSION SPECTROSCOPIC STUDIES

Optical emission spectroscopy (OES) is a powerful non-destructive technique for the estimation of plasma parameters such as electron temperature and number density[38, 113-114].In the present work, OES is recorded for various pressures from $1\times10^{-6}$Torr to $1\times10^{2}$ Torr at two spatial points (2 mm and 4 mm distant from the target surface, onthe plasma plume axis) for both ns and fs irradiations. Figures 4.1 and 4.3show the emission spectra measured for pressures ranging from $1\times10^{2}$Torr to $5\times10^{-2}$Torrfor ns and fs irradiation respectively,measured at 2 mm distance from the target surface.Figures4.2 and 4.4 represent similar data for pressures ranging from $1\times10^{-2}$Torr to $5\times10^{-6}$Torr. The measured variation of emission intensities with respect to the laser pulsewidth and background pressure are discussed below.

### 4.2.1 Effect of pulse width on the emission intensity

Intensity of emission lines is found to depend on the pulsewidth of the laser used. With lasers of sub-picoseconds pulse duration,enhanced emissions are seen from shorter wavelengths due to the large intensity associated with fs laser pulses compared to ns laser pulses.For instance, on fs irradiation, the emission intensity at shorter wavelengths is almost double of that obtained at larger wavelengths,when the pressure is increased from $5\times10^{-6}$Torr to $1\times10^{-2}$ Torr (see figure 4.2). A further increase in pressure from $5\times10^{-2}$ Torr to $1\times10^{2}$ Torr enhances the emission intensity of longer as well as shorter wavelengths, as seen from figure 4.1. The measurement wasdone for both 2 mm and 4 mm distances from the target. Similar trends are observed in both cases except that the intensity of emission is reduced at the 4 mm position due to the expansion of the plume.

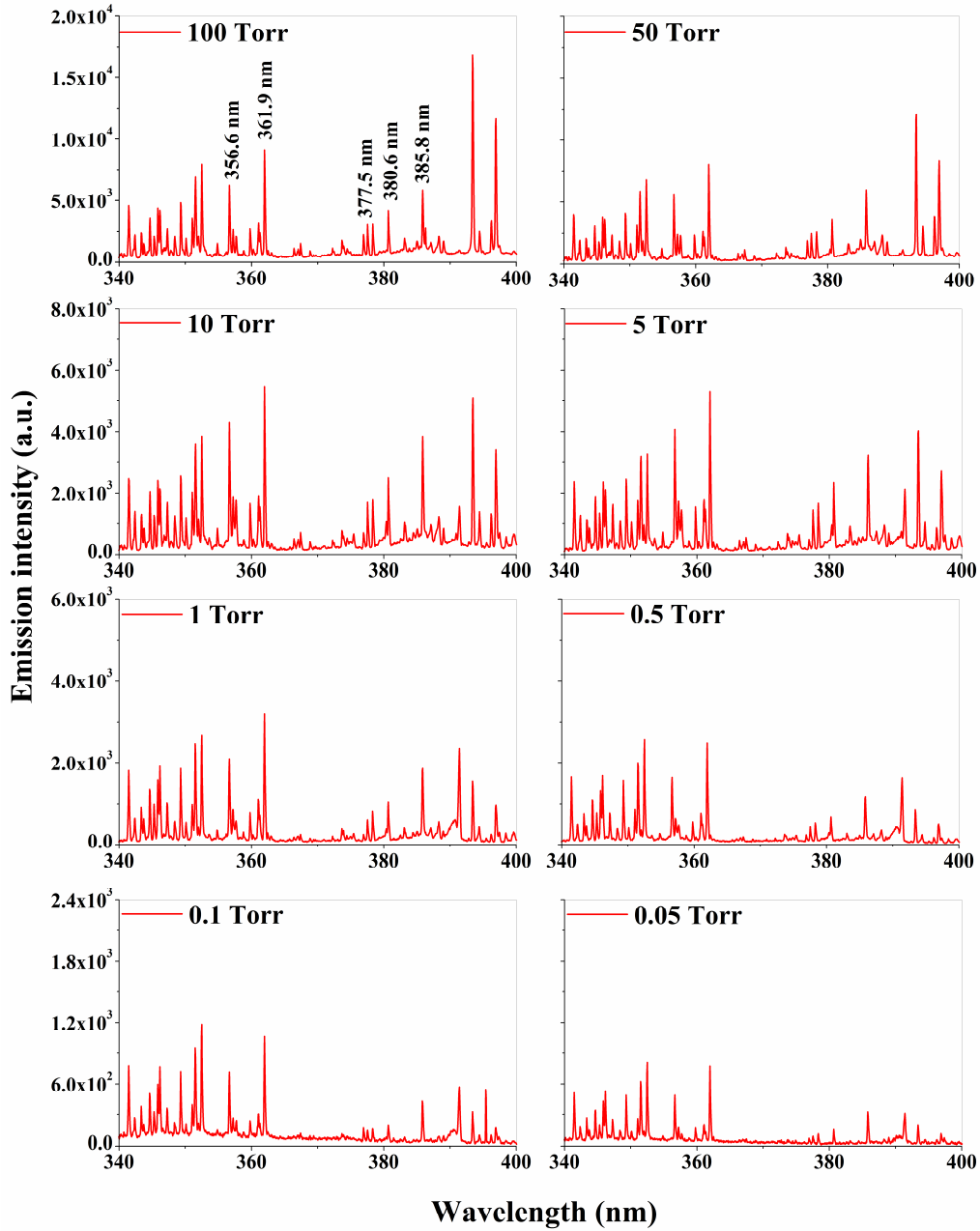

Figure 4.1. Ultrafast (fs) laser produced plasma measured at 2 mm distance from the target surface, in the range of relatively low pressures, from $5\times10^{-2}$ to $1\times10^{2}$ Torr. The emission intensity is found to increase with pressure.

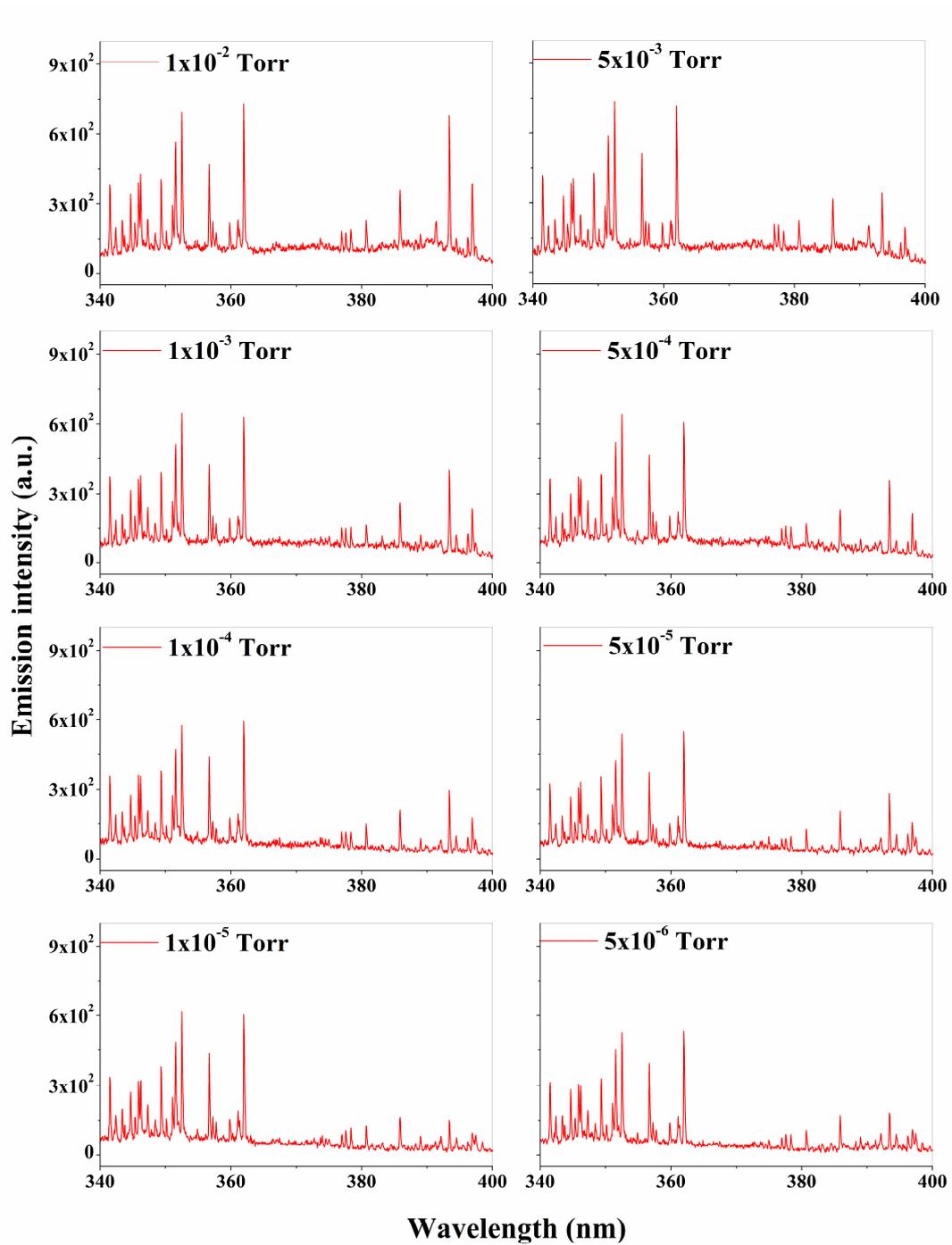

Figure 4.2. Ultrafast (fs) laser produced plasma measured at 2 mm distance from the target surface, in the range of relatively high pressures, from $5\times10^{-6}$ Torr to $1\times10^{-2}$ Torr.

Figures4.3 and 4.4givethe emission spectra measuredby ns irradiation for pressures ranging from $1\times10^{-2}$ Torr to $1\times10^{2}$ Torrand $1\times10^{-6}$ Torr to $5\times10^{-3}$ Torrrespectively. Emission intensity at longer wavelengths is higher compared to shorter wavelengths for ns irradiation, except for pressures between $5\times10^{-3}$ Torrand $1\times10^{-4}$ Torr. At these pressures the emission intensities for longer wavelengths are found to be reduced considerably, which is attributed to the maximum favorable conditions for laser plasma energy coupling which imparts larger kinetic energy to the ablated species, reducingthe probability of collisions among the ablated species. ns LPP is rich with emissions at longer wavelengths while for femtosecond LPPemission at shorter wavelengths dominates.

### 4.2.2  Effect of pressure on emission intensity

Emissions from the excited state of neutral nickel (Ni I) at 356.6 nm, 361.9 nm, 377.5 nm, 378.3 nm, 380.6 nm, and 385.9 nm respectively, and from the ionic state of Nickel (Ni II) at 428.5 nm, were studied using the optical emission spectra. Intensity at these wavelengths as a function of ambient pressure, measured 2 mm away from the target surface, are shown in Figures 4.5a and 4.5b for ns LPP and fs LPP, respectively.Emission intensity is found to be low for pressures less than $1\times10^{-1}$ Torr for both fs and ns irradiation due to the reducedcollisional excitations as plasma expands freely into vacuum: more precisely, an isothermal expansion within the irradiation laser pulse duration followed by an adiabatic expansion. It can also be seen that at low pressures many lines disappear in the spectra. With an increase in pressure from $1\times10^{-1}$ Torr, a sharp increase in the emission intensity from both ionic and neutral species is observed for both irradiations. In general, at low ambient pressures the plume expands more, reducing the number density and collisions, and therefore, the ionization rate. At high pressures the plume is smaller and the collision rate increases leading to an enhanced ionization; however, the quantity of ablated material reduces at still higher pressures due to inefficient laser-target energy coupling due to reflection, scattering etc., and due to laser-plasma energy coupling in ns LPP [20], while beamfilamentation may lead to ineffective laser-target coupling at very high ambient pressures in fs LPP[115]. In addition, for both fs and ns LPP, the higher rate of thermal leak from the plasma plume to the surrounding gas will reduce emission intensity at higher pressures, as thermal leak is dependent on the number density of the background gas. Even though we do not see any reduction in plume intensity up to an ambient pressure of $10^2$Torr, the above effects are likely toreduce plume intensity at still higher pressures.

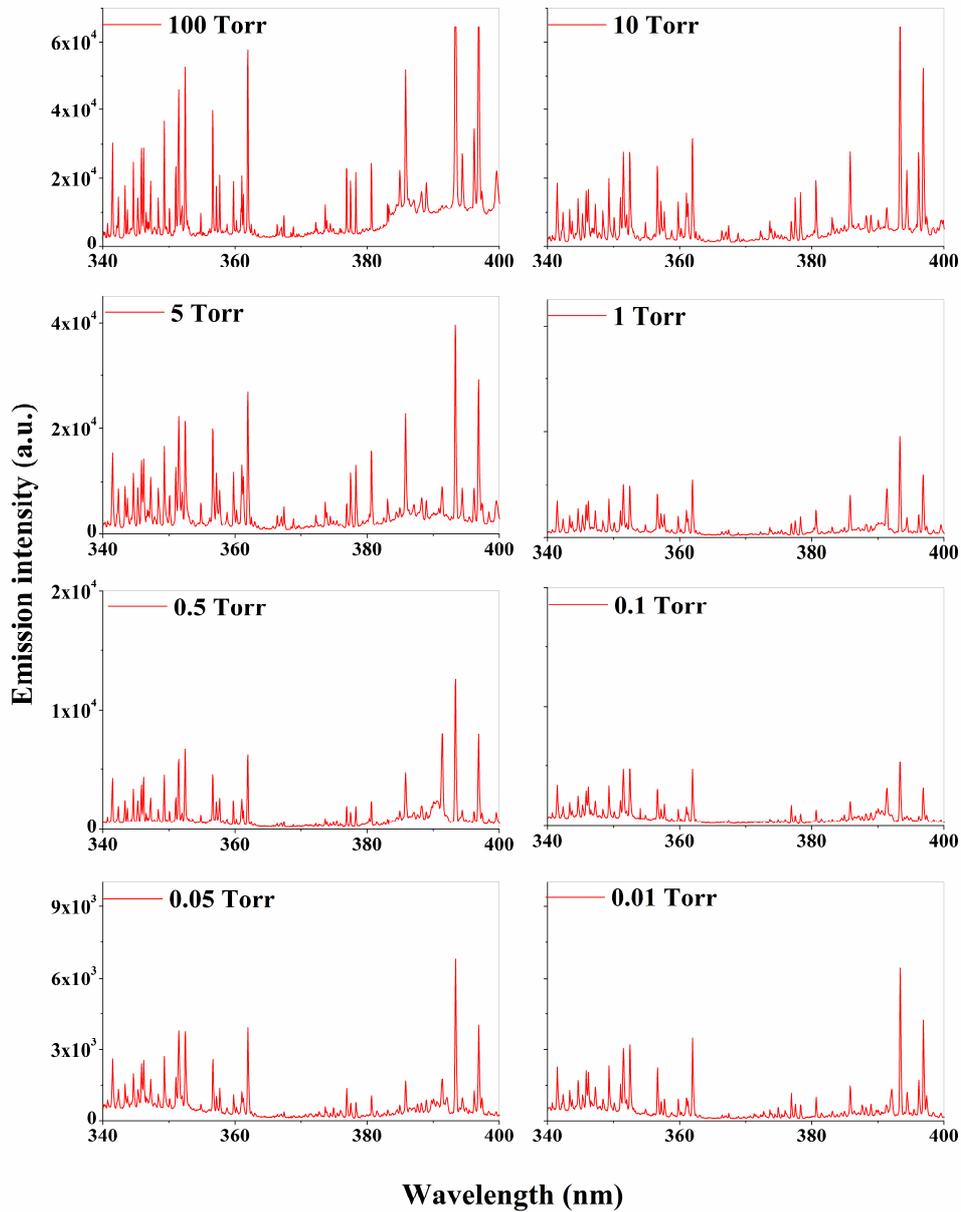

Figure 4.3.Short-pulse (ns) laser produced plasma at 2 mm distance, forrelatively high pressures ranging from $1\times10^{-2}$ to $1\times10^{2}$ Torr. The emission intensity is found to increase with pressure.

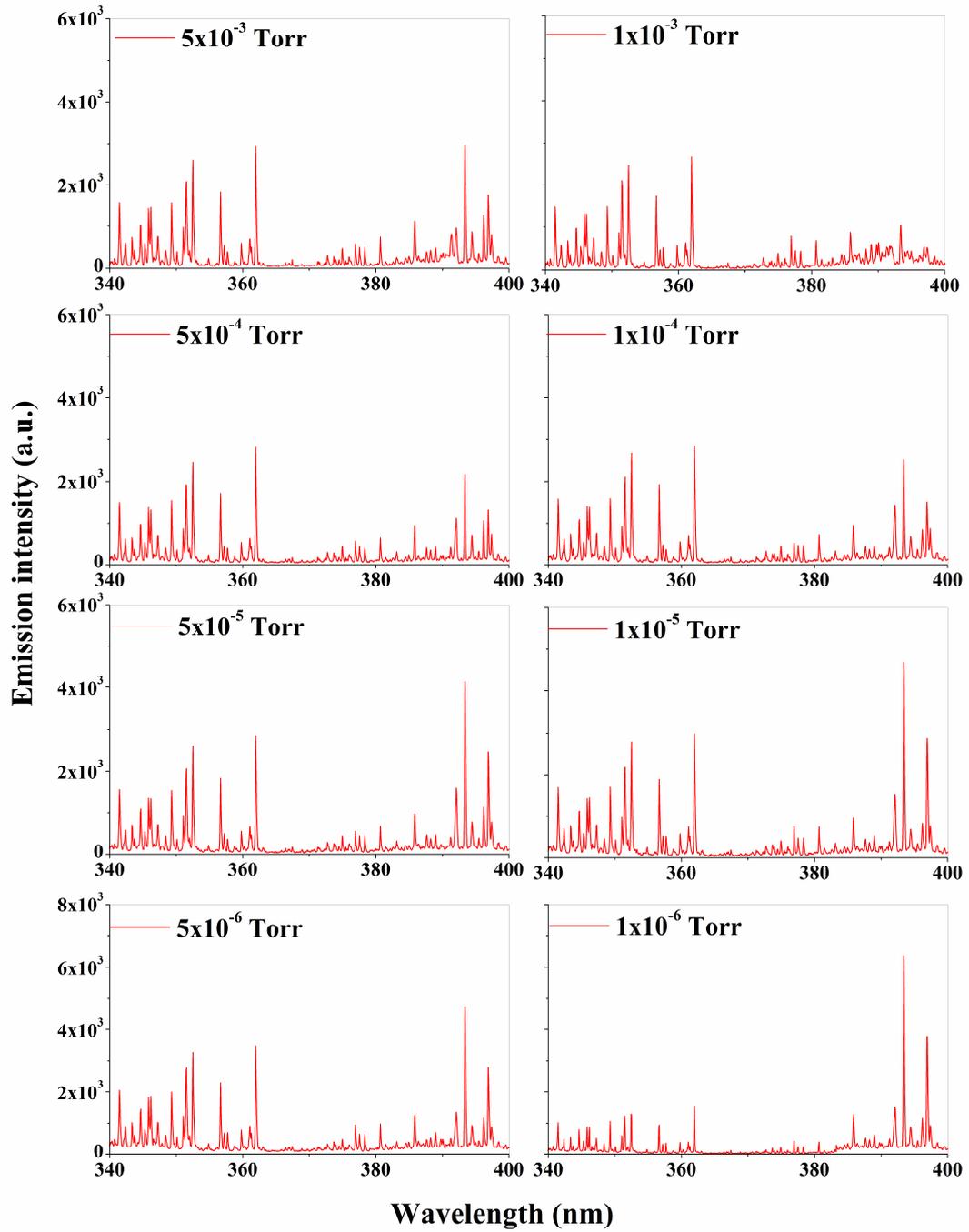

Figure 4.4. Short-pulse (ns) laser produced plasma at 2 mm distance, for relatively low pressures ranging from $1\times10^{-6}$ to $5\times10^{-3}$ Torr.

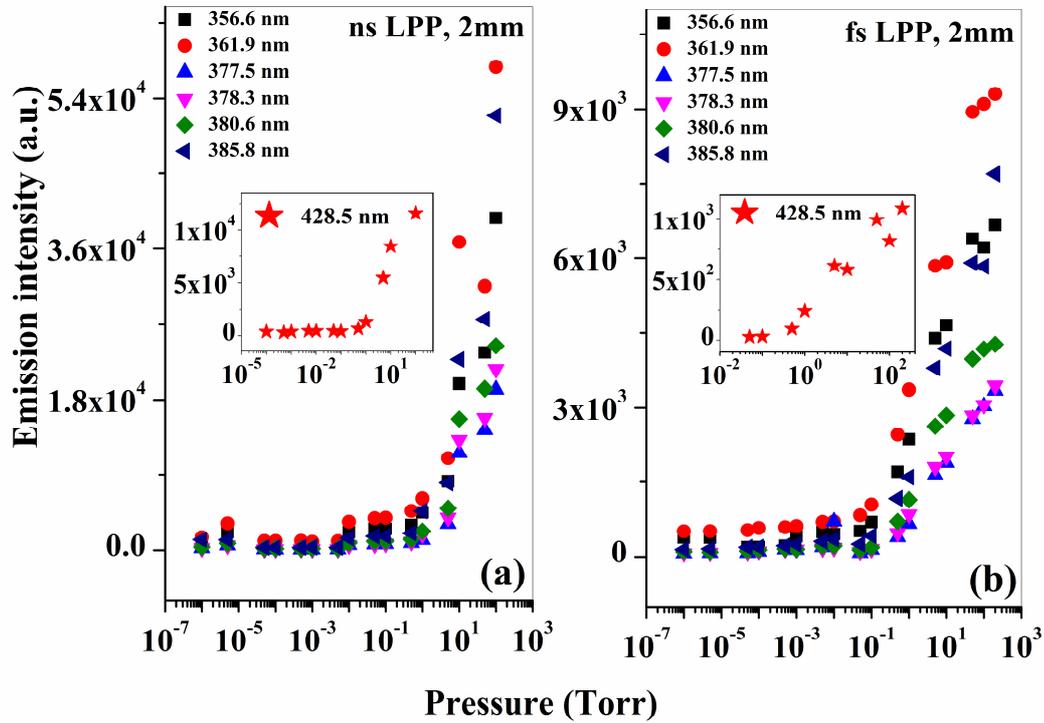

Figure 4.5. Emission intensity of the Ni I spectral lines measured at a distance of 2 mm from the target surface, for (a) 7 ns, and (b) 100 fs irradiation. Insets show the 428.5 nm line emitted by the Ni II species. Intensity rises sharply around an ambient pressure of 1 Torr. ns LPP is more intense compared to fs LPP.

### 4.2.3 Estimation of electron temperature

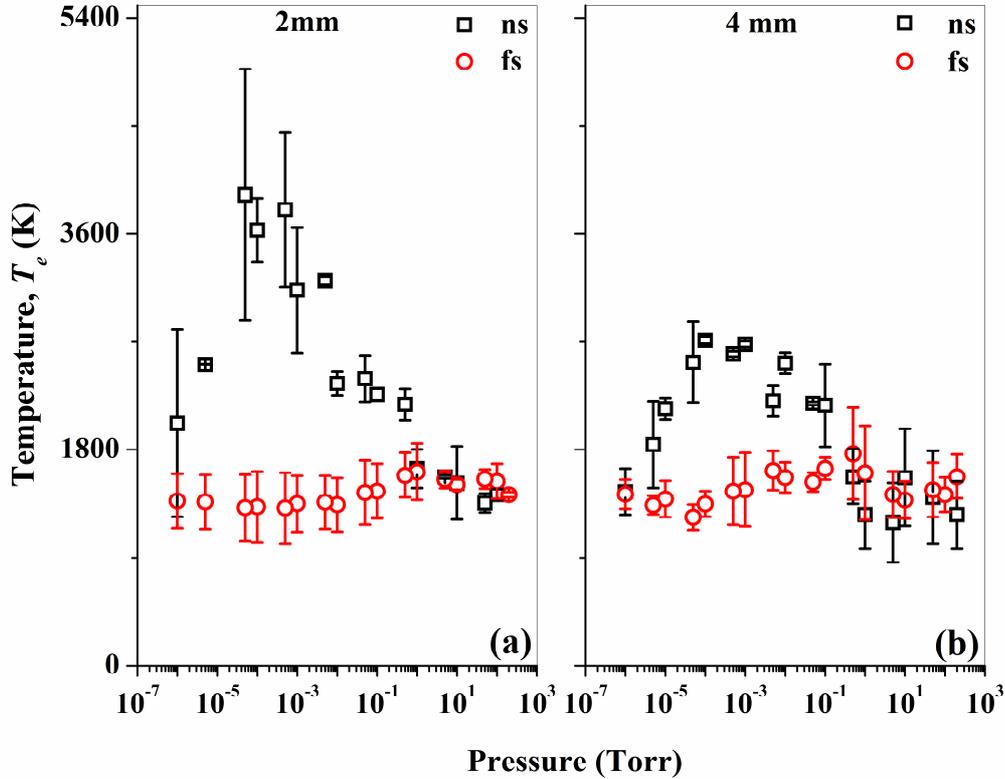

Figure 4.6: Plasma electron temperature $T_e$ calculated for (a) 7 ns and (b) 100 fs irradiations, from plume spectra measured at distances of 2 mm and 4 mm from the target surface.

Emission spectra recorded at distances of 2 mm and 4 mm from the target surface were used for estimating the electron temperature. Electron temperature ($T_e$) is calculated using the ratio of line intensities of Ni I emissions at 361.9 nm and 380.7 nm by assuming local thermodynamic equilibrium (LTE), using Equation 3.1. Figures 4.6a and 4.6b show the pressure dependence of $T_e$ measured for the plasma at different stages of expansion, i.e., at different axial positions in the plume (2 mm and 4 mm respectively) from the target surface. For fs excitation $T_e$ is fairly constant in the region of 1500 K, independent of ambient pressure and measurement position. However, for ns excitation the temperature exhibits both pressure and position dependence. For instance, $T_e$ increases and reaches a maximum of 3900 K at 2 mm, and 2700 K at 4 mm, in the pressure range of $1\times10^{-4}$ to $1\times10^{-3}$ Torr. $T_e$ decreases at higher pressures. This maximization of $T_e$ happens due to the effective heating of the plasma by the trailing edge of the ns laser pulse, which is a pressure dependent process. $T_e$ is found to reduce as the plume expands, which is revealed by the measurements taken at 4 mm distance. At the lowest and highest pressures studied ($10^{-6}$ and $10^2$ Torr respectively) $T_e$ is nearly the same for both fs and ns excitations.

## 4.3 OPTICAL TIME OF FLIGHT MEASUREMENTS

In pulsed laser ablation, the energy absorbed from the laser pulse is utilized for processes such as heat conduction, melting and vaporization of the material surface. Target will be ablated if the light intensity is above the ablation threshold, and the ablated species will interact with the surrounding gas[14, 75, 78]. The ablation threshold, which is defined as the minimum power density required for vaporizing the material, is given by

$$I_{min} = \frac{\rho L_v k^{1/2}}{\Delta t^{1/2}} \tag{4.1}$$

where $\rho$ is the density of the material, $L_v$ is the latent heat of vaporization, k is the thermal diffusivity, and $\Delta t$ is the laser pulse width. $I_{min}$ is calculated to be $8.69 \times 10^{11}$ W/cm$^2$ for fs excitation and $3.89 \times 10^8$ W/cm$^2$ for ns excitation in the present case, with the corresponding laser fluences being $8.69 \times 10^{-2}$ J/cm$^2$ and 2.72 J/cm$^2$ respectively. However, since the applied fluence is higher at 16 J/cm$^2$, and electron-ion collisions and heat conduction occur in the order of picoseconds, material ablation will happen rapidly by the leading edge of the laser pulse itself for ns excitation. Ablation creates a high density region near the target called the Knudsen layer, which shields the rest of the pulse from reaching the target surface. Therefore the evaporated material will absorb energy from the trailing edge of the laser pulse. On the other hand, for fs LPP, the whole energy content of the pulse is directly deposited on the target surface, since the laser pulse is too short for any laser-plasma interaction to occur. Therefore, the dynamics of plume formation and its expansion to the surroundings are entirely different in nature for ns and fs excitations [75, 78].

### 4.3.1 OTOF studies of neutrals

OTOF measurements on the bound-bound transition $3d^9(^2D)\ 4p \rightarrow 3d^9(^2D)\ 4s$ occurring at 361.9 nm (Ni I) were performed to understand the temporal evolution of neutrals in the expanding Ni plasma. Measurements were taken for the 2 mm and 4 mm axial positions in the plume along the expansion direction. Figures 4.7a and 4.7b show the time of flight (TOF) signals observed at 2 mm for ns LPP. The temporal profile shows a double-peak structure (fast peak PK1 and slow peak PK2), which correspond to fast and slow neutrals appearing at different times (~ 80 ns and 210 ns respectively) in the expanding plume. PK1 is attributed to emission from neutral nickel formed by the recombination of fast ionic species with free electrons (which is independently corroborated from ion TOF measurements as discussed later), while PK2 corresponds to the comparatively slower un-ionized neutral atoms present in the ablated material. Intensities of PK1 and PK2 remain relatively constant in the range of $10^{-6}$ Torr to $10^{-3}$ Torr. Between $10^{-2}$ Torr to 1 Torr PK1 gets intensified and PK2 gets diminished. PK2 disappears altogether when pressure increases beyond 1 Torr. This reduction in PK2 intensity at high pressures is due to the combined effects of plasma confinement and weak laser-target energy coupling. Figures 4.8a and 4.8b depict time of flight emission of Ni I species subsequent to fs excitation, measured at the 2 mm position. Unlike ns LPP, fs LPP shows only a single peak in the emission profile for all pressures studied. In addition, the emission lifetime is shorter for fs excitation in comparison to ns excitation. The single peak in the OTOF studies of neutrals in the fs LPP is found to have a sharp leading edge followed by a slowly decaying trailing part indicating that fast recombined neutrals in the plasma are quickly followed by fast moving unionized neutrals in the expanding plume.

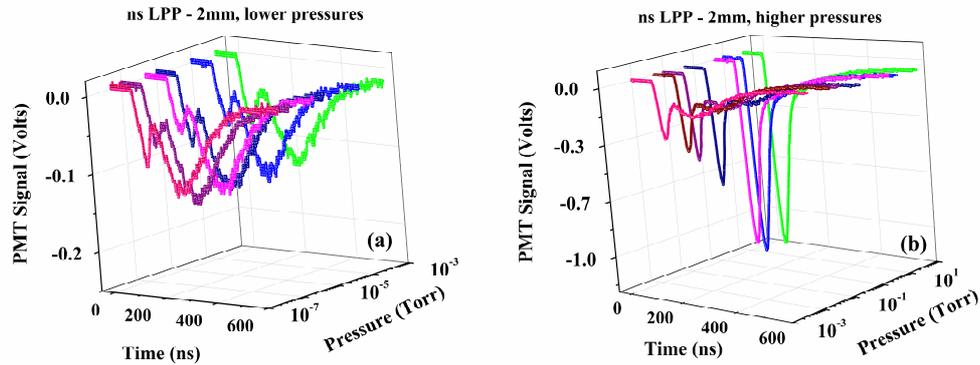

Figure 4.7. Time Resolved emission of the 361.9 nm ($3d^9(^2D)$ $4p \rightarrow 3d^9(^2D)$ $4s$) transition from Ni I, measured in the plasma plume at a distance of 2 mm from the target surface, for 7 ns irradiation. Ambient pressure varies from (a) $10^{-7}$ to $10^{-3}$ Torr, and (b) $10^{-3}$ to $10^2$ Torr.

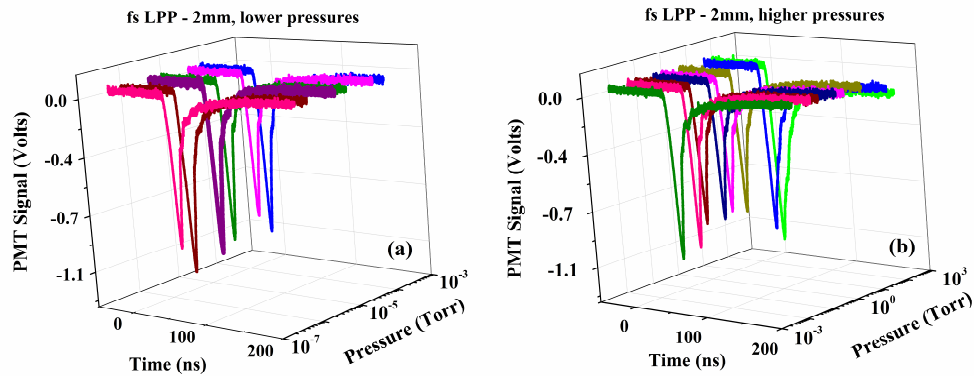

Figure 4.8. Time Resolved emission of the 361.9 nm ($3d^9(^2D)$ $4p \rightarrow 3d^9(^2D)$ $4s$) transition of Ni I, measured in the plasma plume at a distance of 2 mm from the target surface, with 100 fs irradiation. Ambient pressure ranges from (a) $10^{-7}$ to $10^{-3}$ Torr, and (b) $10^{-3}$ to $10^2$ Torr.

### 4.3.2 Velocities of neutrals

As the ns plasma plume expands, velocities of the fast and slow Ni I species (PK1 and PK2 respectively) are found to change, as shown in Figure 4.9. For instance, PK1 speeds up from ~ 30 km/s at 2 mm to ~ 70 km/s at 4 mm (Figure 4.9a). Moreover, while PK1 velocity increases with pressure at 2 mm, it decreases with pressure at 4 mm. PK2 also behaves in a somewhat similar fashion except that its speeding from 2 mm to 4 mm is moderate in comparison to that of PK1. However, the increase in velocity at 2 mm and decrease in velocity at 4 mm with pressure are more pronounced in this case. Even though in general the velocity increases with pressure at 2 mm, a local decrease can be seen in the pressure range of $10^{-4}$ to $10^{-2}$ Torr. Such a fluctuation in velocity arises from the shock wave that propagates back and forth through the ambient gas subsequent to ablation: with a laser fluence of ~ 16 J/cm$^2$, ablation of the target is intense and rapid. While the pressure-dependent velocity increase of PK2 at 2 mm is due to laser-plasmainteraction, velocity decrease at 4 mm is due to the plasma confinement effect. PK2 disappears altogether at higher pressures, indicating that the plume expansion is hindered by plasma confinement.

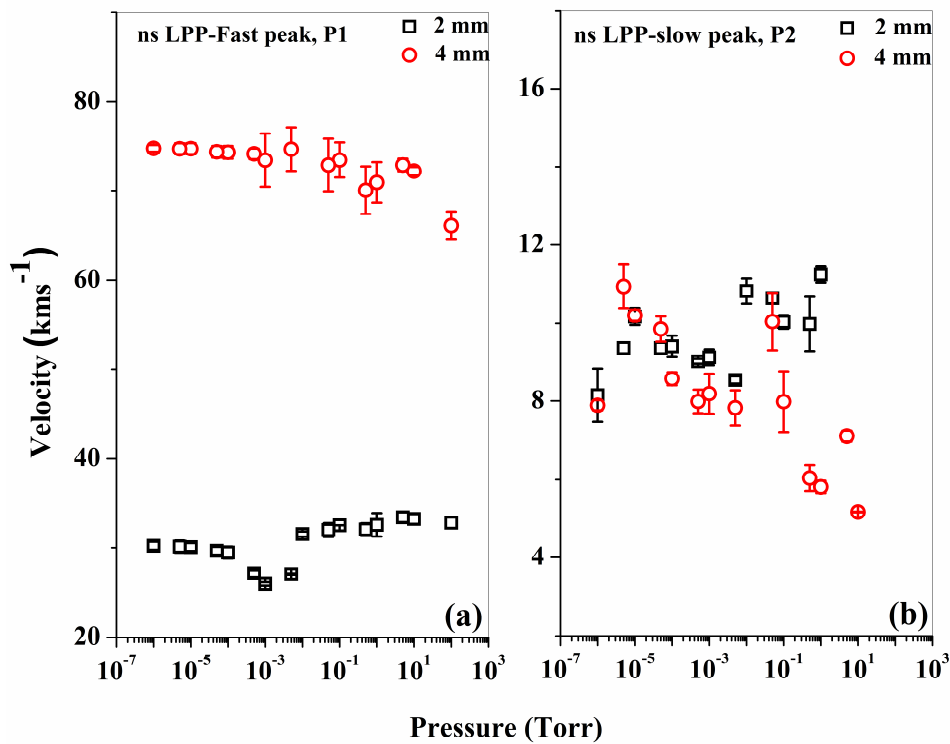

Figure 4.9. Velocities of the (a) fast and (b) slow Ni I species, measured at axial distances of 2 mm and 4 mm from the target surface, for 7 ns irradiation.

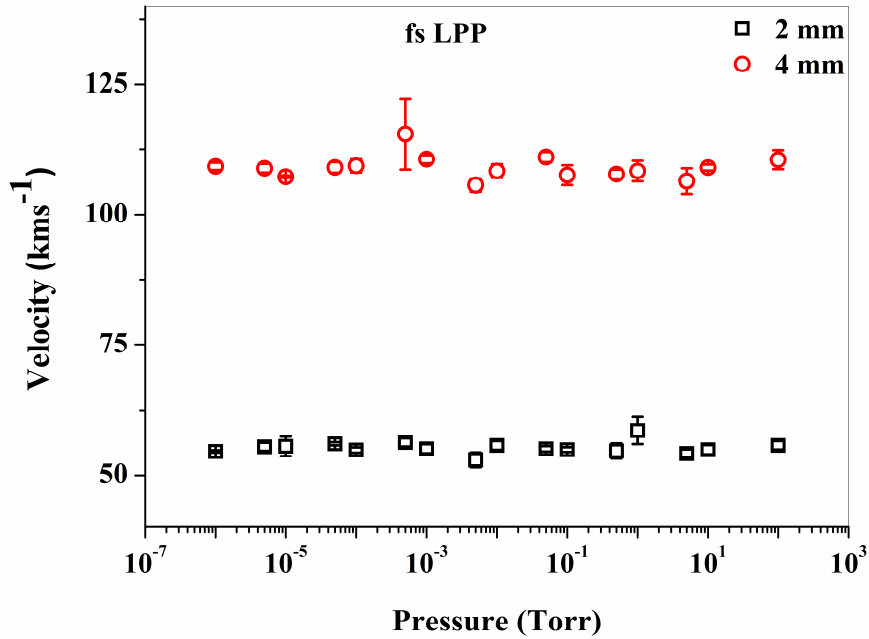

Figure 4.10. Velocities of the Ni I species measured at axial distances of 2 mm and 4 mm from the target surface, for 100 fs irradiation.

On the other hand, the velocities are found to be almost constant and independent of the ambient pressure for fs LPP (Figure 4.10). Absolute velocities are higher compared to the fast species in the ns LPP, but acceleration between the 2 mm and 4 mm positions is less than that seen for ns excitation. While the acceleration seen for fs LPP is due to the pull experienced by ions from the fast electrons moving ahead of them, accelerationobserved in ns LPP arises from laser-plasma interaction as well as the Coulomb pull.

### 4.3.3 OTOF studies of ions

Figures 4.11a and 4.11b show the TOF emission spectra of singly ionized Ni II ions. Emission intensity is relatively lower compared to that of neutrals (see Figure 4.7 and 4.8). Ion emission is less prominent because ion density is reduced at high pressures due to larger recombination rates, and at low pressures due to reduced collisions. Therefore in the present experiments, emission from ionic species could be observed only for certain intermediate pressures. For ns LPP emission could be measured from 0.05 Torr to 100 Torr, but for fs LPP it could be measured only from 10 Torr to 100 Torr. Emission intensity increases as a function of pressure and peaks around 10 Torr for ns excitation. The peak pressure could not be determined for fs excitation due to the limited pressure range available for investigation.

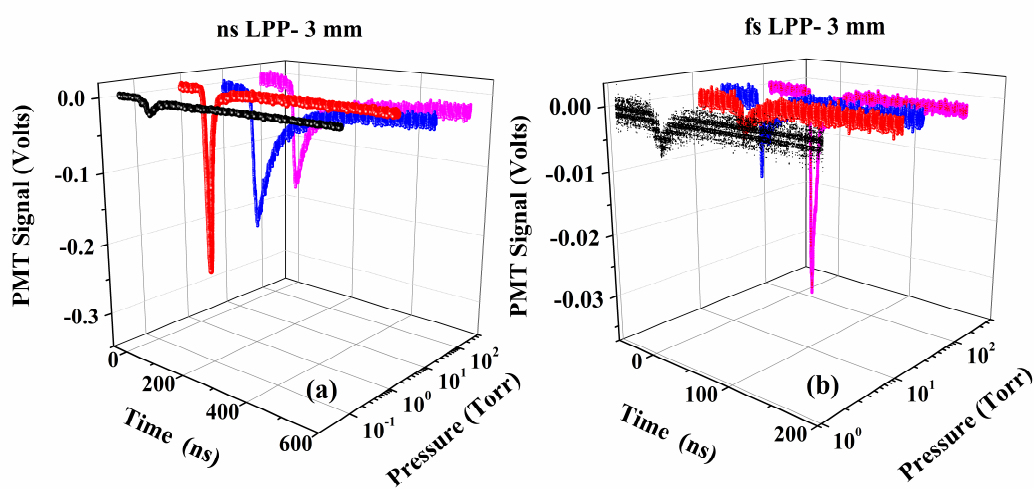

Figure 4.11. Time of flight emission spectra of Ni II at 428.5 nm ($3p^6\ 3d^6\ (^3P) \rightarrow 3p^6 3d^9$) measured in the plasma plume at a distance of 3 mm from the target surface, for (a) 7 ns and (b) 100 fs irradiations.

Figures 4.12a and 4.12b depict the velocity of ionic species measured at distances of 2 and 4 mm, for ns and fs ablations, respectively. For ns irradiation velocity is almost a constant around 90 km/s at 2 mm irrespective of background pressure. Moreover, acceleration is not high at the lower pressures. This result is in agreement withthe fact that fast neutrals are formed from the recombination of fast ions with free electrons. The lighter electrons will travel faster than the ions, and the Coulomb pull on the ions will be inversely proportional to the electron-ion distance. Thus only those ions which are quite fast can recombine with the electrons. Assuming that most of the fast ions undergo recombination, only the slower ions (which also experience less Coulomb pull) will predominantly remain for detection. The acceleration will be smaller for these ions compared to those which went ahead of them. However as we see from the figure acceleration can increase at higher pressures where laser-plasma interaction is stronger. At 4 mm, velocity increases with pressure and reaches a maximum of 175 km/s. On the other hand for fs irradiation, velocity increases linearly with pressure at both the 2 mm and 4 mm positions. Acceleration is higher for ns LPP due to the higher plume heating, and it is maximized around a certain pressure range[89,93] of 10 Torr in our case, where the emission peak intensity from ionic species also is a maximum. The right balance between collisions and recombination is reached at this pressure resulting in maximum ionization, and hence maximum Coulomb pull.

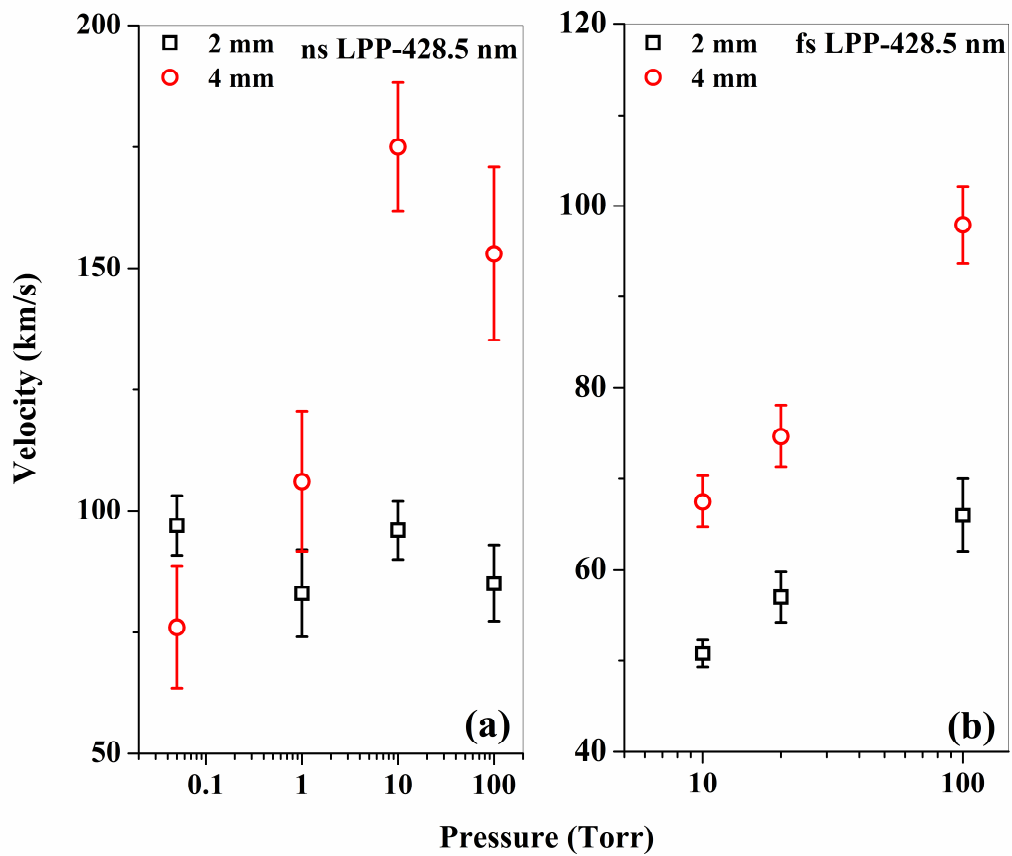

Figure 4.12. Velocity of the Ni II species measured at axial distances of 2 mm and 4 mm from the target surface, for (a) 7 ns and (b) 100 fs irradiation.

Our detailed studies reveal that for the 2 mm position, arrival times of the fast neutral species (33 ns to 70 ns) are similar to the arrival times of the ionized species (23 ns to 40 ns), whereas the arrival of the slow species takes place between 180 ns to 240 ns. This observation further affirms that the generation of fast atomic species in the expanding plume is indeed due to the recombination of fast ions with electrons.

**4.4 EXPANSION OF THE PLASMA PLUME IN TIME**

Time gated imaging using an ICCD is an effective tool toinvestigate the temporal evolution of the LPP plume. An ICCD records the two dimensional image of a three dimensionally expanding plume,featuring the plume propagation dynamics and structure. In general, for each image ICCD gain and gate width may be adjusted to compensate for the reduction in intensity resulting from plasma expansion. Here, the ICCD gain and gate width (5 ns for ns irradiation and 20 ns for fs irradiation) are kept the same for all measurements, to compare the results across different background pressures

Figure 4.13, 4.14 and 4.15 show the ICCD images of the femtosecond laser produced nickel plasma expanding into nitrogen ambient at 100 Torr, 5 Torr, and 0.05 Torr background pressures respectively, for various time delays after the laser pulse hits the target surface. The gate width used is 20 ns. Plume intensity lasts until about 1400 ns, and the plume length remains almost the same at all times, for 100 Torr pressure. The plume moves forward with a certain velocity in the initial stages of expansion and stops around 250 ns time delay, and the plume intensity reduces due to thermal decay at later stages of expansion. At 5 Torr the angular divergence of the expanding plume is found to be less than that measured at 100 Torr. The sharp fast edge seen in the TOF dynamics indicates that the velocity of expanding plume increases at this pressure. The notable axial confinement of the plume is an indication of the presence of giant self-generated magnetic field in the case of fs

excitation. The emission intensity is found to reduce with pressure since the plasma loses its confinement via thermal leak [116].

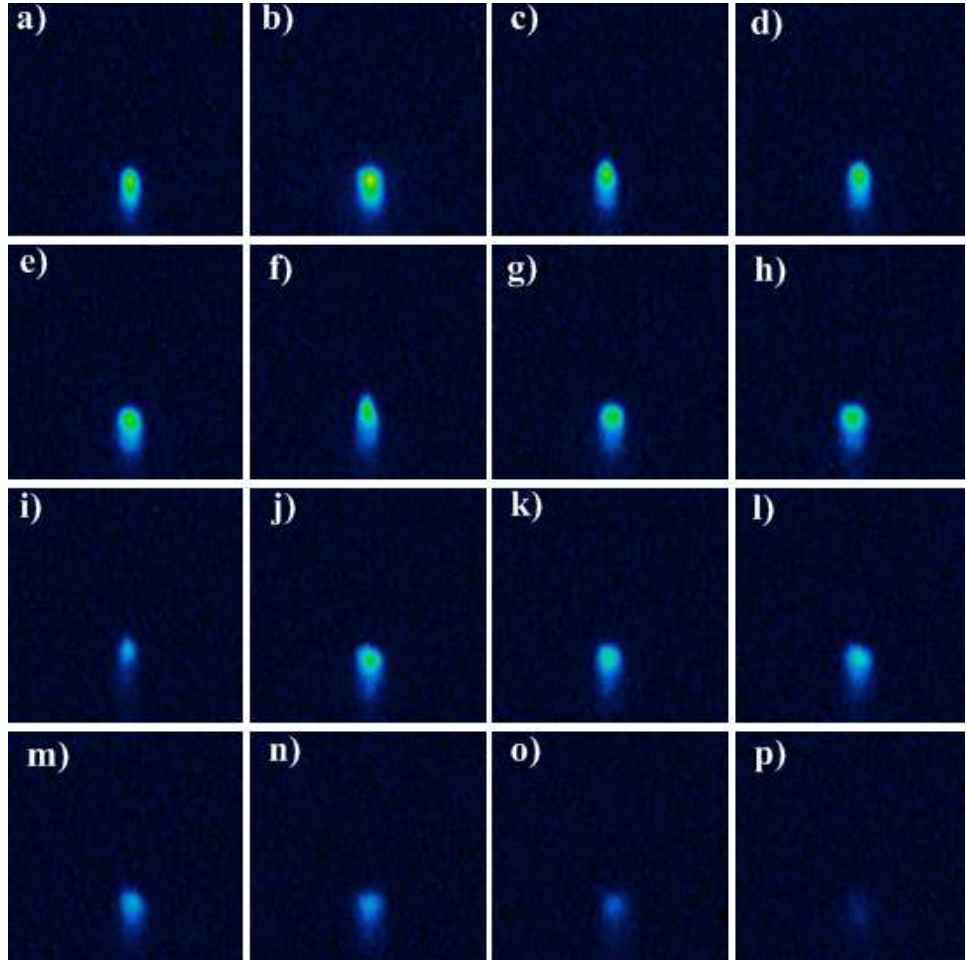

Figure 4.13. Images of the femtosecond laser produced nickel plasma expanding into nitrogen ambient at 100 Torr pressure, measured for time delays of (a) 50 ns, (b) 100 ns,(c) 150 ns, (d) 200 ns, (e) 250 ns, (f) 300 ns, (g) 350 ns, (h) 400 ns, (i) 450 ns,(j) 500 ns,(k) 550 ns,(l) 600 ns,(m) 700 ns,(n) 800 ns,(o) 1000 ns, and(p) 1400 ns (The instant when laser pulse hits the target is taken as zero time delay).An ICCD gate width of 20 ns is used for the measurements. The spatial dimension of the image is 1 cm × 1 cm.

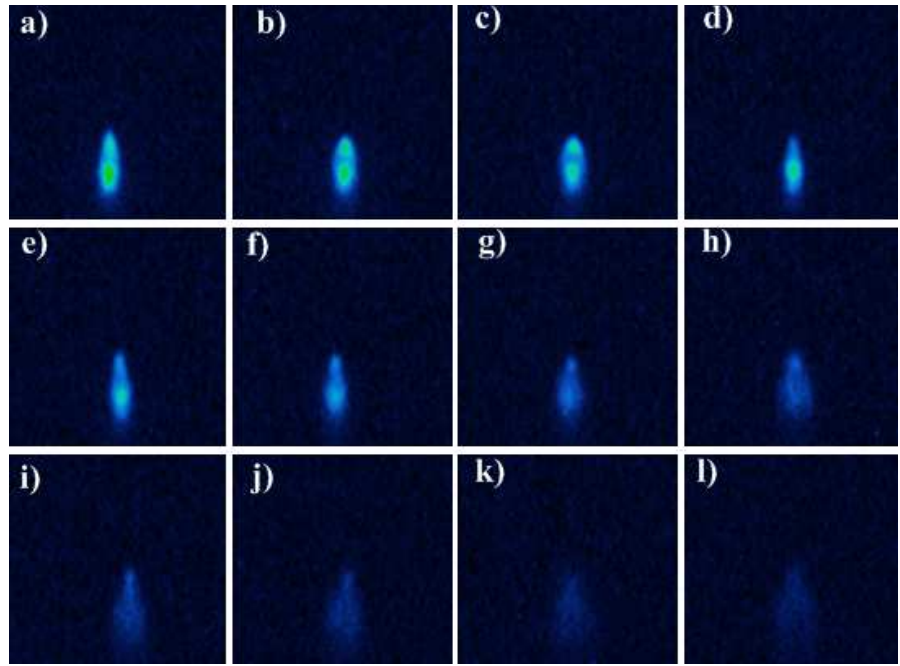

Figure 4.14. Images of the femtosecond laser produced nickel plasma expanding into nitrogen ambient at 5 Torr pressure, measured for time delays of (a) 50 ns, (b) 60 ns, (c) 70 ns, (d) 80 ns, (e) 90 ns, (f) 100 ns, (g) 130 ns, (h) 150 ns, (i) 180 ns, (j) 200 ns, (k) 250 ns, and (l) 290 ns. ICCD gate width used is 20 ns.

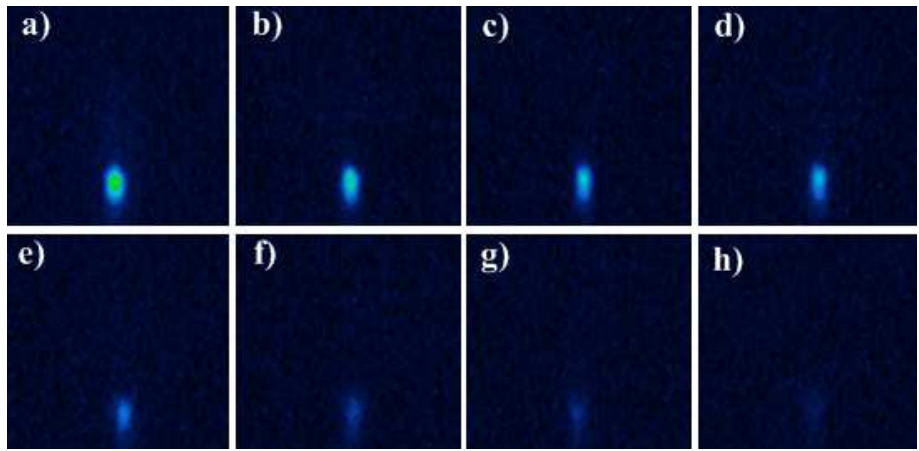

Figure 4.15. Images of the femtosecond laser produced nickel plasma expanding into nitrogen ambient at 0.05 Torr pressure, measured for time delays of (a) 50 ns, (b) 70 ns, (c) 80 ns, (d) 90 ns, (e) 110 ns, (f) 130 ns, (g) 150 ns, and (h) 170 ns. ICCD gate width used is 20 ns.

Figures 4.16, 4.17 and 4.18 give images of the nanosecond laser produced nickel plasma expanding into nitrogen ambient at 100 Torr, 5 Torr, and 0.05 Torr background pressures respectively, for various time delays after the laser pulse hits the target surface. ICCD gate width used is 5 ns. The spatial dimension of the image is 1 cm × 1 cm. Plasma plume intensity is measured up to 150 ns, 350 ns and 750 ns time delays for 0.05 Torr, 5 Torr and 100 Torr pressures respectively. The imaged plasma lasts up to 750 ns for 100 Torr nitrogen pressure, and the plasma is more confined at the higher pressures.

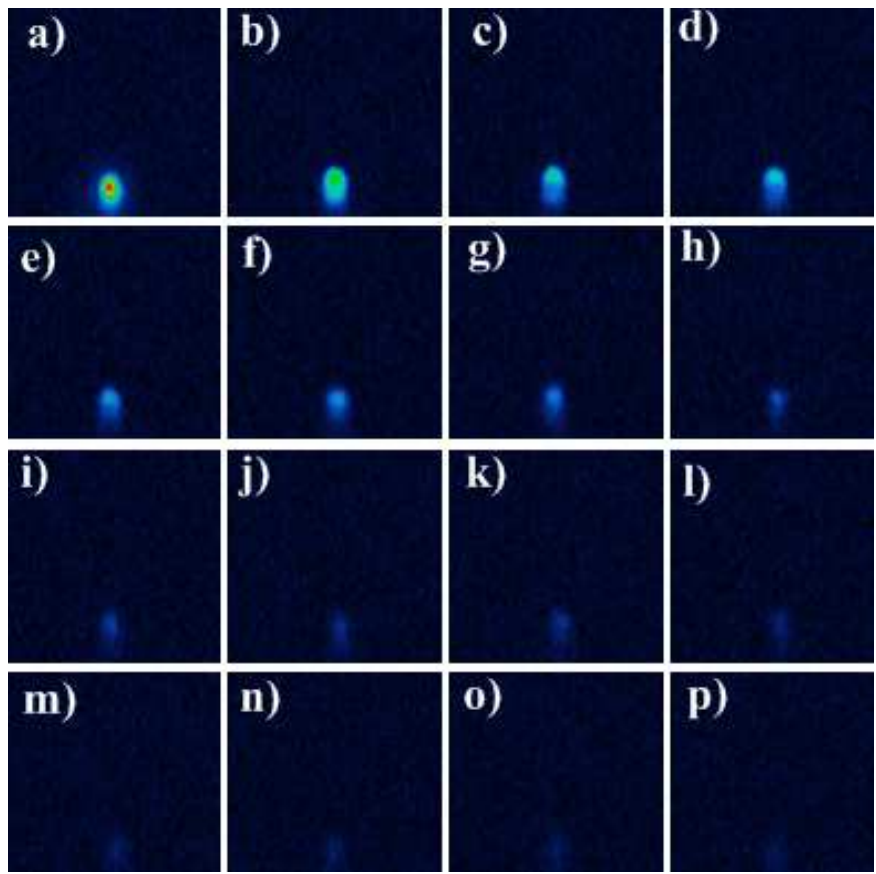

Figure 4.16. Images of the nanosecond laser produced nickel plasma expanding into nitrogen ambient at 100 Torr pressure, measured for time delays of (a) 50 ns,(b) 100 ns,(c) 150 ns,(d) 200 ns,(e) 250 ns,(f) 300 ns,(g) 350 ns,(h) 400 ns,(i) 450 ns,(j) 500 ns,(k) 550 ns,(l) 600 ns,(m) 650 ns,(n) 700 ns,(o) 750 ns,and (p) 800 ns, for anICCD gate width of 5 ns.

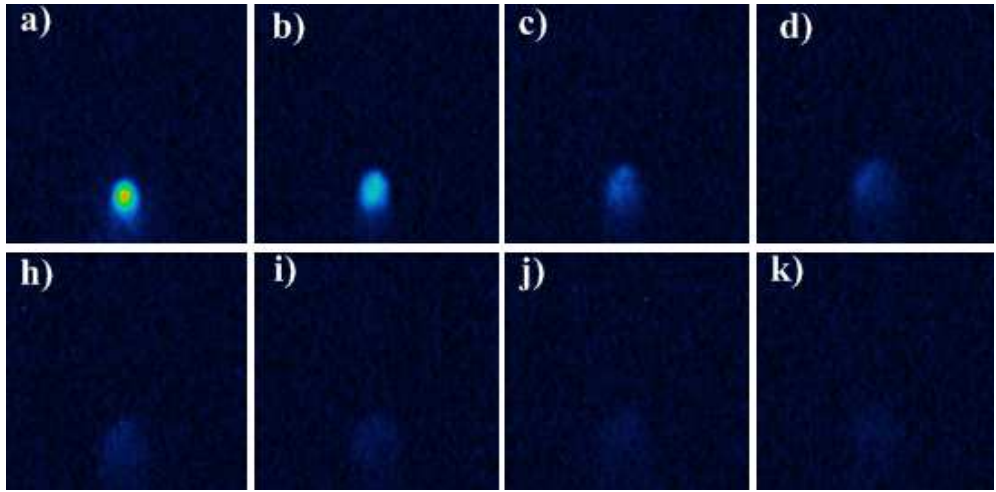

Figure 4.17. Images of the nanosecond laser produced nickel plasma expanding into nitrogen ambient at 5 Torr pressure, measured for time delays of (a) 50 ns,(b) 100 ns,(c) 150 ns,(d) 200 ns,(h) 250 ns,(i) 300 ns,(j) 350 ns,and (k) 400 ns,using an ICCD gate width of 5 ns.

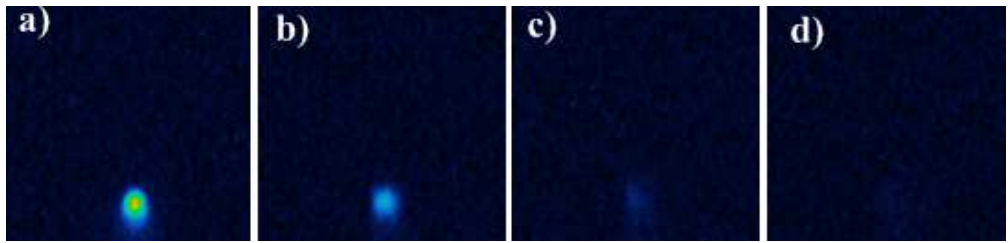

Figure 4.18. Images of the nanosecond laser produced nickel plasma expanding into nitrogen ambient at 0.05 Torr pressure, measured for time delays of (a) 50 ns,(b) 100 ns,(c) 150 ns, and (d) 200 ns, for an ICCD gate width of 5 ns.

## 4.5 CONCLUSION

We have generated and characterized laser produced Ni plasmas using ns and fs laser pulses, and carried out theOTOF measurements of the neutral Ni I emission at 361.9 nm ($3d^9(^2D)$ 4p $\rightarrow$ $3d^9(^2D)$ 4s transition), for a large range of ambient nitrogen pressures from $10^{-6}$Torr to $10^2$Torr. Ionic emission from Ni II at 428.5 nm ($3p^6$ $3d^6$ ($^3P$) $\rightarrow 3p^6 3d^9$), which is weaker in intensity than neutral emission, also has been investigated. Emission and OTOF spectra have been taken from two axial positions, viz. 2 mm and 4 mm away from the target surface, in the expanding plasma plume. Plasma intensity increases with pressure for both ns and fs excitations. While the electron temperature gets maximized around $10^{-4}$ to $10^{-3}$ Torr for ns excitation, it is relatively lower and rather independent of pressure for fs excitation. A double-peak structure is observed in the OTOF spectrum of Ni I under ns excitation where the fast peak corresponds to the neutrals and is formed by ion-electron recombination, and the slow peak corresponds to the un-ionized neutrals ablated from the target. On the other hand, only one peak (fast) is observed in the OTOF spectrum of fs excitation. Measured velocities of the species are relatively higher and independent of pressure for fs excitation, whereas shockwave effects are evident for ns excitation. The measured velocities indicate that the fast species is clearly accelerating in the plume on expansion, at least up to a distance of 4 mm from the target. On the other hand the slow species gets decelerated, particularly at higher pressures. These investigations provide new information on the pressure dependent temporal behavior of Ni plasmas produced by ns and fs laser pulses, which have potential use in applications such as pulsed laser deposition (PLD) and laser-induced nanoparticle generation.

# CHAPTER 5

# INFLUENCE OF AMBIENT PRESSURE AND LASER PULSEWIDTH ON THE SPECTRAL AND TEMPORAL FEATURES OF A LASER PRODUCED ZINC PLASMA

*A detailed study of femtosecond and nanosecond laser produced Zinc plasmas conducted using optical emission spectroscopy and optical time of flight techniques is presented in this chapter. Optical emission spectroscopy allows the estimation of temperature and number density of the plasma, while optical time of flight measurements throw light into the dynamics of ions and neutrals in the plasma. Imaging done using an intensified charge coupled device helps observe features of plasma expansion into the ambient medium with a high time resolution.*

## 5.1 INTRODUCTION

Laser produced plasma (LPP) is transient in nature, and plasma parameters will change rapidly with time and space on expansion. Therefore studies of plasma dynamics are essential to understand the fundamental plasma physics and plasma chemistry relevant at different time scales. In this chapter, OES and OTOF measurements of a laser produced Zinc (Zn) plasma, produced upon irradiating a high purity Zn target (99.99%) using 100 fs (at 796 nm) and 7 ns (at 1064 nm) laser pulses, under similar optical fluences of ~ 16 J/cm$^2$, are presented. The excitation wavelength can play a crucial role in the characteristics of the LPP at fluences just above the threshold fluence, and we have used a larger laser fluence in the present experiments to avoid the same. Moreover, measurements under larger fluences are preferable for the study of the origin and nature of fast neutrals in the expanding plume, since their presence has been generally reported at larger laser fluences. Plasma parameters such as electron temperature and number density have been calculated for a broad range of ambient pressures from $5\times10^{-2}$ Torr to $2\times10^2$ Torr using the relative emission line intensity ratios and the Stark broadening mechanisms respectively for both fs and ns irradiation and are compared. Moreover, an OTOF measurement at 481 nm is performed to determine the dynamics of neutral Zn species in the plume. Atomic emission is found to be prominent in the Zn plasma plume at higher pressures. Spatial information is recorded by scanning the plasma plume along its length, while plume imaging done at various time delays using an ICCD gives a two dimensional view of the expanding LPP. A double-pulse experiment is also performed using fs excitation, to understand the effect of a second pulse on the emission dynamics of fs LPP.

## 5.2 OPTICAL EMISSION SPECTROSCOPY STUDIES

OES of the LPP plume is useful for the estimation of important plasma parameters such as electron temperature ($T_e$) and electron number density ($N_e$). Other parameters such as plasma frequency ($\omega_p$), inverse bremsstrahlung ($\alpha_{ib}$) etc can be deduced from the values of $T_e$ and $N_e$. Therefore, OES of LPP in the visible region has been recorded using a CCD spectrometer. Typical fs and ns LPP emission spectra recorded at 10 Torr nitrogen ambient pressure are presented in the following section in detail. Intensity variations of the emitted spectral lines at different axial positions in the plume, such as 2 mm, 4 mm, 6 mm and 8 mm from the target surface, are recorded for pressures ranging from $5\times10^{-2}$ Torr to $2\times10^{2}$ Torr, and the results are explained.

### 5.2.1 Optical emission spectra

Emission spectra measured in the plume expansion direction at 2 mm, 4 mm, 6 mm and 8 mm distances from the target surface on the plume axis are shown in figures 5.1 and 5.2 respectively. Background pressure is 10 Torr of nitrogen. Even though the wavelength spectrum is essentially unchanged, emission intensity gets reduced by two orders of magnitude upon translation by 4 mm (from the 2 mm to 6 mm position) along the plume axis for fs LPP. At 8 mm no or little line emission is detected for fs LPP whereas emission from ns LPP is still observable. Lines in the emission spectra identified using the standard NIST data base are used for estimating the plasma parameters. Identified spectral lines include emissions from neutral Zn (Zn I) at 330 nm, 334.5 nm, 468 nm, 472 nm, and 481 nm, and emissions from ionized Zn (Zn II) at 491 nm and 492 nm. Emissions from neutral lines are prominent in the LPP, depicting the triplet structure of Zn. Shorter wavelength lines in the emission spectra (such as 334.5 nm) are less intense in ns LPP compared to fs LPP. Moreover, the variation of emission intensity at the shorter and longer wavelengths with pressure is essentially similar to that observed for nickel, as discussed in chapter 4 (section 4.2).

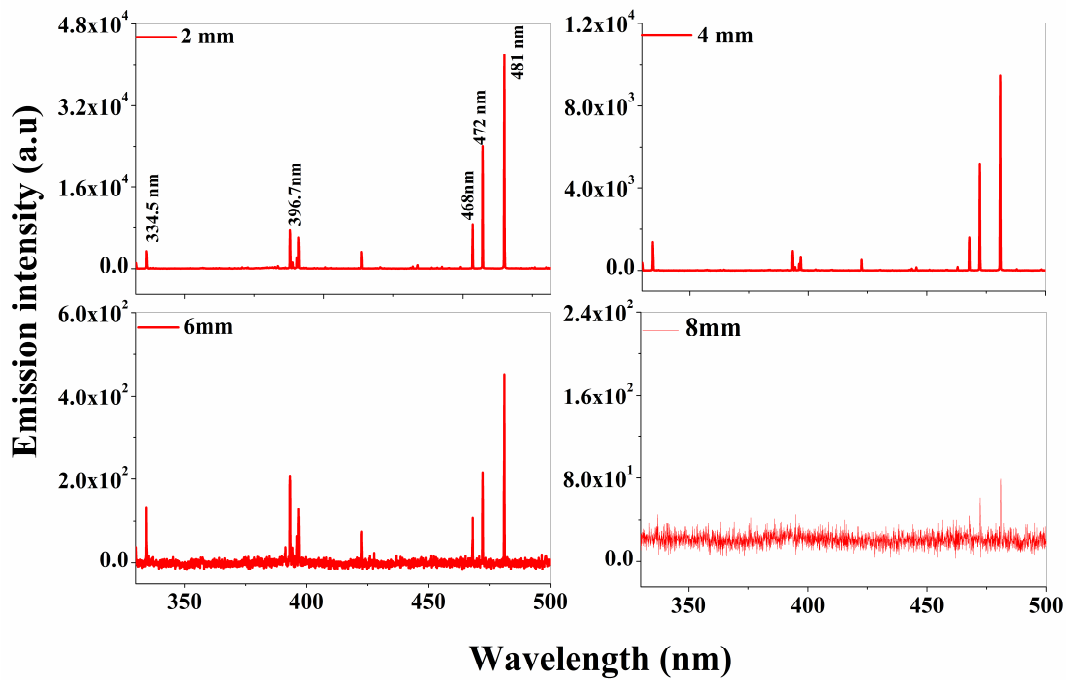

Figure 5.1. Emission in the visible region from an ultrafast Zinc plasma produced by the irradiation of a Zn target by 100 fs laser pulses at 796 nm. Spectra recorded at distances of 2 mm, 4 mm, 6 mm and 8 mm respectively from the target surface, along the expansion direction. Ambient pressure is 10 Torr.

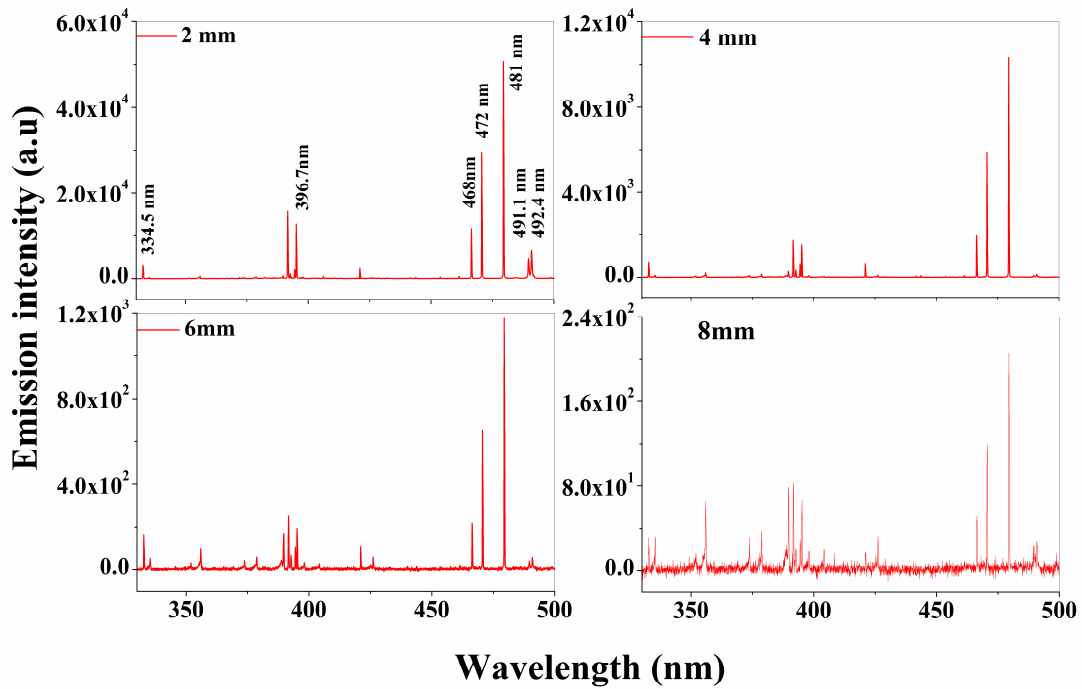

Figure 5.2 Emission in the visible region from a nanosecond Zn plasma produced by the irradiation of a Zn target by 7 ns laser pulses at 1064 nm. Spectra recorded at distances 2 mm, 4 mm, 6mm and 8 mm respectively from the target surface, along the expansion direction. Ambient pressure is 10 Torr.

**5.2.2 Effect of ambient pressure on emission intensity**

Ambient pressure plays a significant role in the intensity and lifetime of laser produced plasmas. At low pressures the plume expands more, reducing the number density and collisions, and therefore, the ionization rate. As the pressure increases self-focusing of the beam and multi-photon ionization of ambient gas will occur, reducing the energy reaching the target [115]. This results in a reduction in the quantity of the ablated material. On the other hand at high pressures the collision rate increases, leading to enhanced ionization of the ablated species. As a result of these competing processes, the emission intensity peaks at a certain intermediate pressure. We measured plasma emission for nitrogen pressures ranging from $5\times10^{-2}$ Torr to $5\times10^{2}$ Torr. The measured intensities of different spectral lines from Zn I, plotted as a function of background pressure for fs and ns irradiations, are shown in Fig 5.3. At a distance of 2 mm from the target surface, maximum emission occurred for the background pressures of 10 Torr and 5 Torr respectively, for ns and fs irradiations.

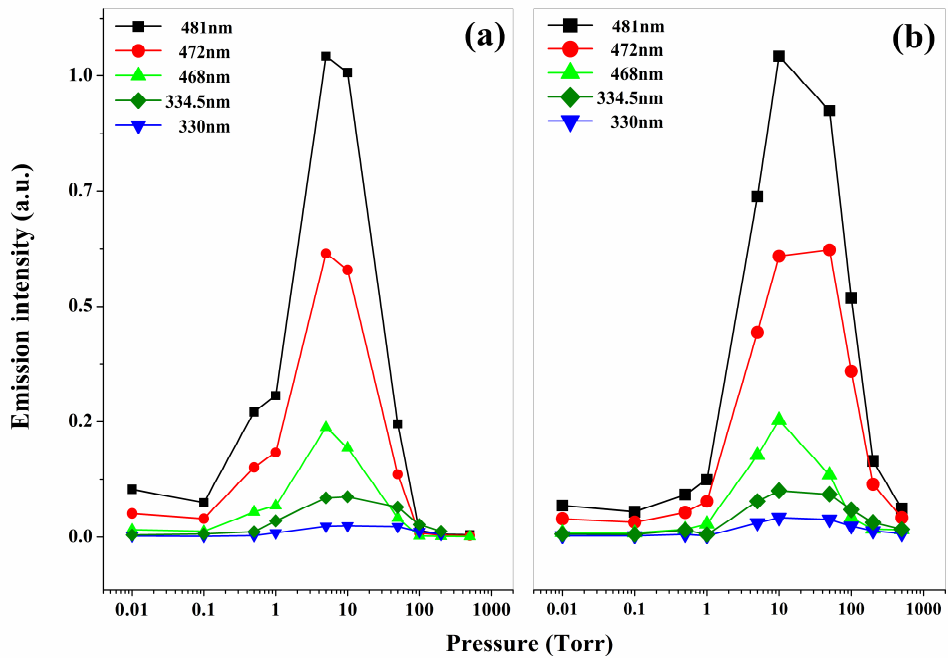

Figure 5.3. Variation of emission intensity with ambient pressure (plotted in the log scale) for different plasma spectral lines of Zn I species, measured 2 mm away from the target surface, for irradiation by (a) 100 fs and (b) 7 ns laser pulses.

### 5.2.3 Dependence of emission intensity on laser pulsewidth

Unlike ns laser pulses, fs laser pulses do not interact with the plasmas produced by them, due to the fact that the electron-ion energy transfer time ($t_{ei}$) and the heat conduction time ($t_h$) are of the order of ps, which is greater than the pulse duration of the fs laser ($t_p$) i.e. $t_{ei} \sim t_h \gg t_p$. Therefore, the entire energy content of the laser pulse deposited on the target results in the generation of a large number of excited neutral species, which intensifies the emission from Zn I in fs LPP at larger laser intensities, as shown in Fig 5.1. However, ionization is found to be stronger in ns LPP compared to fs LPP because of the interaction of the trailing part of the laser pulse with the plasma produced by it, resulting in a large emission yield for Zn II. It is clear from Figure 5.1a that line emissions at shorter wavelengths can be enhanced by fs excitation, illustrating the pulsewidth dependence of emission intensity at shorter wavelengths.

### 5.2.4 Estimation of temperature and number density

Calculation of plasma parameters (electron temperature and number density) was done by conventional methods as follows. Electron temperatures ($T_e$) were evaluated using the line emission intensity ratios of Zn I emissions at 334.5 nm (4s4d $^3D_3$ - 4s4p $^3P_2$) and 481.0 nm (4s5s $^3S_1$ - 4s4p $^3P_2$) by assuming local thermodynamic equilibrium (LTE), using equation (3.1).

The spectroscopic constants required for the calculation ($A_i$ and $g_i$) are taken from a previous work [117], and the estimation of electron number density ($N_e$) was done from Stark broadening of line emission at 481 nm (neglecting ion contribution to line broadening) at each pressure, using equation 3.3 [108]. The theoretical value of $\omega$ was obtained from literature [118]. A Lorentzian fit to the emission profile at 481 nm, measured 2 mm away from target surface at an ambient pressure of 10 Torr, is shown in Figure 5.4. Calculated values of $T_e$ and $N_e$ for various ambient pressures, for different distances in the plume from the target surface, are shown in Table I. The validity of LTE was checked using the McWhirter criterion for each pressure using equation 3.4, which gives a minimum number density of $2.4 \times 10^{15}$ per cubic centimeter in the present case.

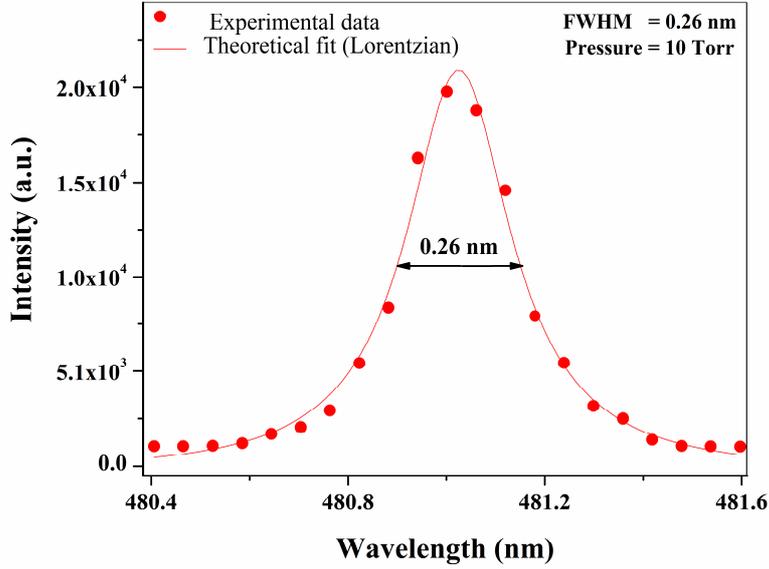

Figure 5.4. Lorentzian fit to the Zn I emission profile at 481 nm recorded at 10 Torr, measured at a distance of 2 mm from the target surface. The linewidth (FWHM) is 0.26 nm.

TABLE 5.1: Electron temperature ($T_e$) and number density ($N_e$) of fs LPP, calculated for different ambient pressures, at increasing distances from the target surface. LTE is found to be valid in all cases.

| Pressure (Torr) | 2mm | | 4mm | | 6mm | |
|---|---|---|---|---|---|---|
| | $T_e$ (K) | $N_e$ (cm$^{-3}$) (x10$^{15}$) | $T_e$ (K) | $N_e$ (cm$^{-3}$) (x10$^{15}$) | $T_e$ (K) | $N_e$ (cm$^{-3}$) (x10$^{15}$) |
| 0.05 | 6826 ± 19 | 3.32 ± 0.16 | 5767 ± 26 | 3.29 ± 0.22 | 5472 ± 31 | 3.16 ± 0.14 |
| 0.1 | 6406 ± 64 | 3.26 ± 0.74 | 4356 ± 30 | 3.17 ± 0.10 | 4005 ± 70 | 3.16 ± 0.15 |
| 0.5 | 4987 ± 82 | 3.56 ± 0.29 | 4725 ± 10 | 3.29 ± 0.22 | 4705 ± 14 | 3.18 ± 0.16 |
| 1 | 5248 ± 16 | 3.82 ± 0.44 | 4900 ± 16 | 3.24 ± 0.12 | 4572 ± 14 | 3.16 ± 0.10 |
| 10 | 8754 ± 69 | 3.88 ± 0.70 | 8844 ± 34 | 3.32 ± 0.42 | 8727 ± 28 | 2.73 ± 0.17 |

TABLE 5.2: Electron temperature ($T_e$) and number density ($N_e$) of ns LPP, calculated for different ambient pressures, at increasing distances from the target surface. LTE is found to be valid in all cases.

| Pressure (Torr) | 2mm | | 4mm | | 6mm | |
|---|---|---|---|---|---|---|
| | $T_e$ (K) | $N_e$ (cm$^{-3}$) ($\times 10^{15}$) | $T_e$ (K) | $N_e$ (cm$^{-3}$) ($\times 10^{15}$) | $T_e$ (K) | $N_e$ (cm$^{-3}$) ($\times 10^{15}$) |
| 0.05 | 8185 ± 165 | 2.61 ± 0.05 | 7991 ± 43 | 2.24 ± 0.02 | 7308 ± 41 | 2.27 ± 0.00 |
| 0.1 | 7658 ± 89 | 2.63 ± 0.02 | 7297 ± 17 | 2.23 ± 0.10 | 6705 ± 50 | 2.22 ± 0.04 |
| 0.5 | 8195 ± 41 | 2.54 ± 0.08 | 7717 ± 56 | 2.24 ± 0.04 | 7703 ± 34 | 2.23 ± 0.02 |
| 1 | 8737 ± 24 | 2.65 ± 0.02 | 8633 ± 37 | 2.26 ± 0.05 | 8009 ± 13 | 2.24 ± 0.01 |
| 10 | 9924 ± 90 | 2.75 ± 0.06 | 9697 ± 48 | 2.37 ± 0.08 | 9363 ± 78 | 2.25 ± 0.07 |

### 5.2.5 Effect of laser pulsewidth on $T_e$ and $N_e$

In laser ablation (LA) by ns pulses the trailing part of the laser pulse will interact with the ablated species, whereas in fs LA the laser pulse duration is significantly short for such interaction to occur. Unlike ns LPP, fs LPP shows a forward axial expansion normal to the target surface, providing more ablated particles to be confined in the axial direction. $T_e$ values calculated from the measured OE spectra plotted as function of pressure for fs and ns irradiations are shown in Fig 5.5a. The maximum and minimum values of Te are found to be 8688 K and 7075 K respectively for ns LPP whereas the corresponding values are 7599 K and 5302 K for fs LPP. The variation of $T_e$ with pressure shows similar trends for both fs and ns LPP at the point of measurement.

The calculated $N_e$ is found to be larger for fs LPP, which increases with pressure at the 2 mm axial position, but little or no variation is seen for ns LPP (Fig 5.5b). While the number density varies from 4.92 x 10$^{15}$ cm$^{-3}$ to 7.47 x 10$^{15}$ cm$^{-3}$ for fs LPP, it is from 2.54 x 10$^{15}$ cm$^{-3}$ to 2.76 x10$^{15}$ cm$^{-3}$ for ns LPP. Since the fs laser pulse ablates more material compared to ns LPP due to the more efficient laser-target energy coupling, the larger $N_e$ values measured close to the target in fs LPP are according to expectation. Away from the target surface the number density falls, leading to diminishing plume size along the axial expansion direction. The evaluated LPP parameters such as $T_e$ and $N_e$ for a given laser fluence reinforces the existing investigations [89, 93].

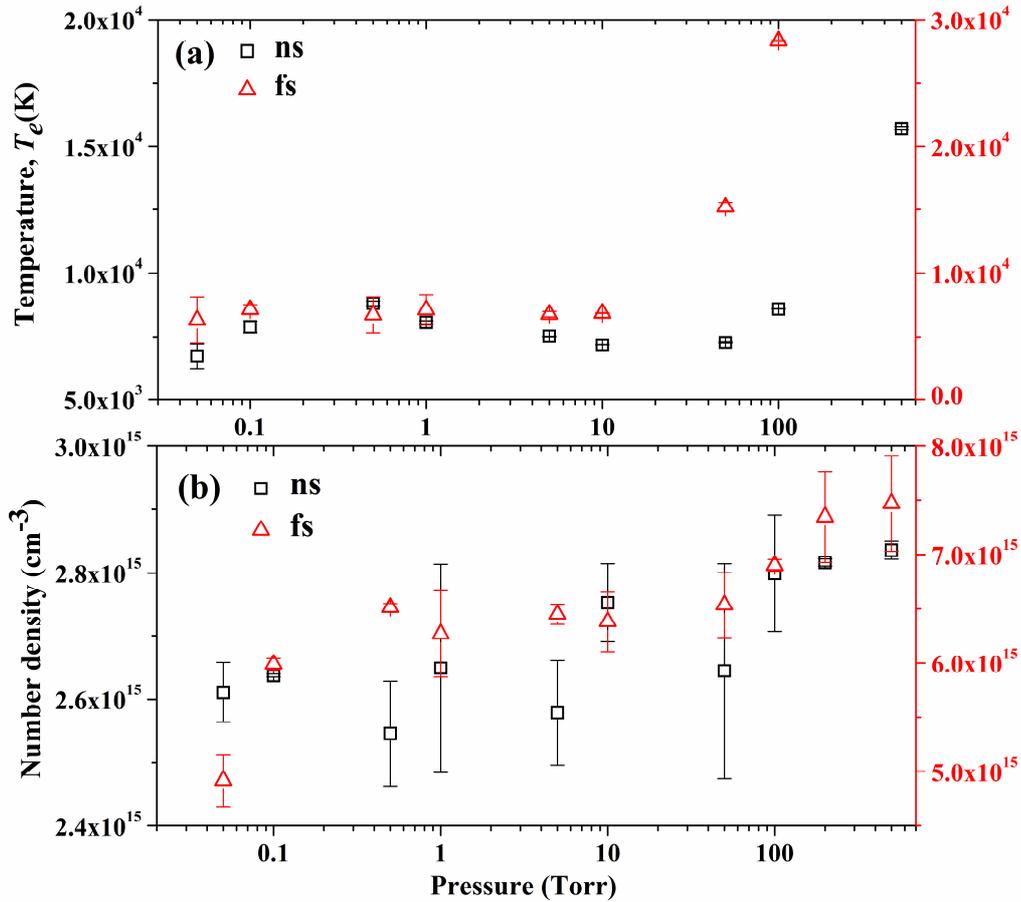

Figure 5.5. (a) Electron temperature and (b) number density calculated from plasma emission measured at 2 mm distance from the target surface, plotted as a function of background pressure, for both nanosecond and femtosecond irradiations. Error bars are estimated from multiple measurements.

## 5.3 OPTICAL TIME OF FLIGHT (OTOF) MEASUREMENTS

OTOF measurements of neutral Zn at 481 nm, for various background pressures, were carried out at a high laser fluence of 16 J/cm$^2$, which is well above the ablation threshold. Two peaks are observed in the OTOF signal, namely, a fast peak and a slow peak, which are contributions from the recombined and un-ionized neutrals respectively. While the time of arrival of fast species is unaffected by pressure variations in the measured range for fs LPP, slow species are affected by the presence of shock waves at higher pressures. On the other hand, for ns LPP, the time of arrival of fast species in the plume is found to be dependent on background pressure. Obviously, velocities of the plasma species play an important role in the emission properties of the plume, which moves rapidly, radiating in the visible, UV, and X-ray spectral regions.

### 5.3.1 Effect of ambient pressure on TOF signals

OTOF measurements were done for distances of 2 mm, 4 mm and 6 mm from the target surface, for an ambient pressure range of $5 \times 10^{-2}$ Torr to $1 \times 10^2$ Torr. Line emission lasts up to ~ 0.9 µs, 5 µs and 3 µs, at the 2 mm, 4 mm and 6 mm positions respectively, for 10 Torr background pressure for fs LPP. The emission lasts long with a duration of ~ 0.9 µs and ~ 2 µs for fs LPP and ns LPP respectively at 10 Torr (the pressure at which emission intensity is maximum).

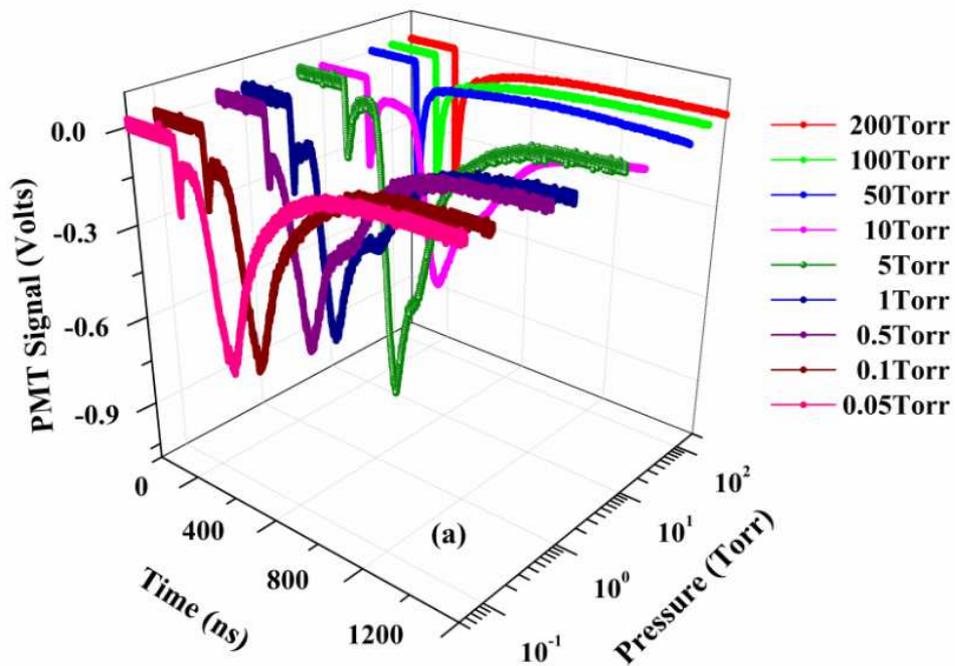

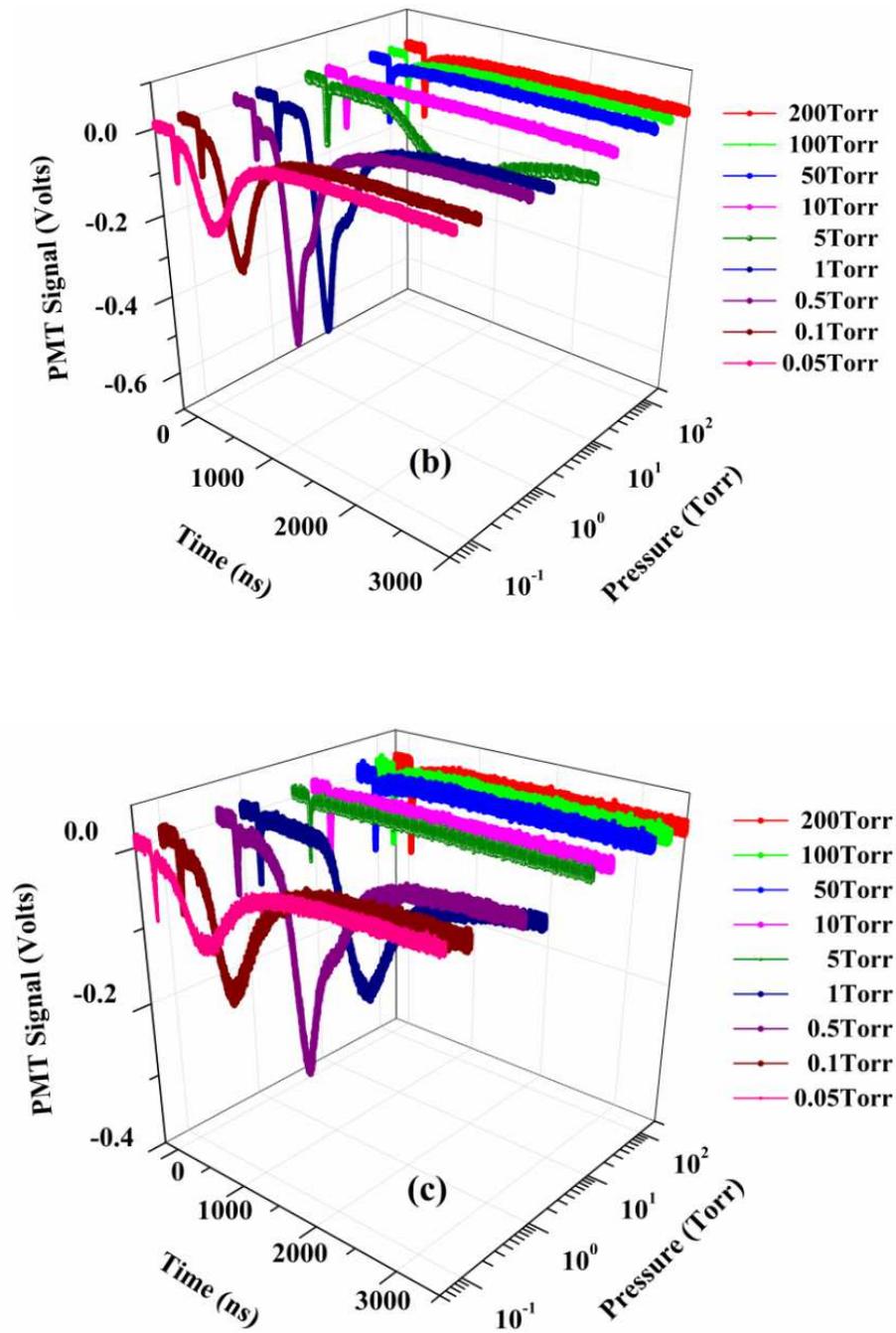

Figure 5.6. Time of flight emission of the 481 nm (4s5s $^3S_1 \rightarrow$ 4s4p $^3P_2$) line of Zn I at different ambient pressures, measured in the plasma plume at distances of (a) 2 mm, (b) 4 mm and (c) 6 mm, respectively, from the target surface, for fs irradiation.

As seen from Fig. 5.3, maximum emission is found at an optimum pressure of 10 Torr for the 2 mm position. Results are shown in Fig. 5.6a, 5.6b and 5.6c respectively. Since the electric field corresponding to the laser pulse intensity is ~ $3.5 \times 10^{10}$ Vm$^{-1}$, which is above the threshold for Coulomb explosion ($1.173 \times 10^{10}$ Vm$^{-1}$), the target surface vaporizes directly upon fs irradiation [111], causing instantaneous (~ 1-100 picoseconds delay) material ablation as the excitation pulse hits the Zn target, and the ejected species expand until the pressure within and outside the plume are balanced.

The processes involved in fs plasma plume formation are simpler compared to those of ns plasma plume formation, but the dynamics of emission is rather complex as revealed by the OTOF signal. It is clear from Fig. 5.6 that the pressure range of $5 \times 10^{-2}$ Torr to $1 \times 10^{1}$ Torr is the regime in which the plasma plume experiences turbulence during expansion. This is because of the competing processes prevalent at these pressures. On the one hand, large number densities ensure collisions thereby maximizing ionization, while on the other hand, recombination of faster ions with electrons enhances optical emission from neutrals at 481 nm. Measurements show that LPP characteristics such as emission life time, emission peak delay and intensity vary significantly with pressure (See Figs 5.6 and 5.7).

The ablation threshold, which is defined as the minimum power density required for vaporizing the material, is given by equation 4.1 [20], and $I_{min}$ can be calculated as $2.77 \times 10^{10}$ W/cm$^{2}$ in the present case (the corresponding laser fluence is $2.77 \times 10^{-3}$ J/cm$^{2}$). As seen from Fig. 5.6, all OTOF spectra exhibit one sharp peak (PK1) first, the intensity of which increases steadily with pressure at the 2 mm position. PK1 is found to have a consistent time delay which is rather independent of the ambient pressure. Therefore it may be suspected that PK1 is a photopeak; however, if the spectrometer setting is changed by 1 nm either to 480 nm or 482 nm, the peak disappears. This confirms that PK1 indeed indicates fast atomic species, which forms the leading edge of the moving plasma plume. In fact this peak arises from the recombination of fast ions with electrons in the early stages of plasma plume evolution. With a laser fluence of ~ 16 J/cm$^{2}$ which is way above the ablation threshold, an efficient ablation of the target material happens rapidly, and a shock wave propagates back and forth through the ambient gas [119], introducing a peak shift at certain pressures in the OTOF. This also results in a small third peak, PK3, which appears as a wing to PK2 at lower pressures as seen in Fig. 5.6. PK2 and PK3 disappear at higher pressures indicating that plume expansion is hindered due to the plasma confinement effect.

As the ambient pressure increases from 0.05 Torr, the integrated line intensity at 481 nm in the TOF shows no or little change initially, but the intensity starts building up after a certain pressure, and reaches a maximum at 5 Torr for fs LPP and 10 Torr for ns LPP (Figs 5.6a and 5.7). A further increase in pressure reduces the emission intensity. Although the emission lifetime is found to be larger for ns LPP, the integrated emission intensity is more for fs LPP. The presence of shock waves play an important role during LPP expansion, causing slow atomic species to split and move with different average velocities within a range of pressures.

### 5.3.2 Effect of pulsewidth on TOF signals

As mentioned before, in ns LPP material ablation occurs by the leading part of the laser pulse and the energy content of the trailing part is absorbed by the ablated species. Ionization followed by radiative recombination makes optical emission feasible up to several microseconds after laser excitation. On the other hand in fs excitation plasma formation happens only after the laser pulse is over, and the absence of laser-plasma interaction results in optical emission of shorter timescales. OTOF measurements are carried out on Zn I species by recording line evolution at 481nm ($4s5s\,^3S_1 \rightarrow 4s4p\,^3P_2$, Zn I) using a fast PMT coupled to the spectrometer for ns irradiation. Measurements done at the 2 mm position is shown in figure 5.7. Intensity variations in the emission of neutrals with time at various ambient pressures is measured for ns as well as fs LPP. The PMT signal is relatively stronger for fs LPP, and maximum signal is measured around 10 Torr for both fs and ns pulses. At low pressures (i.e. for 0.05 Torr to 0.1 Torr) two different peaks are observed in the TOF, namely, the fast peak (PK1) and the slow peak (PK2).

### 5.3.3 Velocities of neutrals in the LPP

When expanded adiabatically, the fast species within femtosecond plasma move with an average speed ~100 kms$^{-1}$ while slow species move at ~1-10 kms$^{-1}$. Ions are ejected normal to the target surface, and measurements show that at 10 Torr pressure, they are accelerating with average velocities of 57 ± 2 kms$^{-1}$, 109 ± 9 kms$^{-1}$, and 154 ± 12 kms$^{-1}$, respectively, at the 2 mm, 4 mm, and 6 mm positions in the plume. This acceleration is caused by the electron cloud which moves faster than the ion cloud in the plasma: the electric field created by the electrons attract the ions so that they follow, and some of these recombine with the electrons to form neutral atoms (a detailed study is given in chapter 6). Velocities measured for different ambient pressures are given in Fig. 5.8 for fs LPP. PK1 is followed by a stronger and broader peak PK2, which, unlike PK1, becomes weaker at higher pressures, and disappears altogether at the highest pressures studied. Similarly, the average velocity of the relatively long-lived PK2 drops from 10 kms$^{-1}$ to 2 kms$^{-1}$ as the pressure is increased from 0.05 Torr to 10 Torr. Adiabatic expansion of the plume consequent to intense fs excitation results in a quick reduction of emission intensity away from the target surface.

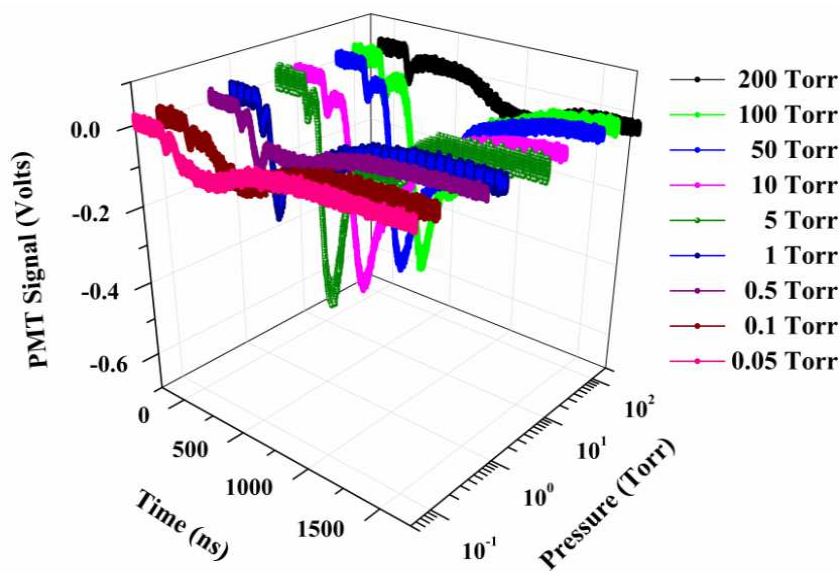

Figure 5.7. Line emission intensity from Zn I at 481 nm for different ambient pressures, plotted as a function of time, for ns irradiation.

PK1 is found to have the same time of arrival at the point of observation for all pressures studied in fs LPP. From fs LPP results it can be assumed that the fast peak arises from the recombination of fast ions with electrons in the early stages of plasma plume evolution [14], with velocities ~ $10^4$ m/s. The velocity of species contributing to PK1 is found to be unchanged on fs irradiation, but it decreases for ns excitation with pressure, as shown in Fig 5.9. This implies that the movement of fast neutrals is highly directed along the axial expansion direction due to the self generated magnetic field. This field pinches the plasma, guiding the ablated species to be confined along the axis in the earlier stages of expansion, as the irradiation intensity is huge for fs LPP. The slow peak, PK2, can be ascribed to the emission from the slower atomic species which moves with a velocity ~$10^3$ m/s. A new peak (PK3) appears in the OTOF from 0.5 Torr to 10 Torr for fs LPP, whereas this additional peak is observed for pressures above 0.1 Torr for ns LPP. Unlike the OTOF of ns LPP, PK2 and PK3 cease to exist above 10 Torr in the fs line emission dynamics. The average velocity of PK3 measured at a distance of 2 mm from the target surface is presented in Table 5.3.

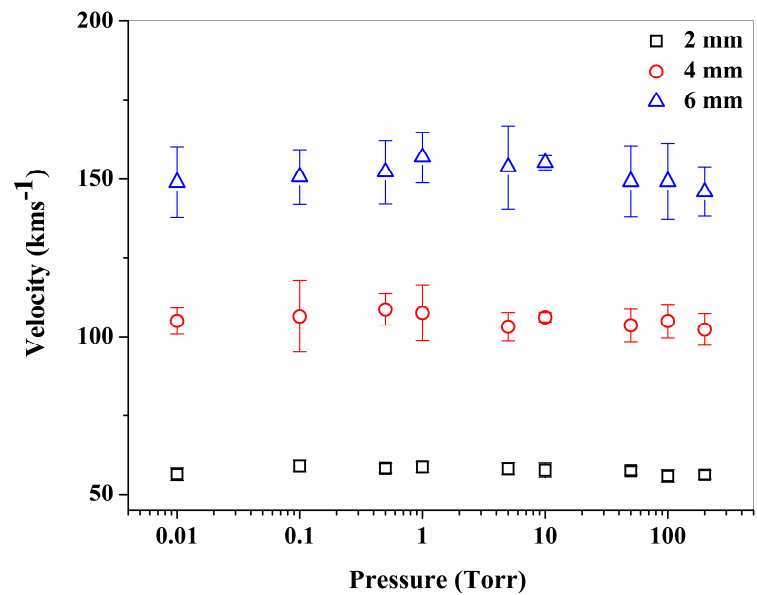

Figure 5.8. Velocity of PK1 at different distances from the target surface, measured for the ambient pressure range of 0.01 to 200 Torr. Error bars are obtained from multiple measurements.

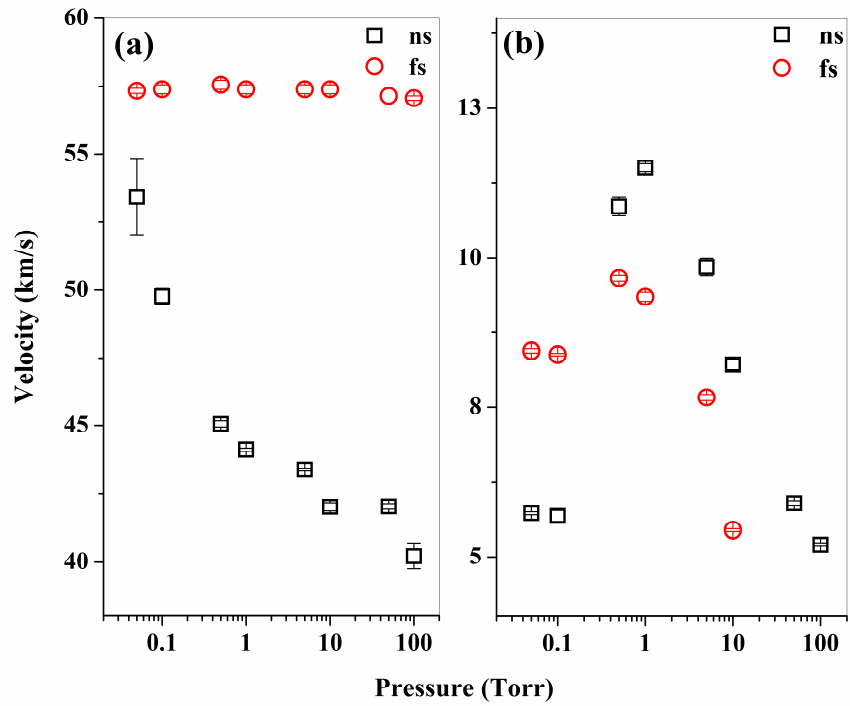

Figure 5.9. Velocities of (a) fast and (b) slow species upon irradiation with ns and fs laser pulses, measured at 2 mm distance from the target surface, for an ambient pressure range of 0.01 to 100 Torr. Error bars are obtained from multiple measurements.

Table 5.3. Average velocities measured for PK3 at 2 mm distance from the target surface, for fs and ns LPP. Error is calculated from multiple measurements (PK3 was not observed for 50 and 100 Torr background pressures).

| Pressure (Torr) | Velocity of PK3 (m/s) | |
| --- | --- | --- |
| | ns LPP | fs LPP |
| 0.5 | 4.41 ± 0.29 | 4.95 ± 0.32 |
| 1 | 4.08 ± 0.17 | 4.82 ± 0.35 |
| 5 | 3.10 ± 0.11 | 5.60 ± 0.15 |
| 10 | 3.58 ± 0.26 | 3.44 ± 0.15 |
| 50 | 3.48 ± 0.12 | --- |
| 100 | 3.38 ± 0.36 | --- |

**5.4. ICCD IMAGING OF THE EXPANDING PLASMA**

Intensified CCD measurements allow characterization of the plasma and the studies on the hydrodynamics of expanding LPP at various time delays after the laser pulse reaches the target surface. Figures 5.10, 5.11 and 5.12 show images of the expanding ns LPP measured for various time delays such as 50 ns, 100 ns, 150 ns etc. for a given gatewidth of 5 ns. The dimension of each image is 1 cm × 1.25 cm for $5\times10^{-2}$ Torr and 5 Torr, and 1 cm × 1 cm for 100 Torr. Larger expansion and shorter lifetime at low pressures is obvious from the measurements. It can be found from the plume intensity in the present case that the emission is present up to 900 ns at lower pressures but at higher pressures (like 5 Torr and 100 Torr) it is visible up to 4000 ns with less expansion. The inference is that lifetime of the plume increases and expansion is hindered by plume confinement at higher pressures. Splitting of the plasma plume at certain delays (see fig 5.11 c and d) are also observed in the ICCD measurements. Even though the plasma lifetime is long due to collisional excitation via plasma confinement, the plume intensity reduces by thermal leak to the surroundings as it is proportional to the density of the background gas.

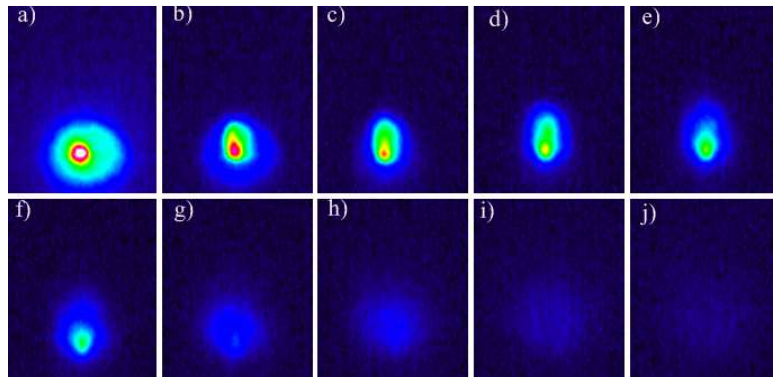

Figure 5.10. Images of nanosecond laser produced zinc plasma expanding into nitrogen ambient at 0.05 Torr pressure, measured at time delays of (a) 50 ns, (b) 100 ns, (c) 150 ns, (d) 200 ns, (e) 250 ns, (f) 300 ns, (g) 400 ns, (h) 500 ns, (i) 700 ns, and (j) 900 ns after the laser pulse, using an ICCD gatewidth of 5 ns. The spatial dimension of the image is 1 cm × 1.25 cm.

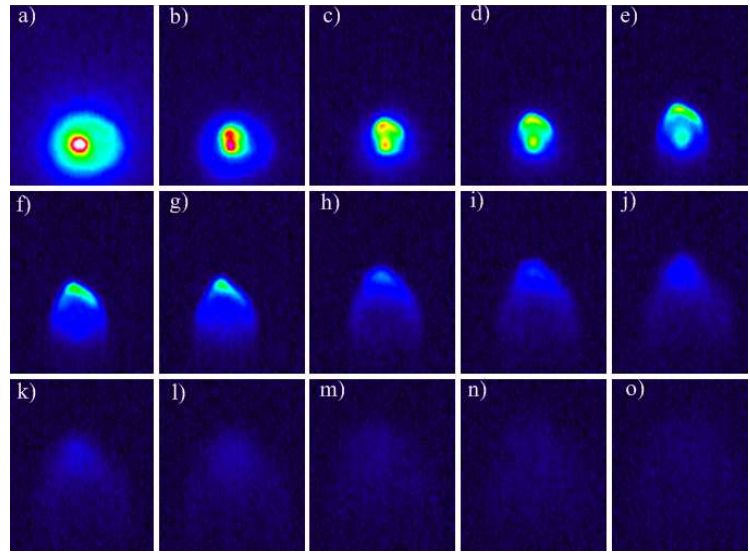

Figure 5.11. Images of nanosecond laser produced zinc plasma expanding into nitrogen ambient at 5 Torr pressure, measured at time delays of (a) 50 ns, (b) 100 ns, (c) 150 ns, (d) 200 ns, (e) 300 ns, (f) 400 ns, (g) 500 ns, (h) 800 ns, (i) 1000 ns, (j) 1200 ns, (k) 1600 ns, (l) 2000 ns, (m) 2500 ns, (n) 3000 ns, and (o) 4000 ns after the laser pulse, using an ICCD gatewidth of 5 ns. The spatial dimension of the image is 1 cm × 1.25 cm.

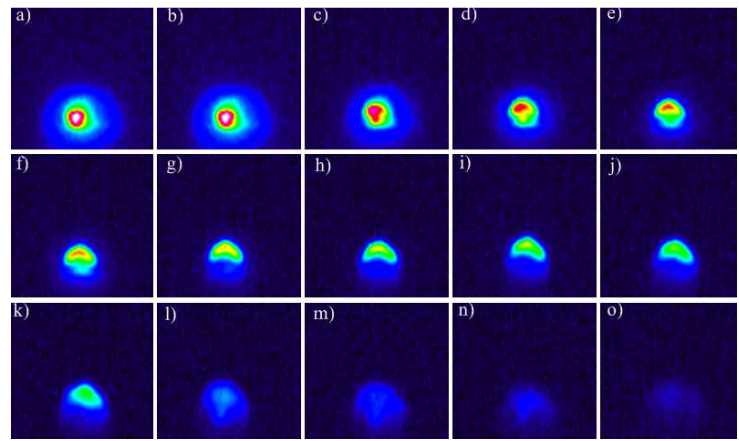

Figure 5.12. Images of nanosecond laser produced zinc plasma expanding into nitrogen ambient at 100 Torr pressure measured for time delays of (a) 50 ns, (b) 100 ns, (c) 150 ns, (d) 200 ns, (e) 250 ns, (f) 300 ns, (g) 400 ns, (h) 500 ns, (i) 600 ns, (j) 700 ns, (k) 1000 ns, (l) 1600 ns, (m) 200 ns, (n) 3000 ns, and (o) 4000 ns after the laser pulse, using an ICCD gatewidth of 5 ns. The spatial dimension of the image is 1 cm × 1 cm.

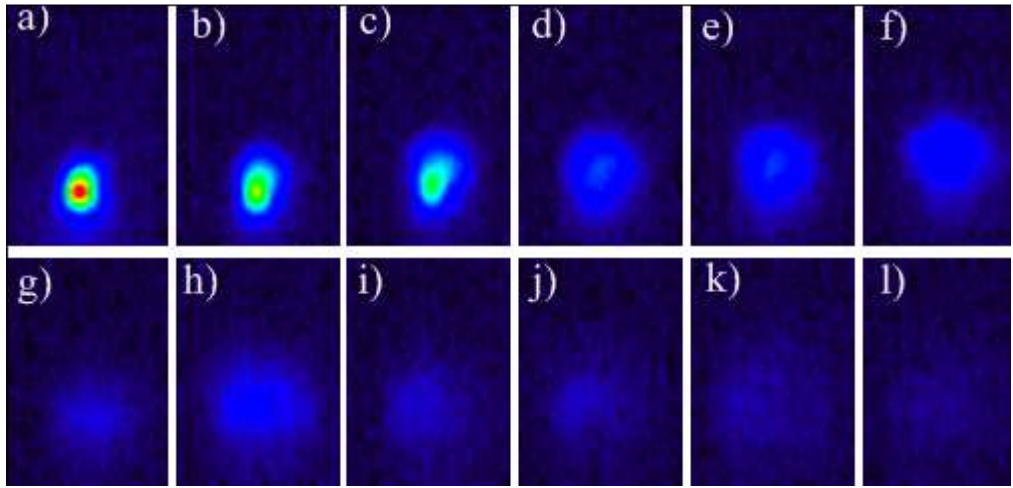

Figure 5.13. Images of femtosecond laser produced zinc plasma expanding into nitrogen ambient at 0.05 Torr pressure measured for time delays of (a) 50 ns, (b) 100 ns, (c) 150 ns, (d) 200 ns, (e) 250 ns, (f) 300ns, (g) 350 ns, (h) 400 ns, (i) 450 ns, (j) 500 ns, (k) 550 ns, and (l) 600 ns after the laser pulse, using an ICCD gate width of 10 ns. The spatial dimension of the image is 0.5 cm × 0.75 cm.

Figures 5.14, 5.15 and 5.16 show the images measured for fs LPP at pressures of 0.05 Torr, 5 Torr and 100 Torr, with a gatewidth of 10 ns, showing a plume which is axially elongated compared to ns LPP excited under similar conditions. Plasma lifetime is also found to be less in comparison to ns LPP due to the absence of laser-plasma interaction, which is a dominating process in ns LPP under larger laser fluence, which increases its lifetime via laser plasma energy coupling. At higher pressures the plasma is found to be more intense compared to lower pressures (for similar delay times) indicating large ionization of background gas due to the emission of high energy radiation from the plasma in the initial stages of expansion. ICCD imaging once again confirms the enhanced confinement of fs LPP along the axial direction, as explained in the previous chapter.

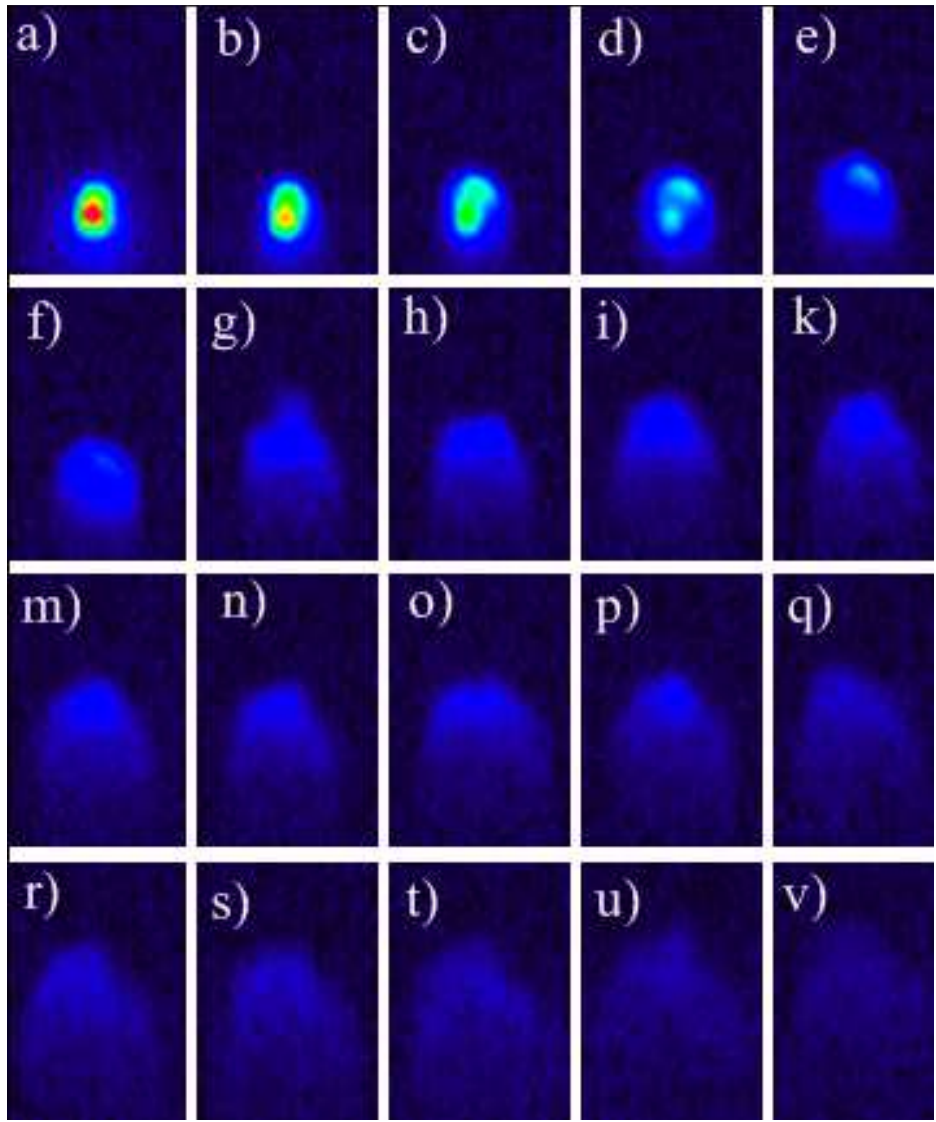

Figure 5.14. Images of femtosecond laser produced zinc plasma expanding into nitrogen ambient at 5 Torr pressure measured for time delays of (a) 50 ns, (b) 100 ns, (c) 150 ns, (d) 200 ns, (e) 250 ns, (f) 300 ns, (g) 350 ns, (h) 400 ns, (i) 450 ns, (j) 500 ns, (k) 550 ns, (l) 600 ns, (m) 650 ns, (n) 700 ns, (o) 800 ns, (p) 900 ns, (q) 1000 ns, (r) 1200 ns, (s) 1400 ns, (t) 1600 ns, (u) 1800 ns, and (v) 2000 ns after the laser pulse, an ICCD gate width of 10 ns. The actual spatial dimension of each image is 0.5 cm × 0.75 cm.

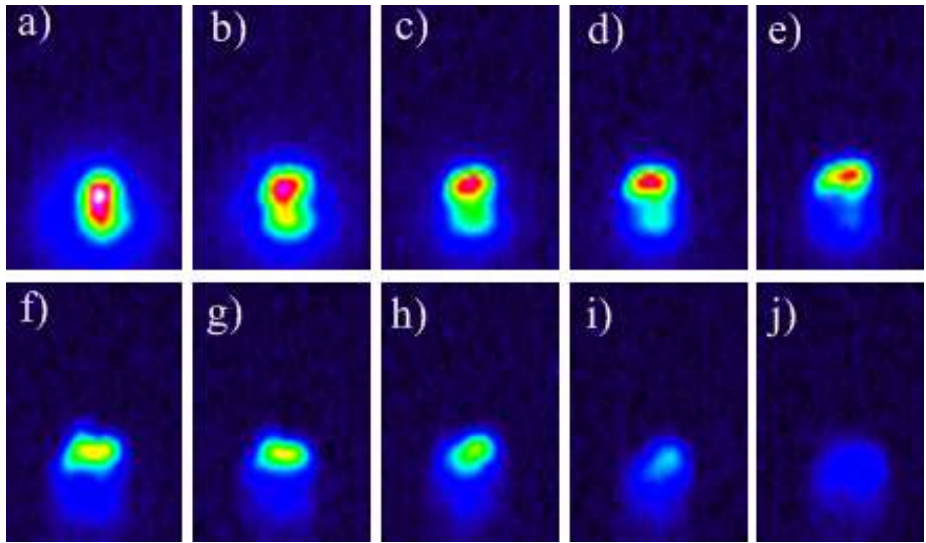

Figure 5.15. Images of femtosecond laser produced zinc plasma expanding into nitrogen ambient at 100 Torr pressure measured for time delays of (a) 50 ns, (b) 100 ns, (c) 150 ns, (d) 200 ns, (e) 300 ns, (f) 400 ns, (g) 500 ns, (h) 600 ns, (i) 1000 ns, and (j) 2000 ns after the laser pulse, an ICCD gate width of 10 ns. Actual spatial dimension of the image is 0.5 cm × 0.75 cm.

## 5.5 DOUBLE PULSE ULTRAFAST EXCITATION

A double pulse (DP) experiment [120-122] in which a 2 mJ pulse was fired into a plasma plume, which was generated 4 ns earlier by a 5 mJ laser pulse, was carried out in the fs irradiation scheme. To generate these pulses the laser was run at full energy (10 mJ) and the beam was divided into two using a 1:1 beam splitter. One of the beams was then sent through a delay stage, and attenuated to 2 mJ using a half-wave plate-polarizer cube combination. The TOF data measured for single pulse (SP) and double pulse (DP) excitations for 5 Torr background, at 2 mm and 4 mm distances from the target surface respectively, are shown in Fig. 9. The TOF dynamics reveals that the relative time of arrival of the fast component (PK1) is unaffected by double pulse excitation: however, the slow species (PK2) is accelerated upon irradiation with the second pulse. DP excitation facilitates coupling of laser energy to the plasma plume in the fs excitation domain: energy is coupled to the fast atomic species with the present delay time of 4 ns, leading to a reduction in its emission intensity. In the DP configuration the TOF signal intensity of PK2 is found to be nearly doubled at the 4 mm position compared to SP excitation, while it is halved for PK1 at the 2 mm and 4 mm positions. The DP plume decays faster than the SP plume as well.

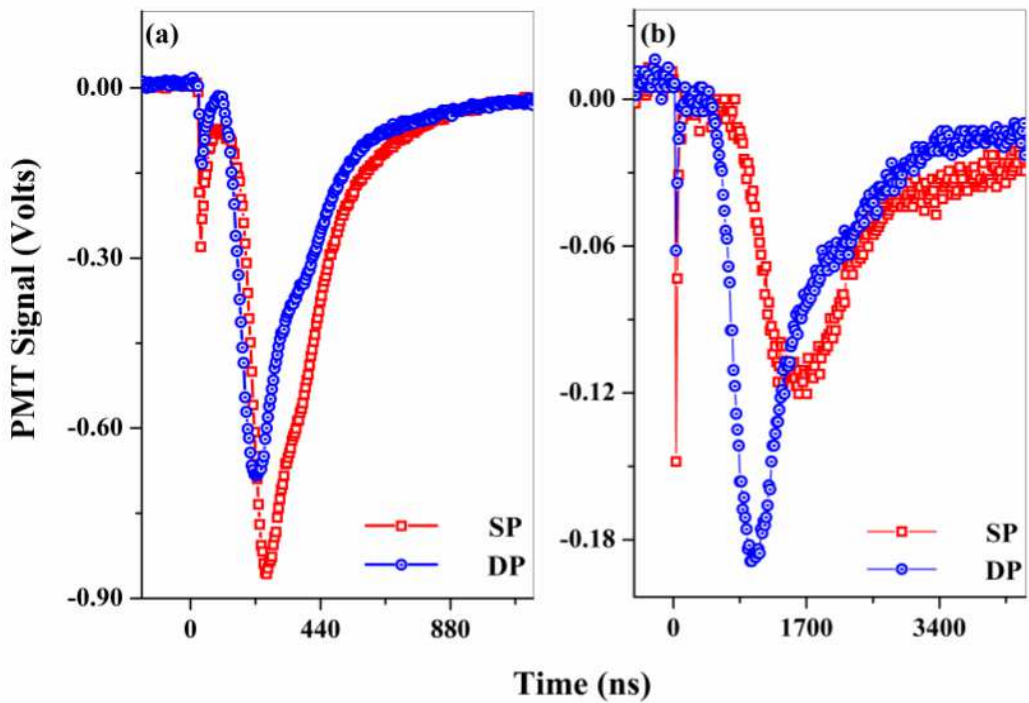

Figure 5.16. TOF spectra of 481 nm emission from Zn I at an ambient pressure of 5 Torr, measured using single pulse (SP) and double pulse (DP) excitations, at distances of (a) 2 mm and (b) 4 mm from the target surface.

**5.6 CONCLUSION**

In conclusion, we have generated fs and ns LPPs from a high quality solid zinc target kept under broad ambient nitrogen pressures from $5\times10^{-2}$ Torr to $5\times10^{2}$ Torr, using 100 fs and 7 ns laser pulses of 800 nm and 1064 nm wavelengths respectively. With larger laser fluence ~16 J/cm$^2$, Zn I and Zn II lines are observed at a distance of 2 mm from the target surface. The temperature and number density are calculated from optical emission spectra and are plotted as a function of background pressure, and analyzed. OTOF of the line at 481 nm shows a double-peak structure, ascribed to the existence of fast and slow atomic species respectively in the LPP. Fast atomic species moves with an average velocity ~$10^4$ m/s whereas slow atomic species moves with a lower velocity ~$10^{-3}$ m/s. Interestingly, the average velocity of the fast atomic component varies with pressure upon ns excitation whereas it remains unchanged for fs excitation. ns LPP from solid Zn shows a longer plume lifetime. Emission from neutral Zn is found to be more intense in both fs and ns LPPs depicting the characteristic triplet structure of Zn. The optimized pressure for maximum emission intensity and life time is found to be ~ 10 Torr, in the present measurements. Furthermore, a double pulse excitation experiment is performed to understand the effect of energy coupling by a delayed ultrafast laser pulse on the emission dynamics of the plasma plume.

# CHAPTER 6
# ACCELERATION OF NEUTRALS IN LASER PRODUCED PLASMAS

*This chapter gives a study of the phenomenon of acceleration of neutrals observed in laser produced plasma from metal targets. Emission from neutrals has been measured to larger axial distances compared to ions in expanding plasmas. The measured times of arrival of ions and neutrals at larger laser fluences confirm the generation of fast neutrals via the recombination of fast ions with electrons. It is also observed that fast neutrals are absent at grater axial distances where emissions from ions are not detected. The estimated peak velocities of fast neutrals and ions are found to increase with an increase in the laser energy. Furthermore, an acceleration of fast neutrals and ions is observed close to the target, which is attributed to the space charge effect. Acceleration of fast neutrals is described in more detail.*

## 6.1 INTRODUCTION

Laser ablation occurs in a few picoseconds after the irradiation pulse, whose intensity is equal to or more than the ablation threshold of the target material. With sufficiently high laser energy, laser-target energy coupling is efficient so that the solid target transforms to the gaseous phase, forming a high temperature, high pressure, partially ionized gas cloud during the first few nanoseconds. This is followed by a rapid expansion leading to the evaporation of neutral atoms, electrons and ions [123] depending on ambient conditions. The velocity of the plasma front depends on the incident laser energy and on the atomic mass number of the target [124-125]. Dynamics of the plasma plume as modeled by R. Sauerbrey et.al. [59], gives an idea about the acceleration of the expanding plasma front during sub-picosecond laser ablation. During the earlier stage of expansion (i.e. in the first few nanoseconds), the inner part of the plasma is opaque to laser radiation due to its high electron density, but it becomes transparent whenever the density falls below the critical value. Therefore, almost the whole energy associated with the laser pulse is coupled to the plasma electrons, which in turn gain enough energy to get excited and further ionize the available atoms and ions in the first few nanoseconds. Besides electron impact ionization, photoionization of species in the plasma by single or multiphoton absorption from the irradiating laser beam causes the generation of more ions, which results in the emission of ultraviolet radiation originating from spontaneous emissions of highly excited atoms and ions and from recombination processes [126-127] which can again ionize the remaining neutral atoms [20, 49, 111].

In 2007, Amoruso.et. al. [14] reported the presence of fast neutrals in a fs laser produced nickel plasma at larger laser fluences, while tuning the laser parameters (such as wavelength and fluence) to optimize the properties of laser ablated nanoparticles and checking the plume atomization in a high vacuum ($\sim 10^{-7}$ Torr). The velocities of fast neutrals in fs laser produced zinc and nickel plasmas have been reported by Smijesh et.al. [109-116] in which acceleration up to a certain distance along the expansion direction (discussed in chapters 4 and 5) is estimated. The origin of fast neutrals and their kinetics in the LPP are studied in detail and presented in this chapter. Acceleration of these species in the expanding plasma under optimized conditions also are studied and discussed.

Experimental setup is similar to that described in chapter 3, except that the irradiation spot size is 300 $\mu$m in the present case. PMT signals recorded on a fast oscilloscope give different temporal emission profiles in the first few shots, but they do not change appreciably for about 50 further shots on the same irradiation point. In order to understand the effect of pressure on the emission dynamics, OTOF has been measured for various background pressures. For a laser produced Ni plasma, the pressure range in which emission from both neutrals as well as ions can be measured is found to be between $5\times10^{-2}$ Torr to $1\times10^{2}$ Torr, and the optimum pressure for obtaining maximum signal in the OTOF in nitrogen background is found to be between 1 Torr and 20 Torr. Measurements of the acceleration of neutrals in a laser produced Nickel plasma and discussion of the results are given below.

## 6.2 EFFECT OF LASER ENERGY

The TOF dynamics of neutrals and ions in the ns laser produced nickel plasma was measured using the 391.6 nm ($3d^9(^2D)$ $4p$ → $3d^9(^2D)$ $4s$) and 428.5 nm ($3p^63d^8$ ($^3P$) → $3p^63d^9$) lines in the plasma plume, respectively. Times of arrival of these species for different laser energies (by keeping the same spot size of irradiation) and background pressures have been measured to identify and confirm the presence of fast neutrals and determine their origin, although their presence was originally reported in a work on femtosecond LPP. OTOF is a non destructive tool for probing the movement of species in an expanding LPP without disturbing them, as the technique relies only on radiations emitted by the plasma. The OTOF signal for Ni-I at 361.9 nm shows two peaks which occur at different times, corresponding to fast neutrals and slow neutrals respectively, which we denote as PK1 and PK2. Peak corresponding to fast neutrals is observed only at larger laser energies.

OTOF studies of Ni II at 428.5 nm display a single but broader peak (compared to the OTOF of Ni I). From Fig 6.1a and 6.1b, it can be seen that signals corresponding to fast neutrals and ions are absent at relatively a lower irradiation energy of ~ 10 mJ, indicating low ionization and recombination in the plume, whereas signals recorded from slow neutrals confirm the presence of unionized neutrals at a lower laser energy. With an increase in laser energy, signals from fast neutrals and ions appear at almost similar times in comparison with the atomic species. From Fig 6.1a, it is found that the velocity of fast neutrals increases and slow neutrals remain unaltered with increase in laser energy. One interesting aspect to be noted here is that the time of arrival of fast neutrals falls well within the time of arrival of ions, substantiating the recombination mechanisms involved. In short, the fast atomic component is absent if the number density of ions is low, which happens at lower irradiation energies. Moreover the velocity of fast neutrals is found to increase with increase in irradiation energy.

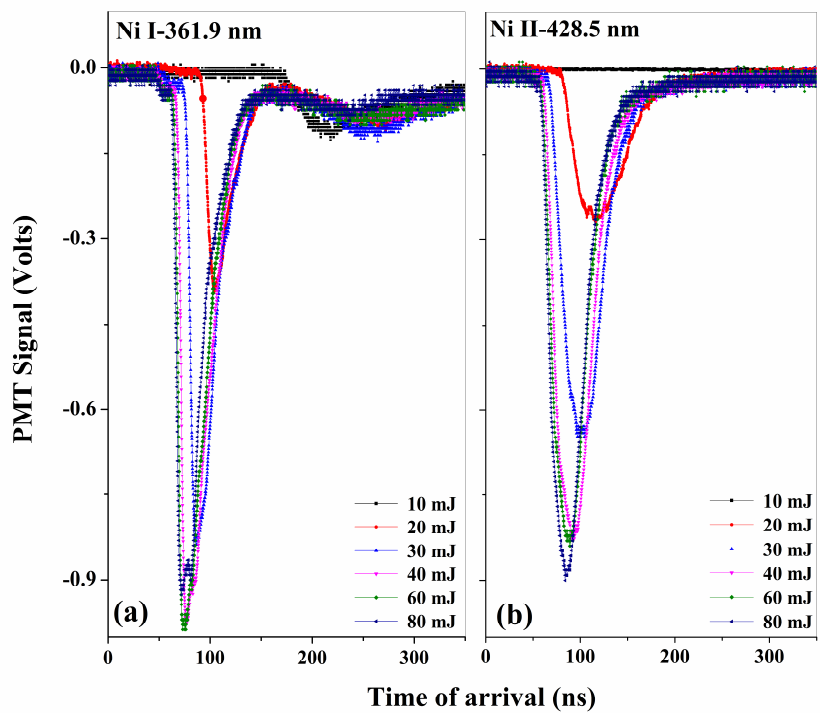

Figure 6.1. Time-resolved emission of the (a) 361.9 nm ($3d^9(^2D)\ 4p \rightarrow 3d^9(^2D)\ 4s$) transition from Ni I and (b) 428.5 nm ($3p^6 3d^8\ (^3P)\ 4s \rightarrow 3p^6 3d^9 4s$)) transition from Ni II, measured in the plasma plume at 2mm distance from the target surface, for 7 ns irradiation. Ambient pressure is 5 Torr.

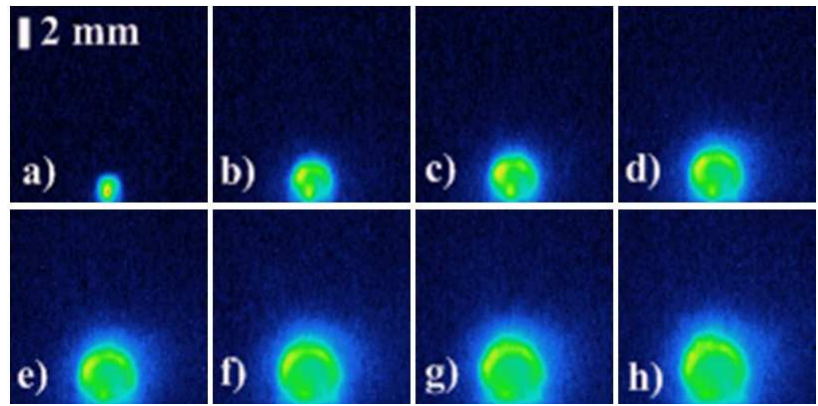

Figure 6.2. ICCD images of an expanding Laser produced nickel plasma for the laser energies (a) 10 mJ,(b) 20 mJ,(c) 30 mJ,(d) 40 mJ,(e) 50 mJ,(f) 60 mJ,(g) 70 mJ, and (h) 80 mJ, measured using a gate delay of 5 ns. The measurement is done for a time delay of 100 ns. Ambient pressure is 5 Torr.

Fig 6.2 shows ICCD images of LPP recorded at a time delay of 100 ns, with a gate width of 5 ns, for energies ranging from 10 mJ to 80 mJ, measured at a pressure of 5 Torr. The presence of two peaks at larger laser energy (i.e., more than 10 mJ in the present case) in an expanding LPP is once again confirmed by ICCD images. An increase in the separation between the times of arrival of fast and slow species with laser energy can also be observed in the ICCD images. LPP expands more at higher irradiation energies: i.e., when the energy increases from 10 mJ to 80 mJ, the plasma front reaches up to 9 mm from 3mm, when measured from the target surface. The plume expansion velocity increases from 30 km/s to 90 km/s simultaneously.

In order to understand the effect of laser energy on the arrival of ions and fast peaks in an expanding LPP, an experiment was performed by varying the laser energy, while the irradiation spot size was kept a constant. Material ablation happens once the irradiation intensity crosses the ablation threshold of the material, which can be calculated using Equation 4.1. $I_{min}$ is calculated to be $8.69 \times 10^{11}$ W/cm$^2$ for fs excitation and $3.89 \times 10^8$ W/cm$^2$ for ns excitation, with the corresponding laser fluences being $8.69 \times 10^{-2}$ J/cm$^2$ and 1.94 J/cm$^2$ respectively. The measured variation in OTOF for Ni I (361.9 nm) and Ni II (428.5 nm) for different distances from the target, at the laser fluence of 10 J/cm$^2$, is plotted in Fig 6.3.

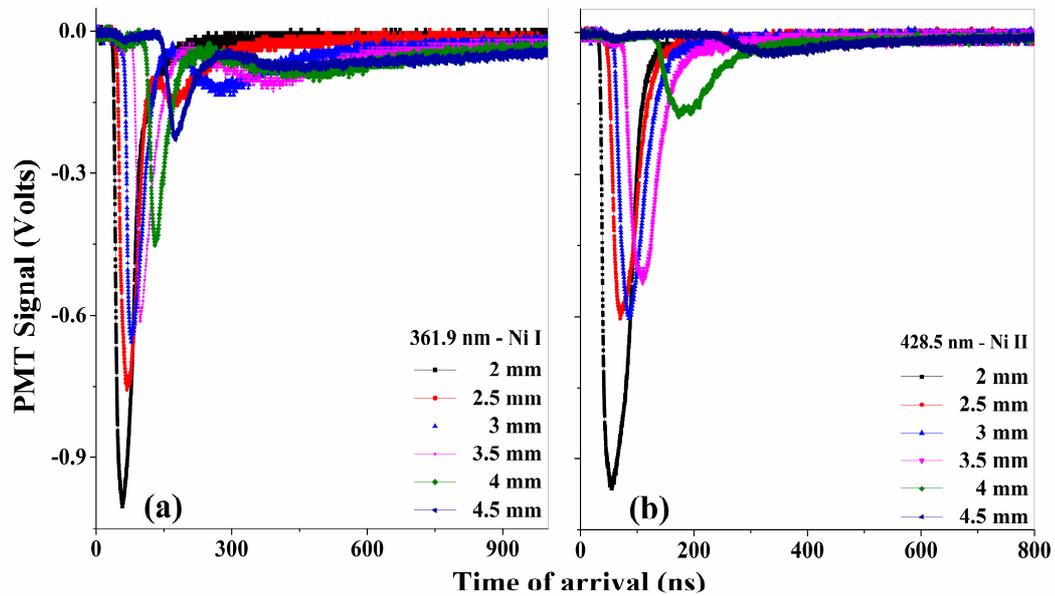

Figure 6.3. OTOF of (a) neutral (Ni I) and (b) ionic (Ni II) transitions measured in the plasma plume at various distances from the target surface, for 7 ns irradiation. Ambient pressure is 5 Torr.

It is clear from Figs.6.1 and 6.2 that at larger fluences an acceleration of fast neutrals and ions happens in the ns laser produced expanding Ni plasma. The acceleration of ions happens due to the Coulomb pull of space charges present in the expanding plume. Reduction in ion velocities seen at larger distances is due to the reduction in the number density of the plasma plume on expansion. The acceleration of fast atomic species seen here for ns LPP is hitherto unreported: we find that it occurs only for certain optimum ranges of pressure and at larger fluences. The times of arrival of peaks corresponding to fast neutrals and ions are almost the same in the present ns LPP, similar to the observation reported by Amoruso et. al. for fs LPP of nickel in vacuum ~ 10$^{-7}$ Torr.

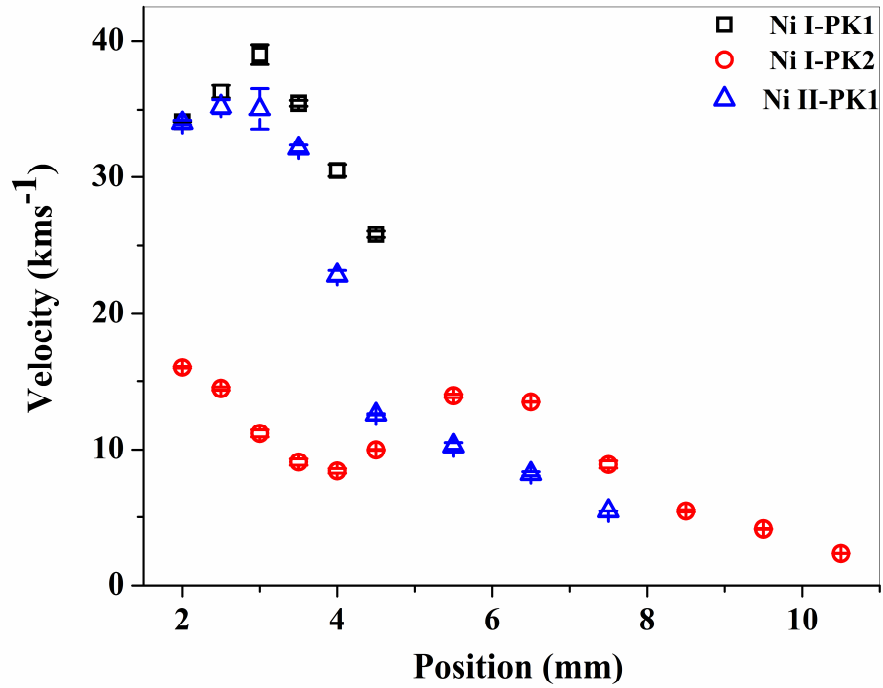

Figure 6.4. Calculated velocities of fast peaks corresponding to a transition from Ni I and a transition from Ni II, measured in the plasma plume at various distances from the target surface, for 7 ns irradiation. Ambient pressure is 5 Torr, and laser pulse energy is 80 mJ.

Figure 6.4 shows the calculated velocities of fast neutrals, slow neutrals and ions, measured at various distances from the target surface in the plasma plume, generated by ns irradiation. The velocity of ions and fast neutrals increase up to a distance of 3 mm from the target surface, indicating acceleration. Ion velocity increases from 33 km/s to 36 km/s over a movement of 1 mm (from 2 mm to 3 mm), while for fast neutrals the velocity increases from 33 km/s to 39 km/s over a similar movement of 1 mm. Measurements at larger distances (after 3 mm)show that the velocity reduces, indicating deceleration. On the other hand, the velocity of un-ionized slow neutrals initially decreases with distance up to 4 mm, then increases up to 6.5 mm, and then decreases again. Fast neutrals lose their energy on the go and slow down, and they are quickly followed by slow neutrals; this leads to a single peak in the OTOF in between 4 mm to 6.5 mm. This observed single peak may be suspected as an acceleration of slow neutrals at these spatial points, which is not really the case. At the position where slow neutrals are found to accelerate, fast and slow neutrals cannot be resolved in time and space. Hence this increase in velocity of the PK2 cannot be attributed to acceleration of slow neutrals between 4 mm to 6.5 mm.

## 6.3 EFFECT OF PRESSURE

Background gas and its pressure affect the expansion of LPP as explained in chapters 4 and 5. At lower pressures plasma expands adiabatically with fewer collisions among species which reduces the number density. Plasma lifetime increases due to plume confinement at higher pressures; however, thermal leak due to collisions among the plasma species and with the surrounding gas will reduce plasma life time at the same time. Therefore the number density will be maximum at certain intermediate pressures, where maximum emission from the ions can be observed. To understand and compare the dynamics of both ions and neutrals in the LPP, we made the measurements in a broad range of pressures from $5\times10^{-2}$ Torr to $1\times10^{2}$ Torr.

Emission lines at 391.6 nm and 428.5 nm were studied to understand the expansion dynamics of neutrals and ions, respectively. The PMT signal for Ni-I at 391.6 nm shows two peaks (PK1 and PK2) occurring at different times. The fast peak (PK1) is ascribed to the emission from recombined neutrals while the slow peak (PK2) corresponds to the relatively slow moving un-ionized neutrals (this is explained in the section 6.2). OTOF for Ni II at 428.5 nm displays a single but broader peak due to the larger Columbic force experienced by ions in the plume, compared to that of Ni I. The spatial extent of emission from ions along the axis is less compared to that of neutrals, indicating the recombination of ions on the go. Details of OTOF recorded for different pressures, axial positions, and laser energies are as follows.

The dynamics of neutrals in the LPP was measured for the pressures 0.05 Torr, 1 Torr, 20 Torr and 100 Torr at the larger laser fluence of 10 J/cm$^2$ using the 361.9 nm ($3d^9(^2D)$ $4p$ → $3d^9(^2D)$ $4s$) line at different spatial points along the expansion direction of the plume. Figure 6.5 shows OTOF signals obtained from these measurements. The presence of fast and slow neutrals (PK1 and PK2) is observed close to the target but only a single, broader peak is observed at larger distances. This is because the kinetic energy of the fast species reduces as they travel, and they become slow enough so that un-ionized neutrals catch up with them. Signals from neutral species have maximum spatial extent at 1 Torr, compared to other pressures. Figure 6.6 shows OTOF signals corresponding to the ions, which help substantiate the concept of acceleration of neutrals via recombination.

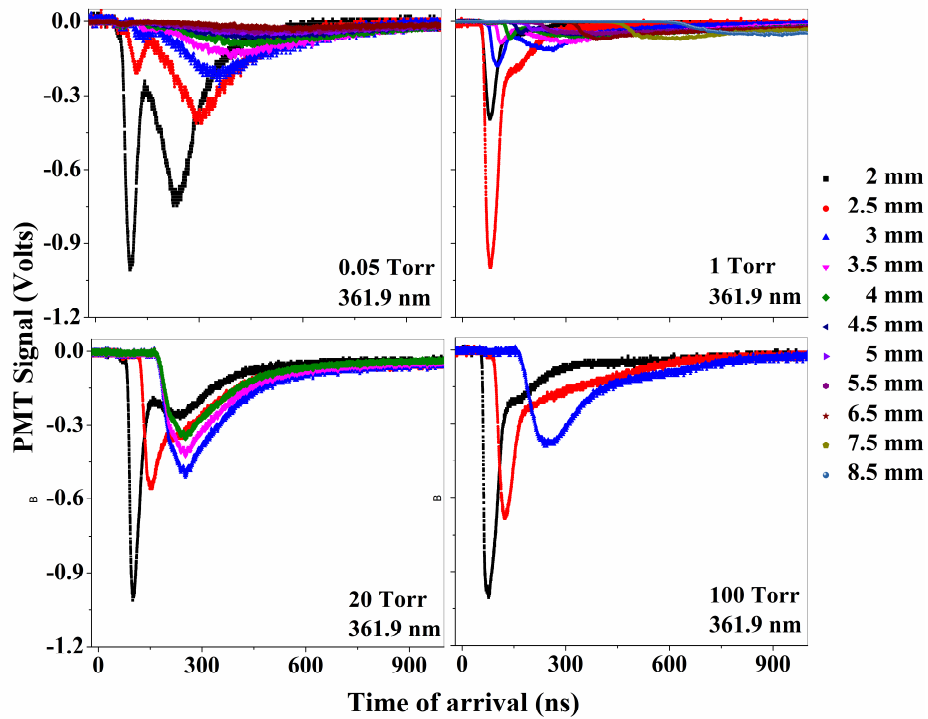

Figure 6.5. Optical time of flight (OTOF) measurement of the 361.9 nm ($3d^9(^2D)$ $4p \rightarrow 3d^9(^2D)$ $4s$) Ni I (neutral) transition at various axial positions in the plasma, for 0.05 Torr, 1 Torr, 20 Torr and 100 Torr ambient pressures. The arrival of fast and slow neutrals can be seen distinctly at certain pressures and positions.

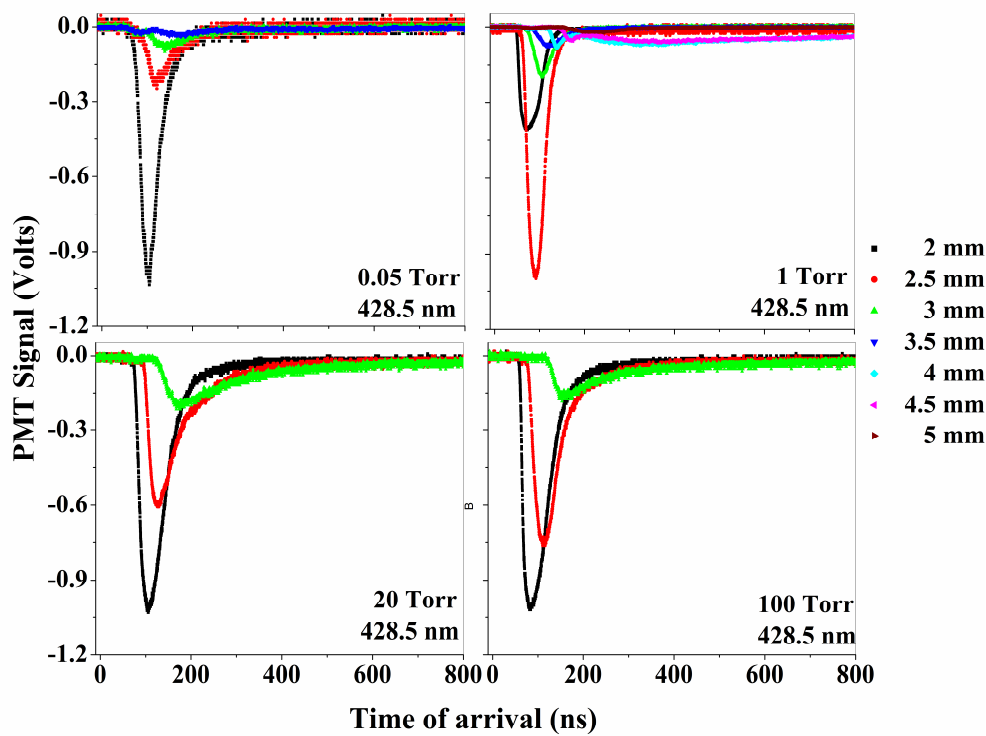

Figure 6.6. Optical time of flight (OTOF) measurement of the 428.5 nm ($3p^63d^8$ ($^3P$)) $4s \rightarrow 3p^63d^94s$Ni II (ion) transition at various axial positions in the plasma, measured for the ambient pressures 0.05 Torr, 1 Torr, 20 Torr and 100 Torr.

The velocities of both Ni I and Ni II species were measured on the axial positions until the signal could be detected, for a range of pressures. Velocities of fast neutrals and ions calculated from the OTOF signals (Figures 6.5 and 6.6 respectively) are shown in Figs.6.7a and 6.7b respectively. Velocity of the fast species is found to increase with axial distance up to 3.5 mm for the lowest pressure studied (~ 0.05 Torr), and it decreases thereafter. The signal could not be observed after 4 mm. At a higher pressure of 1 Torr the fast atomic species accelerates up to 4 mm before getting decelerated. The emission could be observed up to 9.5 mm distance in this case. On further increase in pressure fast neutrals decelerate abruptly with distance. Moreover, plume confinement limits the emission to a shorter axial distance at higher pressures. From Fig. 6.6 it is found that the optimum pressure to obtain emission to longer spatial extent from neutrals and ions along the length of the plume is 1 Torr. Ions as well as fast neutrals accelerate to larger distances at this pressure. In short, at $5\times10^{-2}$ Torr the ions accelerate and fly away, and cannot be detected after 4.5 mm. At 1 Torr the ions accelerate from 2 mm to 4 mm along the plume axis, but get decelerated at larger distances due to the reduced Coulomb pull, and the emission can be detected up to 6.5 mm. At the highest pressures studied (20 Torr and 100 Torr) the OTOF signal is very weak beyond an axial distance of 3.5 mm.

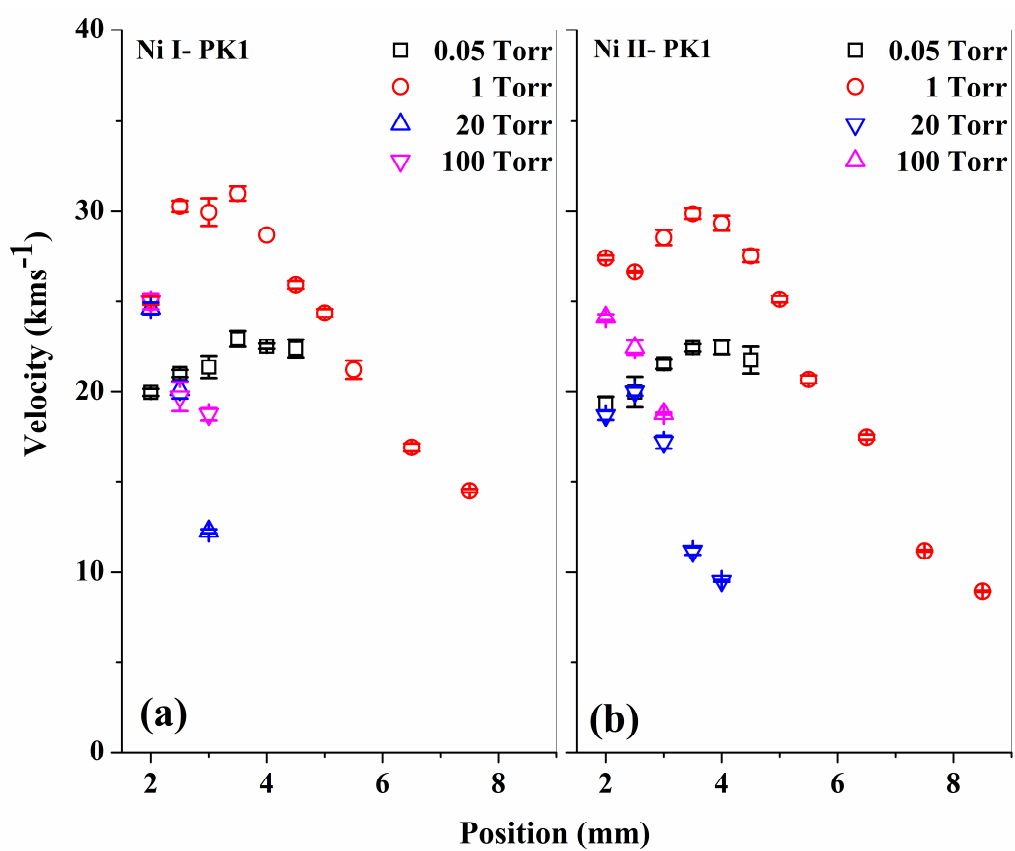

Figure 6.7. Velocities of the (a) fast Ni and (b) Ni II species, measured at various axial distances from the target surface, for 7 ns irradiation. Ambient pressures vary from 0.05 Torr to 100 Torr.

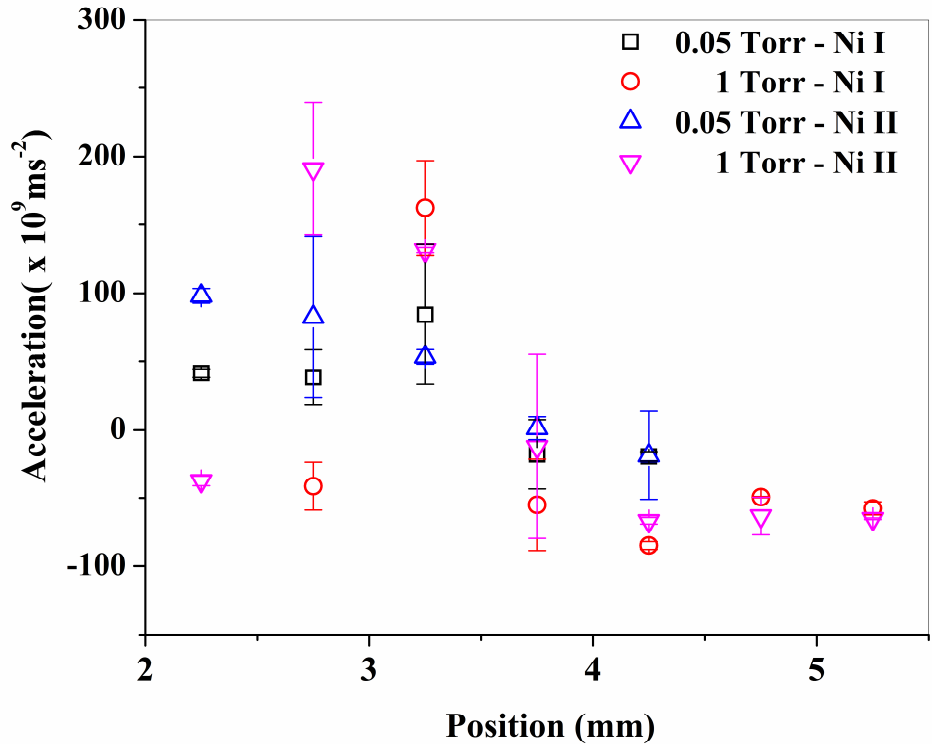

Figure 6.8. Accelerations of Ni I and Ni II species, measured at various axial distances from the target surface, for 7 ns irradiation. Acceleration of fast neutrals and ions is observed to a certain distance along the axis of the plume.

### 6.4 MEASUREMENTS IN A Zn TARGET

We obtained similar results when a Zinc (Zn) target was irradiated with fs laser pulses at the fluence of 10 J/cm$^2$. Acceleration of neutrals was observed, once again giving experimental evidence for this phenomenon in LPP from metal targets. The velocities calculated are plotted in Figure 6.8 for various axial distances at different pressures. It is found that the ion velocity increases up to a certain distance and decreases thereafter, irrespective of ambient pressures, in the measured range. The velocity of neutrals doubles its value on a movement of 2 mm (from 1 mm to 3 mm) for 50 Torr pressure, while that of the ions increases from 11 km/s to 22 km/s on a movement of 1.5 mm (from 1 mm to 2.5 mm), indicating acceleration of both the species. On expansion ion velocity increases, reaches a maximum and then slows down, and neutrals follow the same trend even though they are moving at relatively lower velocities. The measured velocity variation is similar for both ions and neutrals, which can be explained as follows. Ion acceleration occurs due to space charge effects, Coulomb pull of the bunch of electrons moving ahead in the plume, and push of the ions present near the emitters in the plume. Acceleration of neutrals may be attributed to the recombination of fast ions with free electrons as discussed above for the case of laser produced nickel plasma.

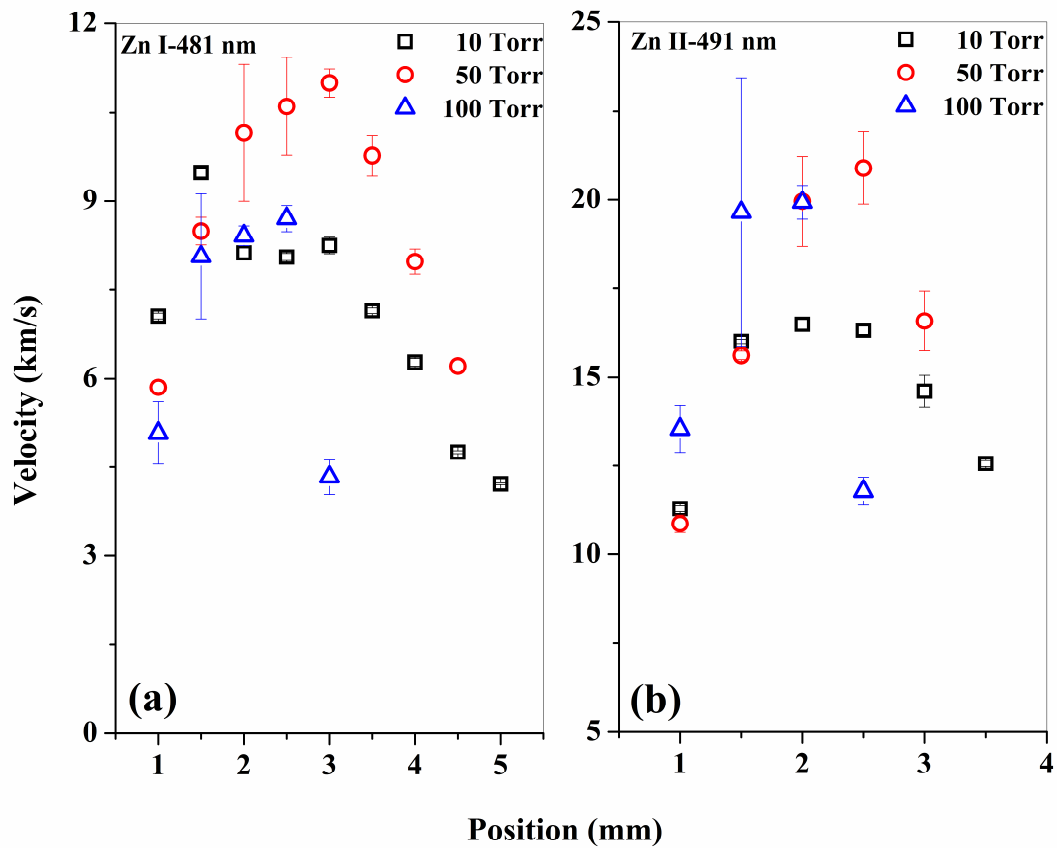

Figure 6.9: Calculated velocities of the peaks in the OTOF signals corresponding to (a) neutral and (b) ionic species in the plume, plotted against various axial positions in a laser produced expanding Zn plasma, for different ambient pressures. The error bar is calculated from multiple measurements.

**6.5 CONCLUSION**

A detailed investigation of the acceleration of neutrals in laser produced plasmas, observed in the LPP from solid metal targets, is given in this chapter. Studies of the acceleration of fast neutrals in a nanosecond laser produced Ni plasma are given in detail, along with similar observations made in a femtosecond laser produced Zn plasma. These studies confirm that the acceleration of neutrals from the LPP of metallic targets is indeed possible in the case of nanosecond laser excitation. Detailed theoretical calculations should be able to provide more insights into the phenomenon of neutral acceleration in nanosecond laser LPPs.

# CHAPTER 7
# OPTICAL GAIN IN A LASER-PRODUCED ALUMINUM PLASMA: ASYNCHRONOUS PUMP-PROBE MEASUREMENTS

*Optical pump-probe technique is the best choice for tracking fast processes which demand high time resolutions, where traditional fast detectors (except the streak camera) fail. The temporal dynamics of laser produced plasma can be monitored by a probe beam as a function of time, by which it is possible to obtain information on the decay of the plume. We report a novel asynchronous pump-probe experimental scheme, employed here to understand the dynamics and transient nature of ns LPP generated by a Q-switched Nd: YAG laser (1064 nm, 7 ns). The plasma is probed using a very low energy 100 fs, 80 MHz pulse train, which allows interrogation of the plasma at every 12.54 ns interval. Measurements reveal that the plasma can act as an optical gain medium and amplify the probe beam, under suitable conditions of irradiation energy and background pressure. Experimental details and results obtained are described in detail in this chapter.*

## 7.1 INTRODUCTION

Optical pump-probe spectroscopy is a powerful tool for the investigation of fast dynamic processes occurring in the optical domain [128]. Pump-probe technique is the best choice for the detection of very fast processes which demand a time resolution better than 100 picoseconds, where all other fast detectors (except the streak camera) fail. Pump-Probe spectroscopy was initially developed to measure collisional relaxation in liquids [129], electronic relaxation in semiconductors [130-132], femtosecond transition state dynamics [133-135], real time observations of molecular vibrations, etc. In general, pump–probe measurements can provide useful information on ultrafast phenomena. Here, the excitation or modification produced in a sample by a pump pulse is monitored using a probe signal as a function of time, by which it is possible to obtain information on the decay of the generated excitation. The temporal resolution of the experiment is fundamentally limited only by the pulse duration. Two colour pump-probe measurements are also possible using two different synchronized lasers, which are generally used for studying the relaxation of saturable absorbers after excitation. In this chapter, the dynamics of a ns laser produced expanding LPP in Aluminum is studied using a mode-locked train of 100 fs laser pulses, having a repetition rate of 80 MHz. Since the temporal separation between two consecutive pulses in the mode-locked pulse train is ~ 12.54 ns, the plasma can be sampled every 12.54 ns by this scheme. Interestingly, we observe that the plasma can act as an optical gain medium under suitable conditions, capable of amplifying the probe radiation.

## 7.2 EXPERIMENTAL

The pump-probe measurement is performed by altering some of the equipment used in the experimental setup given in chapter 3. Plasma is generated by irradiating a solid aluminum target using Q-switched Nd: YAG (1064 nm, 7 ns) laser. The plasma is probed by an 80 MHz train of 100 fs pulses obtained from a femtosecond laser oscillator. The pump laser works in the single shot mode as explained in chapter 3. Intensity of the probe beam after passing through the transient plasma is monitored using a fast photodiode. Since the pulses repeat every 12.54 ns (time period corresponding to 80 MHz repetition rate), variations in the probe intensity as it passes through the expanding plasma can be recorded every 12.54 ns. The probe beam is aligned parallel to the target surface and perpendicular to the plume axis, and is measured using a fast photodiode. This detector is a biased fast silicon PIN diode having a rise time $< 300$ ps (*ET-2030*, Electro Opic Technology Inc., USA). Signals from the photodiode are recorded on a fast oscilloscope (~ 3.5 GHz, DPO 7354, Tektronix USA) having a rise time of ~120 ps.

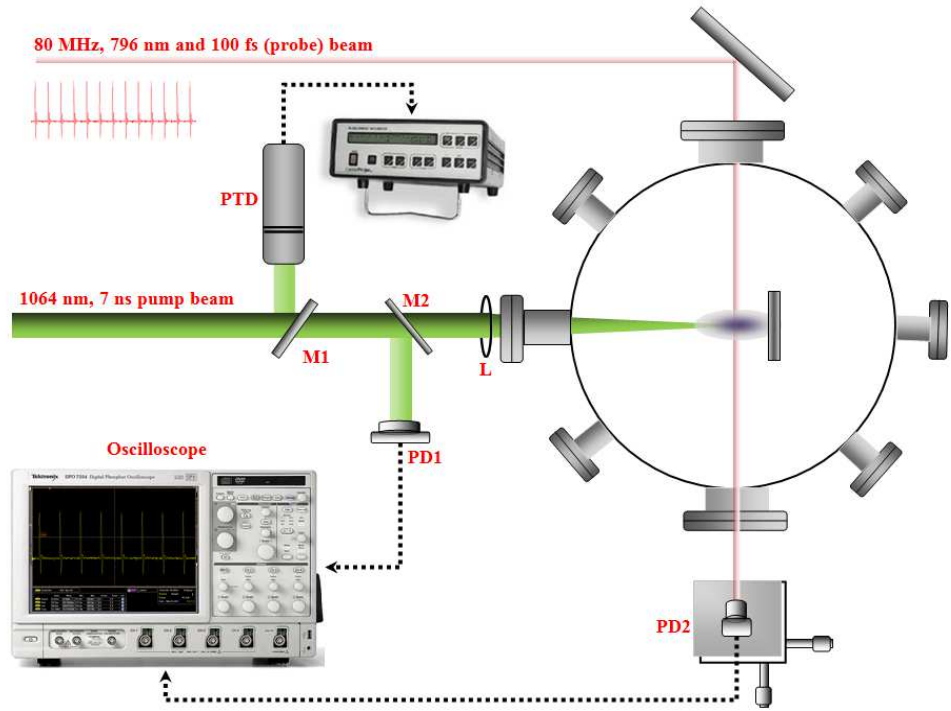

Figure 7.1 Schematic of asynchronous pump-probe measurements on an expanding laser produced plasma. PD2 is placed at a distance of 84 cm from the plasma plume.

## 7.3 DETECTOR CALIBRATION

Before doing the experiment, the probe beam needs to be characterized appropriately. The spatial profile of the beam is measured using a CCD beam profiler (*Beamlux II*, Metrolux, USA), and is found to have a Gaussian intensity profile with an average diameter ~1.2 mm in both vertical and horizontal cross sections, as shown in Fig. 7.2.

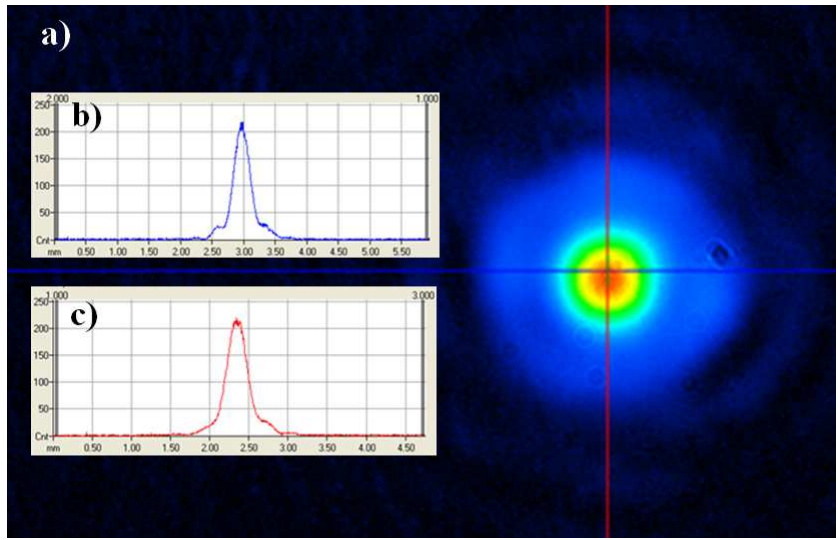

Figure 7.2. Spatial profile of the *Tsunami* (100 fs, 80 MHz) laser pulses measured using a CCD beam profiler. Insets (b and c) show the intensity distribution measured along the horizontal and vertical directions respectively, on the beam cross section.

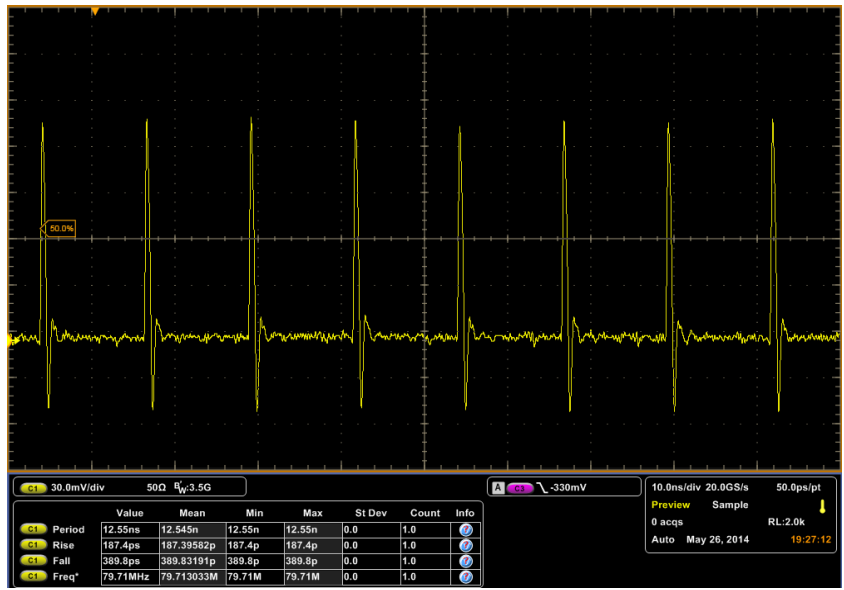

Figure 7.3. Probe laser pulses of 100 fs FWHM measured using a fast photodiode connected to fast oscilloscope. The measured FWHM is larger (risetime of ~187.4 ps and falltime of ~ 389.8 ps) due to the finite response times of the photodiode and oscilloscope. The mean repetition rate is accurately measured, which is found to be 80 MHz (i.e., period is ~ 12.54 ns).

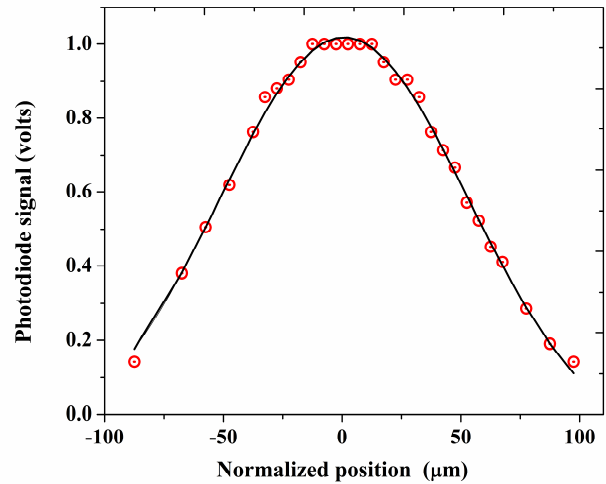

Figure 7.4. Measurement of the spatial intensity distribution of the probe beam (of diameter 1.2 mm) using the small area photodiode (of 0.4 mm diameter active area). Signal voltage is normalized to the maximum value measured at the centre of the probe beam. This measurement is done at a distance of 84 cm from the plasma plume.

Due to its small active area (diameter 0.4 mm) the fast Silicon PIN photodiode used in the present experiments gives good spatial resolution in a typical measurement, as demonstrated by a spatial scan of the probe beam (shown in Figure 7.4). The photodiode is sensitive to the range of 670 nm - 860 nm, and has a responsivity of about 0.4 A/W for a DC bias of 9 V. The photodiode is mounted on a XZ translator and moved across the probe beam, measuring the intensity at several points along the beam diameter. We found that only about 25 $\mu$m of the 0.4 mm active area gives constant voltage, when the photodiode is translated across the gaussian probe beam diameter. Therefore to measure probe variations occurring in very short timescales, especially if beam deflections also exist in the measurements, the photodiode needs to be aligned very carefully. Large area detectors like PMTs are a possible alternative, but due to the relatively large continuum emission, it will be impossible to extract signals in the earlier stages of plasma expansion, if PMT is used.

**7.4 RESULTS AND DISCUSSION**

We used a continuous train of mode-locked ultrafast (100 fs) laser pulses, which repeat every 12.54 ns, to probe the changes that occur with time in the transient plasma plume. The pulses pass through the LPP perpendicular to the expansion direction (i.e., parallel to the target), and are detected by the fast photodiode. We did a preliminary measurement to confirm that the probe beam is indeed deflecting due to transient refractive index variations as it passes through the plasma. Results of this measurement are given in Figure 7.5. Here the fast photodiode detector is placed at different positions with respect to the centre of the probe beam, and the change in beam intensity at the instant of plasma formation is monitored. From the results it can be concluded that the probe beam is deflecting towards the right side (as seen by the detector) at the instant of plasma formation. It may be noted that intensity change is a minimum when measured at or near the beam centre, because of the spatially Gaussian distribution of intensity.

Interestingly, it was found that if the detector is aligned to the beam centre and the intensity variation due to plasma formation is measured for a large range of background pressures and laser pulse energies, there are some situations where the beam intensity actually increases, rather than decreases. With central alignment the expectation is that of a decrease even if causes such as absorption, refraction, scattering etc. are considered in addition to beam deflection. Since the photodiode is aligned perfectly to the centre of the probe beam, a possible reason for this increase is self focusing of the beam caused by refractive index variations in the plasma. However we found that the beam is actually getting amplified by the plasma, and details of this interesting observation are given in the following sections.

**7.4.1  Dependence on irradiation energy**

Figure 7.6 shows probe transmission measured as a function of pulse energy used for generating the LPP. At lower energies (up to about 90 mJ) a dip followed by peak appears in the probe signal. Neglecting chances of beam reflection, the observed dip in the probe beam in the initial stages of expansion could be either due to absorption or scattering by the plasma species, if the number density is less than the critical density ~ $1.72 \times 10^{21}$ cm$^{-3}$ (for 796 nm). On further increase of irradiation energy to 106 mJ, only the peak appears. In the time window of 200 - 300 ns after plasma formation, the amplitude of this peak exceeds the normal amplitude of the probe beam ('normal amplitude' is the amplitude measured in the absence of the plasma). This observation clearly indicates that the ultrafast laser pulses are getting amplified by a single pass through the plasma. The use of plasma as a gain medium has been discussed earlier in literature [136-141], and the present results are a validation of this possibility. At larger irradiation energies a peak followed by a dip is seen, and at the highest energies used only the dip is seen. In general, for a given pressure, as the irradiation energy is increased, the features in the probe signal vary as: dip - peak, peak only, peak - dip, dip only. Moreover, at higher energies the features occur at earlier times.

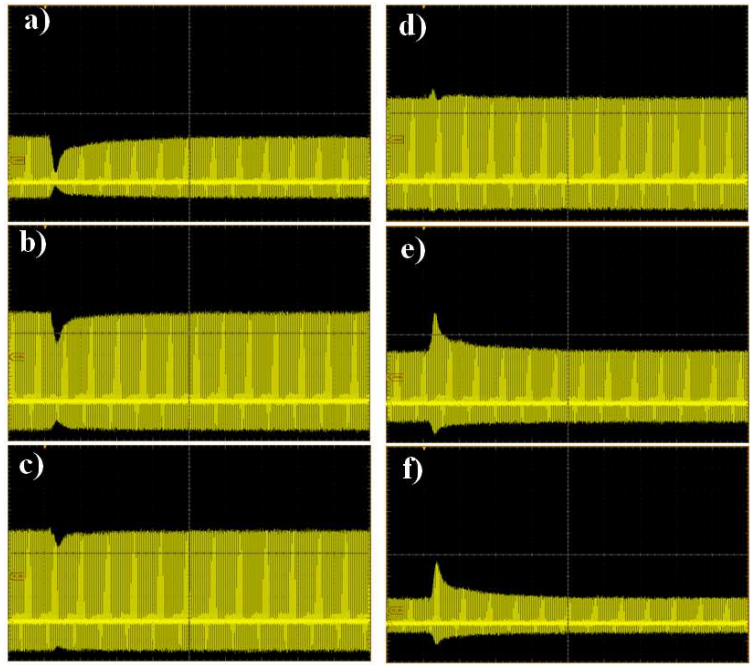

Figure 7.5. The probe beam (train of 100 fs pulses from the ultrafast oscillator, *Tsunami*) measured using a fast photodiode, at different positions along the beam diameter, at a distance of 84 cm from the centre of the plasma. LPP background pressure is 50 Torr and laser pulse energy is ~120 mJ. Measurement positions from the centre of the probe beam are, (a) 70 μm left, (b) 30 μm left, (c) 0 μm (centre), (d) 10 μm right, (e) 30 μm right, and (f) 60 μm right. Observed intensity variations show that the probe beam is deflecting in the rightward direction (as seen by the detector) at the instant of plasma formation.

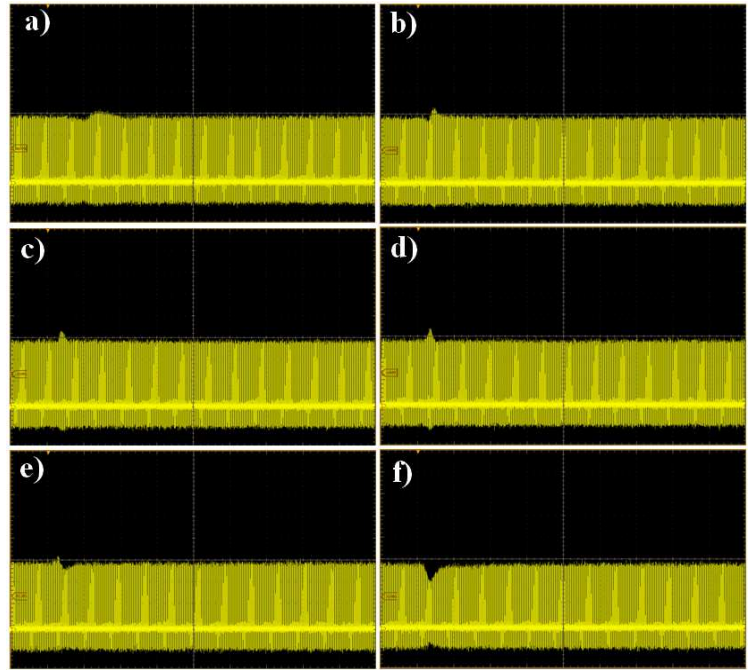

Figure 7.6. Variation of the probe beam signal with the pulse energy used for generating LPP: (a) 50 mJ, (b) 88 mJ, (c) 97 mJ, (d) 106 mJ, (e) 114 mJ, and (f) 120 mJ. Photodiode detector is aligned in line with the centre of the probe beam.

### 7.4.2  Dependence on ambient pressure

To optimize the background pressure for maximum amplification, the experiment was repeated in the pressure range of 0.05 Torr to 150 Torr, at an irradiation energy of 105 mJ. The probe signal showed neither peak nor dip at 0.05 Torr indicating an adiabatic expansion. As the pressure is increased plume confinement increases, and the dip starts appearing in the probe signal, the width of which increases with pressure. The increased width, which remains fairly constant from 25 Torr to 50 Torr, indicates a longer lifetime for the plasma. In the range of 50 to 80 Torr a peak followed by dip is the observed trend, which changes to dip-peak-dip from 85 to 150 Torr. Above 100 Torr the amplitude of the peak is found to increase. In general, with increase in pressure, the features consecutively observed in the probe signal are: no peak & no dip, dip, dip with increased width, peak followed by dip, and finally, dip and peak followed by dip. This behaviour is obvious from figure 7.7.

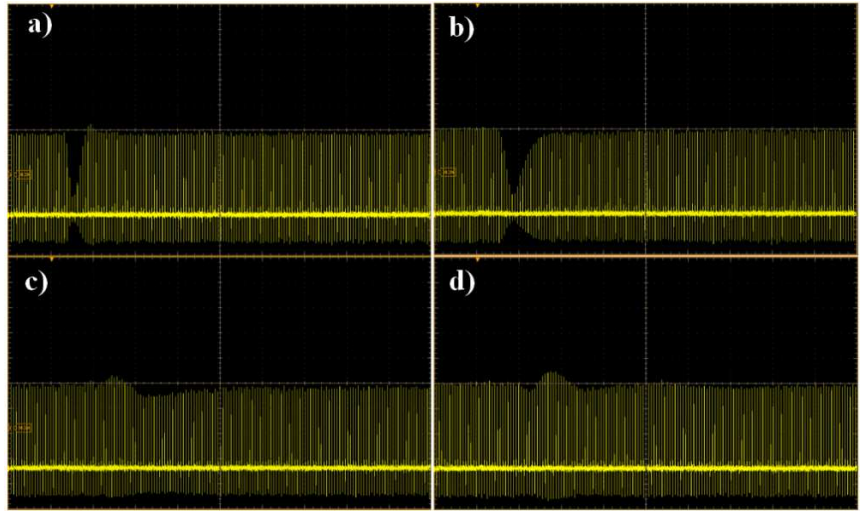

Figure 7.7. Probe beam signal measured for an LPP energy of 105 mJ, for various ambient pressures: (a) 5 Torr, (b) 20 Torr, (c) 60 Torr, and (d) 85 Torr.

In general, dips in the probe signal may be due to losses from scattering, reflection, absorption, etc. while peak can occur due to self focusing and/or laser–plasma energy coupling and amplification. To estimate the contribution of self-focusing, if any, the experiment was repeated for different distances from the target such as 48 cm, 66 cm and 84 cm. These measurements show similar trends, indicating that the plasma does not focus the probe beam at the measured distances. For 45 Torr, 75 Torr and 110 Torr background pressures, features of the probe signal become steady after the first 3 or 4 laser shots, up to about 50 shots (the first few laser shots typically etch and clean the target surface). At 70 Torr pressure and 105 mJ energy it is found that the probe signal rises above the input signal, for time delays from 250 ns to 520 ns from the time of irradiation. In the absence of self focusing, this confirms that the probe beam (ultrafast laser pulses) is getting amplified on a single pass through the plasma gain medium. For the same target location, increase in the photodiode output voltage is found to be ~ 21% for the 5th to 15th shots, and ~ 15% for the next 40 shots. Thus it is clear that under optimized conditions ultrafast laser pulses can be amplified by a factor of 15% to 20% on a single pass through the present ns Al LPP. The experiment was repeated 5 times, giving similar results. In general, amplification occurs earlier in time with increase in pressure. Moreover, maximum amplification occurs in the space of 2 mm to 4.5 mm from the target surface, along the plume expansion direction.

## 7.5 CONCLUSION

We have carried out a novel asynchronous pump-probe experiment to characterize transient refractive index variations in a ns LPP generated in Aluminum, employing a mode-locked pulse train of 80 MHz, 100 fs laser pulses as the probe. The technique has a temporal resolution of 12.54 ns (inverse of the pulse repetition frequency). Measurements reveal amplification of the probe pulses under optimal LPP energy and pressure conditions, indicating that the plasma is acting as an ultrafast photon gain medium. Detailed studies can reveal more information on the nature of this amplification, which may have potential practical applications.

# CHAPTER 8
# CONCLUSION AND FUTURE SCOPE

*An account of the major findings of the present study is given in this chapter. A brief overview of the scope for future work in the direction of the current work is also included.*

In a nutshell, the work reported in this thesis is on the generation and temporal characterization of laser produced plasmas from metallic targets. Nanosecond (ns-short-pulse) and femtosecond (fs-ultrafast pulse) Laser Produced Plasmas (LPP) were generated using 7 ns and 100 fs laser pulses respectively, by irradiating Nickel and Zinc targets. The studies show that the nature of the generated LPPs is different between these irradiations, as mentioned in literature. A Q-Switched Nd:YAG laser, emitting 1064 nm,7 ns pulses, and a regeneratively amplified Ti:Sapphire laser, emitting 800 nm,100 fs pulses, were used to generate the ns and fs LPPs respectively. Even though the system generates 10 Hz pulses, single pulses were selected from the pulse train to facilitate single shot measurements, as described in Chapter 3. Proper synchronization using required electronic circuits was done to synchronize the instruments used in the experiment with the incident laser pulse, to make accurate measurements. Once the plasma was formed it was analyzed by a CCD (to study the optical emission spectra), PMT (to study the dynamics of the species) and an ICCD (to image the plasma). Data acquisition was done through a PC installed with appropriate softwares.

The plasma was found to be composed of neutrals, ions and electrons which are referred to as plasma species in general. Optical Emission Spectroscopy (OES) was carried out to identify the species in the generated plasma, by matching the observed spectral lines with the NIST standard database. The identified lines were then categorized and the lines with relatively higher probability were chosen and sent to the PMT to investigate the dynamics (via Optical Time of Flight (OTOF measurements)) of that particular species in the plasma. Furthermore, imaging using ICCD was done which provided images of the expanding plasma at different delay times, revealing interesting trends in the dynamics of the species in the expanding plasma.

Nickel plasma produced using ns and fs irradiations were characterized using OES and OTOF measurements and ICCD imaging. Emissions at 361nm ($3d^9(^2D)$ 4p $3d^9(^2D)$ 4s transition) for the neutrals and 428.5nm ($3p^63d^8(^3P)$ 4s $3p^63d^9$ 4s) for ions were investigated for a large range of ambient nitrogen pressures varying from $10^{-6}$ Torr to $10^2$ Torr. OES and OTOF measurements were done at two different spatial points whereas imaging was done at different time delays at specific gatewidths to observe the dynamics of the expanding plasma. Plasma intensity was found to increase with pressure for both ns and fs excitations. While the electron temperature is maximized around $10^{-4}$ Torr to $10^{-3}$ Torr for ns excitation, it was relatively lower and rather independent of pressure for fs excitation. A double-peak structure was observed in the OTOF spectrum of Ni I under ns excitation where the fast peak was due to the neutrals formed by ion-electron recombination, and the slow peak was due to the un-ionized neutrals ablated from the target. On the other hand, for fs excitation, only one peak (fast) was observed in the OTOF spectrum. Measured velocities of the species were found to be relatively higher and independent of pressure for fs excitation, whereas shockwave effects were evident for ns excitation. The measured velocities clearly indicate acceleration of fast species in the plume on expansion, at least up to a distance of 4 mm from the target, whereas the slow species is found to decelerate, particularly at higher pressures. These investigations provide new information on the pressure dependent temporal behavior of Ni plasmas produced by ns and fs laser pulses.

Similarly, plasma generated by fs and ns LPPs from a high quality solid zinc target kept under broad ambient nitrogen pressures from $5 \times 10^{-2}$ Torr to $5 \times 10^2$ Torr were also studied. The Zn I ($4s5s\ ^3S_1 - 4s4p\ ^3P_2$) line was observed at a distance of 2 mm from the target surface with a laser fluence of ~16 J/cm$^2$. Temperature and number density were calculated from the optical emission spectra and plotted as a function of pressure. OTOF of the line at 481 nm showed a double-peak structure, ascribed to the existence of fast and slow atomic species respectively in the LPP. Fast atomic species was found to move with an average velocity of ~$10^4$ m/s whereas slow atomic species was found to move with a velocity of ~$10^3$ m/s. Interestingly, the average velocity of the fast atomic component was found to vary with pressure upon ns excitation, whereas it remained unchanged for fs excitation. ns LPP from solid Zn showed a longer plume lifetime. Emission from neutral Zn was found to be more intense in both fs and ns LPPs for similar fluences, depicting the characteristic triplet structure of Zn. The optimum pressure for maximum emission intensity and lifetime was experimentally found to be ~ 10 Torr. Furthermore, a double pulse excitation measurement was performed to understand the effect of laser-plasma energy coupling on the emission dynamics of the plasma plume.

Acceleration of neutrals was calculated for both ns and fs LPPs from the velocity data. Data obtained from OTOF spectra are well supported by the ICCD images. Occurrence of the phenomenon of acceleration of neutral species is verified and confirmed by the experiments carried out in this work. Another important result obtained is the amplification of the 796 nm probe beam by the plasma, which shows that laser produced plasma is acting as an optical gain medium.

The results obtained from the present work lay a strong basis for the fundamental study of laser produced plasmas from solid metal targets kept under a nitrogen background, for both ns and fs laser irradiations. The obtained results will be useful to understand the dynamics of various plasma species, nature of expansion, and various phenomena associated with the plasma, which find potential applications in the fields of pulsed laser deposition, nanoparticle generation, production of fourth generation light sources, plasma based accelerators etc. Detailed studies with respect to various ambient gases, various other targets etc. can be performed to get a better idea about the fundamental physics of LPPs.

The observed acceleration of neutrals in the LPP of metal targets opens up new possibilities for further research, which can lead to fundamental studies focusing on plasma based accelerators for neutral species. The novel asynchronous pump-probe experiment described in chapter 7 is essentially a new and effective technique for probing highly transient plasmas with high temporal resolution. Possibilities of optical gain and amplification in LPP, which is an important outcome of the experiments performed, is a very promising result obtained from these investigations, which may pave the way for potential applications in future.

# REFERENCES

[1]  D. H. Lowndes, D. B. Geohegan, A. A. Puretzky, D. P. Norton, and C. M. Rouleau, Synthesis of novel thin-film materials by pulsed laser deposition, Science, Vol. 273, No. 5277, August 1996, pp. 898-903.

[2]  C. Thaury and F. Qu´er´e, High-order harmonic and attosecond pulse generation on plasma mirrors: basic mechanisms, Journal of Physics B: Atomic, Molecular and Optical Physics, Vol. 43, October 2010, pp. 21300-1-33.

[3]  R. A. Ganeev, High-order harmonic generation in a laser plasma: a review of recent achievements, Journal of Physics B: Atomic, Molecular and Optical Physics, Vol. 40, October 2007, pp. R213-R253.

[4]  S. Varró and F. Ehlotzky, Higher harmonic generation at metal surfaces by powerful femtosecond laser pulses, Physical Review A, Vol. 54, No. 4, October 1996, pp 3245-3249.

[5]  U. Teubner and P. Gibbon, High- order harmonics from laser-irradiated plasma surfaces, Reviews of Modern Physics, Vol. 81, No.2, April 2009, pp 445-479.

[6]  C. Chenais-Popovics, O. Rancu, P. Renaudin, and J. C. Gauthier, X-ray spectroscopy of laser-produced hot dense plasmas, Physica Scripta, Vol. T65, 1996, pp-163-167.

[7]  B. C. Fawcett, A. H. Gabriel, F. E. Irons, N. J. Peacock, P. A. H. Saunders, Extreme ultra-violet spectra from laser-produced plasmas, Proceedings of the Physical Society, Vol. 88, May 1966, pp 1051-1053.

[8]  D. D. Burgess, B. C. Fawcett and N. J. Peacock, Vacuum ultra-violet emission spectra from laser-produced plasmas, Proceedings of the Physical Society, Vol. 92, July 1967, pp 805-816.

[9]  J. D. Kmetec, Ultrafast Laser Generation of Hard X-Rays, IEEE Journal of Quantum Electronics, Vol. 28, No. 10, October 1992, pp. 2382- 2387.

*References*

[10]  M. Anand, C. P. Safvan, M. Krishnamurthy, Hard X-ray generation from microdroplets in intense laser fields, Applied Physics B, Vol. 81, July 2005, pp 469-477.

[11]  S. Amoruso, G. Ausanio, R. Bruzzese, M. Vitiello, and X. Wang, Femtosecond laser pulse irradiation of solid targets as a general route to nanoparticle formation in a vacuum, Physical Reveiw B, Vol. 71, January 2005, pp. 033406-1-4.

[12]  S. Amoruso, R. Bruzzese, X. Wang, N. N. Nedualkov, and P. A. Atanasov, femtosecond laser ablation of nickel in vacuum, Journal of Physics D: Applied Physics, Vol. 40, January 2007, pp 331-340.

[13]  S. Amoruso, G. Ausanio, A. C. Barone, R. Bruzzese, C. Campana, X. Wang, Nanoparticles size modifications during femtosecond laser ablation of nickel in vacuum, Applied Surface Science, Vol. 254, July 2007, pp 1012-1016.

[14]  S. Amoruso, R. Bruzzese, C. Pagano, X. Wang, Features of plasma plume evolution and material removal efficiency during femtosecond laser ablation of nickel in high vacuum, Applied Physics A, Vol. 89, August 2007, pp 1017-1024.

[15]  D. R. Nicholson, Introduction to Plasma Theory, © Wiley, New York, 1983.

[16]  M. N. Rosenbluth and R. Z. Sagdeev, Handbook of plasma Physics, Volume 1: Basic Plasma Physics, © North-Holland Publishing Company, 1983.

[17]  F. F. Chen, Introduction to plasma physics and controlled fusion © Plenum Press New York, 1984.

[18]  C. Uberoi, Introduction to unmagnetized plasmas, © PHI Pvt. Ltd. New Delhi 1988.

[19]  http://farside.ph.utexas.edu/teaching/plasma/lectures/node9.html.

[20]  D. A. Cremers and L. J. Radziemski, Handbook of laser induced breakdown spectroscopy, © John Wiley & Sons Ltd, 2006.

[21]  E. Woryna, P. Parys, J. Wołowski and W. Mróz, Corpuscular diagnostics and processing methods applied in investigations of laser-produced plasma as a source of highly ionized ions, Laser and Particle beams, Vol. 14, Issue 3, September 1996, pp. 293-321.

[22]   T.N. Hansen, J. Schou, J.G. Lunney, Langmuir probe study of plasma expansion in pulsed laser ablation, Applied Physics A: Materials Science and processing, Vol. 69 [Suppl.], December 1999, pp.S601–S604.

[23]   R. L. Merlino, Understanding Langmuir probe current-voltage characteristics, American Journal of Physics, Vol. 75, No. 12, December 2007, pp 1078-1083.

[24]   D. E. T. F. Ashby and D. F. Jephcott, Measurement of plasma density using a gas laser as an infrared interferometer, Applied Physics Letters, Volume 3, No.1, July 1963, pp-13-16.

[25]   L. C. Johnson and T. K. Chu, Measurements of electron density evolution and beam self-focusing in a laser-produced plasma, Physical Review Letters, Vol. 32, No. 10, March 1974, pp. 517-520.

[26]   R. Rajesh, B. Ramesh Kumar, S. K. Varshney, Manoj Kumar, Chhaya Chavda, Aruna Thakkar, N. C. Patel, Ajai Kumar and ADITYA team, PRAMANA- journal of physics, Vol. 55, No. 5&6, December 2000, pp 733-740.

[27]   I.M. Hutchinson. Principles of plasma diagnostics. © Cambridge University Press, Cambridge, 1987.

[28]   K. L. Lancaster, J. Pasley, J. S. Green, D. Batani, S. Baton, R. G. Evans, L. Gizzi, R. Heathcote, C. Hernandez Gomez, M. Koenig, P. Koester, A. Morace, I. Musgrave, P. A. Norreys, F. Perez, J. N. Waugh, and N. C. Woolsey, Temperature profiles derived from transverse optical shadowgraphy in ultraintense laser plasma interactions at $6 \times 10^{20}$ W cm$^{-2}$, Physics of Plasmas, Vol. 16, May 2009, pp-056707-1-5.

[29]   W. I. Linlor, Some properties of plasma produced by laser giant pulse, Physical Review Letters, Vol. 12, No. 14, 1964, pp 383-385.

[30]   S. A. Ramsden, and P. Savie, A radiative detonation model for the development of a laser-induced spark in air, Nature, Vol. 203, 1964, 1217-1219.

[31]   R. G Meyerand, Jr., and A. F. Haught, Optical-energy absorption and high density plasma production, Physical Review Letters, Vol. 13, No. 1, 1964, pp 7-9.

[32]  S. A. Ramsden, and W. E. R. Davies, Radiation scattered from the plasma produced by a focused ruby laser beam, Physical Review Letters, Vol. 13, No. 7, 1964, pp 227-229.

[33]  V. V. Korobkin and R. V. Serov, Investigations of the magnetic field of a spark produced be focusing laser radiation, zHETFPis'ma, Vol. 4, No. 3, 1966, pp 103-106.

[34]  S. S. Harilal, Riju C. Issac, C. V. Bindhu, V. P. N. Nampoori and C. P. G. Vallabhan, Temporal and spatial evolution of $C_2$ laser induced plasma from graphite target, Journal of Applied Physics, Vol. 80, No. 6, July 1996, pp 3561-3565.

[35]  S. S. Harilal, C. V. Bindhu, Riju C. Issac, V. P. N. Nampoori, and C. P. G. Vallabhan, Electron density and temperature measurements in a laser produced plasma, Journal of Applied Physics, Vol. 82, No. 5, May1997, pp 2140-2146.

[36]  S. S. Harilal, Riju C. Issac, C. V. Bindhu, V.P.N. Nampoori and C. P. G. Vallabhan, Emission characteristics and dynamics of $C_2$ from laser produced graphite plasma, Journal of Applied Physics, Vol. 81, No. 8, April 1997, pp 3637-3643.

[37]  S. S. Harilal, Beau O' Shay and M. S. Tillack, Spectroscopic characterization of laser-induced tin plasma, Journal of Applied Physics, Vol. 98, July 2005, pp 013306-1-7.

[38]  S. S. Harilal, T. Sizyuk, A. Hassanein, D. Campos, P. Hough, and V. Sizyuk, The effect of excitation wavelength on dynamics of laser-produced tin plasma, Journal of Applied Physics, Vol. 109, March 2011, pp 063306-1-9.

[39]  D. Campos, S. S. Harilal, and A. Hassanein, Laser wavelength effects on ionic and atomic emission from tin plasma, Applied Physics Letters, Vol. 96, April 2010, 151501-1-3.

[40]  S. S. Harilal, T. Sizyuk, A. Hassanein, D. Campos, P. Hough, and V. Sizyuk, The effect of excitation wavelength on dynamics of laser-produced tin plasma, Journal of Applied Physics, Vol. 109, March 2011, pp 063306-1-9.

[41]  C. S. Suchand Sandeep, Investigations of Nonlinear Optical Effects and Ultrafast laser induced plasma in Nano structured Media, March 2010 (http://dspace.rri.res.in/bitstream/2289/4315/1/SuchandSandeep_Thesis.pdf).

[42]   Jeffrey L. Krause, Kenneth J. Schafer, and Kenneth C. Kulander, High-order harmonic generation from atoms and ions in the high intensity regime, Physical Review Letters, Vol. 68, No. 24, June 1992, pp. 3535–3538.

[43]   J. J. Macklin, J. D. Kmetec, and C. L. Gordon, III , High-order harmonic generation using intense femtosecond pulses, Physical Review Letters, Vol. 70, No.6, February 1993, pp. 766–769.

[44]   B. M. Jovanović, Optimum conditions for second harmonics generation in a magnetized plasma, ICPP & 25$^{th}$ Conference on Controlled Fusion and Plasma Physics, Praha, ECA Vol. 22C, 29 June – 3 July 1998, pp 2415-2418.

[45]   A. Giulietti, C. Beneduce, T. Ceccotti, D. Giulietti, L.A. Gizzi and R. Mildren, Study of second harmonic emissions for characterization of laser–plasma X-ray sources, Laser and Particle Beams, Volume 16, No. 2, June 1998, pp 397-404.

[46]   J. Parashar and A. K. Sharma, Second-harmonic generation by an obliquely incident laser on a vacuum-plasma interface, Europhysics Letters, Vol. 41, No.4, February 1998, pp 389-394.

[47]   J. Parashar and H. D. Pandey, Second- Harmonic Generation of Laser Radiation in a Plasma with a Density Ripple, IEEE Transactions on Plasma Science, Vol. 20, No.6, December 1992, pp 996-999.

[48]   T. Engers, W. Fendel, H. Schüler, H. Schulz and D. Von der Linde, Second-harmonic generation in plasmas produced by femtosecond laser pulses, Physical Review A, Vol. 43, No. 8, April 1991, pp 4564-4567.

[49]   R. M. More, Laser interaction with atoms, solids and plasmas © Plenum Press, New York and London, 1994.

[50]   C. Chenais-Popovics, 0. Rancu, P. Renaudin and J. C. Gauthier, X-ray spectroscopy of laser-produced hot dense plasmas, Physica Scripta. Vol. T 65, November 1996, pp. 163-167.

[51]   Y. F. Lu, M. Takai, S. Komuro, T. Shiokawa and Y. Aoyagi, Surface cleaning of metals by pulsed laser irradiation in air, Applied Physics A: Solids and Surfaces, Vol. 59  May 1994, pp. 281-288.

[52] A. V. Bulgakov and N. M. Bulgakova, Dynamics of laser-induced plume expansion into an ambient gas during film deposition, Journal of Physics D: Applied Physics, Vol. 28, May 1995, pp 1710-1718.

[53] X. Wang, S. Amoruso, M. Armenante, A. Boselli, R. Bruzzese, N. Spinelli, R. Velotta, Pulsed laser ablation of borocarbide targets probed by time-of-flight mass spectroscopy, Optics and Lasers in Engineering, Vol. 39, 2003, pp 179-190.

[54] L. Torrisi, F Caridi, D. Margarone, and L. Giuffrida, Nickel Plasma Produced by 532-nm and 1064-nm Pulsed Laser Ablation, Plasma Physics Reports, Vol. 34, No. 7, 2008, pp 547-554.

[55] S. J. Henley, J. D. Carey, S. R. P. Silva G. M. Fuge, M. N. R. Ashfold, D. Anglos Dynamics of confined plumes during short and ultrashort pulsed laser ablation of graphite, Physical. Review. B, Vol. 72, November 2005, pp. 205413-1-3.

[56] F. E.Irons, R. W. P. Mc Whirter and N. J. Peacock, The ion and velocity structure in a laser produced plasma, Journal of Physics B: Atomic and Molecular Physics, Vol. 5, October 1972, pp 1975-1987.

[57] R. Dinger, K. Rohr, and H. Weber, Ion distribution in laser produced plasma on tantalum surfaces at low irradiances, Journal of Physics D: Applied Physics, Vol. 13, 1980, pp 2301-2307.

[58] K. Nakajima, D. Fisher, T. Kawakubo, H. Nakanishi, A. Ogata, Y. Kato, Y. Kitagawa, R. Kodama, K. Mima, H. Shiraga, K. Suzuki, K. Yamakawa, T. Zhang, Y. Sakawa, T. Shoji, Y. Nishida, N. Yugami, M. Downer and T. Tajima, Observation of Ultrahigh Gradient Electron Acceleration by a Self-Modulated Intense Short Laser Pulse, Physical Review Letters, Vol. 74, No. 22, May 1995, pp 4428-4431.

[59] R. Sauerbrey, Acceleration in femtosecond laser produced plasmas, Physics of Plasmas, Vol. 3, No. 12, December 1996, pp 4712-4716.

[60] S. Amoruso, G. Ausanio, M. Vitiello, X Wang, Infrared femtosecond laser ablation of graphite in high vacuum probed by optical emission spectroscopy, Applied Physics A: Material Science and Processing Vol. 81, January 2005, pp. 981-986.

[61]  F. Claeyssens, S. J. Henley, Michael and M. N. R. Ashfold, Comparison of the ablation plumes arising from ArF laser ablation of graphite, silicon, copper and aluminium in vacuum, Journal of Applied Physics, Vol 94, No 4, August 2003, pp. 2203-2211.

[62]  X. Zeng, X. L. Mao, R. Greif, R. E. Russo, Experimental investigation of ablation efficiency and plasma expansion during femtosecond and nanosecond laser ablation of silicon, Applied Physics A, Vol. 80, 2005, pp 237-241

[63]  S. Amoruso, G. Ausanio, A. C. Barone, R. Bruzzese, L. Gragnaniello, M. Vitiello and X. Wang, Ultrashort laser ablation of solid matter in vacuum: a comparison between the picoseconds and femtosecond regimes, Journal of Physics B: Atomic, Molecular and Optical Physics, Vol. 38, September 2005, pp L329-L338.

[64]  L. A. Emmert, R. C. Chinni, D. A. Cremers, C. R. Jones, and W. Rudolph, Comparative study of femtosecond and nanosecond laser induced breakdown spectroscopy of depleted uranium, Applied Opticcs, Vol. 50, No. 3, January 2011, pp 313-317.

[65]  B. Verhoff, S. S. Harilal, J. R. Freeman, P. K. Diwakar and A. Hassanein, Dynamics of femto-and nanosecond laser ablation plumes investigated using optical emission spectroscopy, Journal of Applied Physics, Vol. 112, November 2012, pp 093303-1-9.

[66]  X. Bai, Q. Ma, V. Motto-Ros, J. Yu, D. Sabourdy, L. Nguyen, and A. Jalocha, Convoluted effect of laser fluence and pulse duration on the property of a nanosecond laser-induced plasma expanding into an argon ambient gas at the atmospheric pressure, Journal of Applied Physics, Vol. 113, January 2013, pp 013304-1-10.

[67]  J. R. Freeman, S. S. Harilal, P. K. Diwakar, B.Verhoff, A. Hassanein, Comparison of optical emission from nanosecond and femtosecond laser produced plasma in atmosphere and vacuum conditions, Spectrochimica Acta Part B, Vol. 87, May 2013, pp 43-50.

[68]   M. E. Shaheen, J. E. Gagnon and B. J. Fryer, Laser ablation of iron: A comparison between femtosecond and picoseconds laser pulses, Journal of Applied Physics, Vol. 114, August 2013, pp 083110-1-8.

[69]   S. Hafeez, N. M Shaikh and M.A Baig, Spectroscopic studies of Ca plasma generated by the fundamental, second and third harmonics of an Nd:YAG laser, Laser and Particle Beams Vol. 26, January 2008, pp. 41-50.

[70]   D. Campos, S.S. Harilal and A. Hassanien, The effect of laser wavelength on emission and particle dynamics of Sn plasma, Journal of Applied Physics, Vol. 108, December 2010, pp. 113305-1-7.

[71]   S. Amoruso, Modelling of laser produced plasma and time-of-flight experiments in UV laser ablation of aluminium targets, Applied Surface Science, Vol. 138-139, 1999, pp 292-298.

[72]   A. E. Hussein, P. K. Diwakar, S. S. Harilal and A. Hassanein, The role of laser wavelength on plasma generation and expansion of ablation plumes in air, Journal of Applied Physics, Vol. 113, April 2013, pp 143305-1-10.

[73]   S. S. Harilal, Influence of spot size on propagation dynamics of laser-produced tin plasma, Journal of Applied Physics, Vol. 102, December 2007, pp. 123306-1-6.

[74]   J. E. Crow, P. L. Auer and J. E. Allen, The expansion of a plasma into a vacuum, Journal of Plasma Physics, Vol. 14, Part 1, 1975, pp 65-76.

[75]   A. Neogi, A. Mishra, and R. K. Thareja, Dynamics of laser produced carbon plasma expanding in low pressure ambient atmosphere, Journal of Applied Physics, Vol. 83, No. 5, March 1998, pp 2831-2834.

[76]   E. G. Gamaly, A. V. Rode, and B. Luther-Davies, V. T. Tikhonchuk, Ablation of solids by femtosecond lasers: Ablation mechanism and ablation threshold for metals and dielectrics, Physics of Plasmas, Vol. 9, No. 3, March 2002, pp 949-957.

[77]   H. Matsuta and K. Wagatsuma, Emission Characteristics of a Low –Pressure Laser- Induced Plasma: Selective Excitation of Ionic Emission Lines of Copper, Applied Spectroscopy, Vol. 56, No. 9, May 2002, pp 1165-1169.

[78]   S. S. Harilal, C. V. Bindhu, M. S. Tilack, F. Najamabadi, and A. C. Gaeris, Internal structure and expansion dynamics of laser ablation plumes into

ambient gases, Journal of Applied Physics, Vol. 93, No. 5, March 2003, pp 2380-2388.

[79] C. G. Parigger, J. O. Hornkohl, A. M. Keszler and L. Nemes, Measurement and analysis of atomic and diatomic carbon spectra from laser ablation of graphite, Applied Optics, Vol.42, No. 30, October 2003, pp 6192-6198.

[80] A. K. Sharma, R. K. Thareja, Plume dynamics of laser-produced aluminium plasma in ambient nitrogen, Applied Surface Science, Vol. 243, 2005, pp 68-75.

[81] J. S. Cowpe, R. D. Pilkington, J. S. Astin and A. E. Hill, The effect of ambient pressure on laser-induced silicon plasma temperature, density and morphology, Journal of Physics D: Applied Physics, Vol. 42, July 2009, pp 165202-1-8.

[82] H. C. Joshi, V. Prahlad, R. K. Singh, Ajai Kumar, Emission analysis of expanding laser produced lithium plasma plume in presence of ambient gas, Physics Letters A, Vol. 373, Issue 37, September 2009, pp 3350-3353.

[83] A. J. Effenberger, Jr. And J. R. Scott, Effect of Atmospheric Conditions on LIBS Spectra, Sensors, Vol.10, April 2010, pp 4907-4925.

[84] T. Donnelly, J. G. Lunney, S. Amoruso, R. BRuzzese, X. Wang, and X. Ni, Angular distributions of plume components in ultrafast laser ablation of metal targets, Applied Physics A, Vol. 100, June 2010, pp 569-574.

[85] T. Donnelly, J. G. Lunney, S. Amoruso, R. BRuzzese, X. Wang, and X. Ni, Dynamics of laser produced by ultrafast laser ablation of metals, Journal of Applied Physics, Vol. 108, August 2010, pp 043309-1-13.

[86] N. Farid, S. Bashir and K. Mahmood, Effect of ambient gas conditions on laser-induced copper plasma and surface morphology, Physica Scripta, Vol. 85, December 2011, pp 015702-1-7.

[87] P. E. Nica, M. Agop, S. Gurlui, C. Bejinariu, and C. Focsa, Characterization of Aluminum Laser Produced Plasma by Target Current Measurements, Japanese Journal of Applied Physics, Vol. 51, October 2012, pp 106102-1-10.

[88] K. F. Al- Shboul, S. S. Harilal and A. Hassanein, Emission features of femtosecond laser ablated carbon plasma in ambient helium, Journal of Applied Physics, Vol. 113, April 2013, pp 163305-1-9.

[89]     N. Farid, S. S. Harilal, H. Ding, and A. Hassanein, Emission features and expansion dynamics of nanosecond laser ablation plumes at different ambient pressures, Journal of Applied Physics, Vol. 115, January 2014, pp 033107-1-9.

[90]     B Rethfeld, K. Sokolowski-tinten, D. Von der Linde and S. I. Anisimov, Timescales in the response of materials to femtosecond laser excitation, Applied Physics A: Material Science and Processing, Vol. 79, March 2004, pp. 767–769.

[91]     S. S. Harilal, N. Farid, A. Hassanein, and V. M. Kozhevin, Dynamics of femtosecond laser produced tungsten nanoparticle plumes, Vol. 114, 2013, pp 203302-1-7.

[92]     S. Amoruso, G. Ausanio, C. de. Lisio, V. Iannotti, M. Vitiello, X. Wang, L. Lanotte, Synthesis of nickel nanoparticles and nanoparticles magnetic films by femtosecond laser ablation in vacuum, Applied Surface Science, Vol. 247, February 2005, pp 71-75.

[93]     N. Farid, S. S. Harilal, H. Ding, and A. Hassanein, Dynamics of ultrafast laser plasma expansion in the presence of an ambient, Applied Physics Letters, Vol. 103, November 2013, pp 191112-1-5.

[94]     S. S. Harilal, C. V. Bindhu, M. S. Tilack, F. Najamabadi, and A. C. Gaeris, Plume splitting and sharpening in laser-produced aluminium plasma, Journal of Physics D: Applied Physics, Vol. 35, November 2002, pp 2935-2938.

[95]     F. Garrelie, J. Aubreton and A. Catherinot, Monte Carlo simulation of the laser-induced plasma plume expansion under vacuum: Comparison with experiments, Journal of Applied Physics Vol. 83, No. 10, May 1998, pp. 5075-5082.

[96]     S. R. Franklin, R. K Thereja, Monte Carlo simulation of laser ablated plasma for thin film deposition, Applied Surface science, Vol. 177 January 2001, pp. 15-21.

[97]     Ya. B. Zel'dovich and Yu. P. Raizer, Physics of Shock Waves and High Temperature Hydrodynamic Phenomena, Academic, New York, 1966, Vol. 1, pp. 94.

[98]   M . A Liberman and A. L. Velikovich, Physics of Shock Waves in Gases and Plasmas, Springer, Berlin, 1986.

[99]   T. E. Itina, J.Hermann, P. Delaporte and M.Sentis, Laser-generated plasma plume expansion: Combined continuous-microscopic modeling, Physical Review E, Vol. 66, No. 6, December 2002, pp. 066406-1-12.

[100]  L. Fornarini, R. Fantoni, F. Colao, A. Santagata, R. Teghil, A. Elhassan and M. A. Harith, Theoretical modeling of laser ablation of quaternary bronze alloys: Case studies comparing femtosecond and nanosecond LIBS experimental, data Journal of Physical Chemistry A, Vol 113, No. 52, September 2009, pp.14364-14374.

[101]  J. F. Kielkopf, Spectroscopic study of laser-produced plasmas in hydrogen, Physical Review E, Vol. 52, No.2, August 1995, pp 2013-2024.

[102]  S. I. Anisimov, B. L. Kapeliovich and T. L. Perel'man, Electron emission from metal surfaces exposed to ultrashort laser pulses, Soviet Physics-Journal of Experimental and Theoretical Physics, Vol. 39, No. 2, August 1974, pp 375-377.

[103]  A.J. McAlister and E.A. Stern, Plasma resonance absorption in thin metal films Physical Review Vol. 132, No. 4, November 1963, pp. 1599-1602.

[104]  Michael I. Zeifman, Barbara J. Garrison, and Leonid V. Zhigilei, Combined molecular dynamics–direct simulation Monte Carlo computational study of laser ablation plume evolution, Journal of Applied Physics, Vol. 92, No. 4, August 2002, pp. 2181-2193.

[105]  T. A. Witten and L. M. Sander, Diffusion-Limited Aggregation, a Kinetic Critical Phenomenon, Physical Review Letters, Vol. 47, No. 19, November 1981, pp. 1400-1403.

[106]  D. W. Hahn and N. Omenetto, Laser Induced Breakdown spectroscopy (LIBS), Part I: Review of Basic diagnostics and plasma-particle interactions: Still-challenging issues within the analytical plasma community, Applied Spectroscopy, Vol. 64, No. 12, September 2010, pp. 335A-366A.

[107]  M. Polek, S. S. Harilal, and A. Hassanein, Two-dimensional mapping of the electron density in laser-produced plasmas, Applied Optics, Vol. 51, No. 4. February 2012, pp 498-503.

[108] H. R. Griem, Principles of plasma spectroscopy © Springer- Verlag, Berlin, Heidelberg, 2009.

[109] N. Smijesh and Reji Philip, Emission dynamic of an expanding ultrafast-laser produced Zn plasma under different ambient pressures, Journal of Applied Phyiscs, Vol. 114, September 2013, pp. 093301-1-5.

[110] R. W. P. McWhirter, Plasma Diagnostic Techniques, edited by R. H. Huddlestone and S. L. Leonard (Academic Press, New York, 1965), Chap. 5

[111] C. Phipps, Laser Ablation and its Applications, Springer Science & Business media LLC, 2007.

[112] http://physics.nist.gov/PhysRefData/ASD /lines_form.html.

[113] S. S. Harilal, Riju C. Issac, C. V. Bindhu, V. P. N. Nampoori and C. P. G. Vallabhan, Optical emission studies of $C_2$ species in laser-produced plasma from carbon, Journal of Physics D: Applied Physics, Vol. 30, 1997, pp 1703-1709.

[114] A. M. El Sherbini, A. A. Saad Al Amer, A. T. Hassan, T. M. El Sherbini, Measurement of plasma electron temperature utilizing magnesium lines appeared in laser produced Aluminum plasma in air, Optics and Photonics Journal, Vol. 2, December 2012, pp 278-285.

[115] A. Couairon and A. Mysyrowics, Femtosecond filamentation in transparent media, Physics. Reports, Vol. 441, February 2007, pp. 47-189.

[116] N. Smijesh, K. Chandrasekharan, Jagdish C Joshi and Reji Philip, Time of flight emission spectroscopy of a laser produced expanding nickel plasma: short-pulse and ultrafast excitations, Journal of Applied Phyiscs, Vol. 116, July 2014, pp. 013301-1-7.

[117] N. M. Shaikh, B. Rashid, S. Hafeez, Y. Jamil, and M. A. Baig, Measurement of electron density and temperature of a laser-induced zinc plasma, Journal of Physics D: Applied Physics, Vol. 39 March 2006, pp. 1384–1391.

[118] M. S. Dimitrijevic and S. Sahal-Bréchot, Stark Broadening of Neutral Zinc spectral lines Astronomy and Astrophysics Supplementary Series, Vol. 140, No. 2, December 1999, pp. 193-196.

[119] D. Devaux, R. Fabbro, L. Tollier, and E. Bartnicki, Generation of shock waves by laser-induced plasma in confined geometry, Journal of Applied Physics, Vol. 74, No. 4, August 1993, pp 2268-2273.

[120] V. S. Burakov, A. F. Bokhonov, M. I. Nedel'ko, N. V. Tarasenko, Change in the ionization state of a near-surface laser-produced aluminium plasma in double-pulse ablation modes, Quantum Electronics, Vol. 33, No. 12, 2003, pp 1065-1071.

[121] S. S. Harilal, P. K. Diwakar, J. R. Freeman and A. Hassanein, Role of laser pre-pulse wavelength and inter-pulse delay on signal enhancement in collinear double-pulse laser-induced breakdown spectroscopy, Spectrochimica Acta Part B, Vol. 87, May 2013, pp 65-73.

[122] S. S. Harilal, P. K. Diwakar and A. Hassanein, Electron-ion relaxation time dependent signal enhancement in ultrafast double-pulse laser-induced breakdown spectroscopy, Applied Physics Letters, Vol. 103, July 2013, pp 041102-1-4.

[123] T. Y. Chang and C. K. Birdsall, Laser-induced emission of electrons, ions, and neutrals from ti and ti-d surfaces, Applied Physics Letters, Vol. 5, No. 9, November 1964, pp. 171-172.

[124] P.A.H. Saunders, P. Avivi, W. Millar, Laser Produced Plasmas from solid hydrogen targets, Physics Letters A, Vol. 24, No. 5, February 1967, pp. 290-291.

[125] N. G. Basov, O. N. Krokhin, and G. V. Sklizkov Laser Application for the Production and Diagnostics of Pulsed Plasma, Applied Optics, Vol. 6, No. 11, 1967, pp. 1814-1817.

[126] M. Mattioli, D. Véron, Electron-ion Recombination in laser produced plasma, Plasma Physics, Vol.11, 1969, pp 684-686.

[127] G. J. Tallents, An experimental study of recombination in a laser-produced plasma, Plasma Physics, Vol. 22, 1980, pp 709-718.

[128] Xin Zhao, Zheng Zheng, Lei Liu, Qi Wang, Haiwei Chen, and Jiansheng Liu, Fast, long-scan-range pump-probe measurement based on asynchronous sampling using a dual wavelength mode-locked fiber laser, Optics Express, Vol. 20, No.23, November 2012, pp 25584-25589.

[129] F. Stienkemeier, F. Meier, A. Hägele, and H. O. Lutz, E. Schreiber, C. P. Schulz, and I. V. Herte, Coherence and Relaxation in Potassium-Doped Helium Droplets Studied by Femtosecond Pump-Probe Spectroscopy, Physical Review Letters, Vol. 83, No. 12, September 1999, pp. 2320-2323.

[130] C. Manzoni, A. Gambetta, E. Menna, M. Meneghetti, G. Lanzani, and G. Cerullo, Intersubband Exciton Relaxation Dynamics in Single-Walled Carbon Nanotubes, Physical Review Letters, Vol. 94, No. 20, May 2005, pp. 207401-1-4.

[131] W. J. H. Leyland, G. H. John, R. T. Harley, M. M. Glazov, E. L. Ivchenko, D. A. Ritchie, I. Farrer, A. J. Shields, and M. Henini, Enhanced spin-relaxation time due to electron-electron scattering in semiconductor quantum wells, Physical Review B, Vol. 75, No. 16, April 2007, pp. 165309-1-8.

[132] V. I. Klimov, A. A. Mikhailovsky, D. W. McBranch, C. A. Leatherdale, and M. G. Bawendi, Mechanisms for intraband energy relaxation in semiconductor quantum dots: The role of electron-hole interactions, Physical Review B, Vol. 61, No. 20, May 2000, pp. R13 349-R13 352

[133] M. Drescher, M. Hentschel, R. Kienberger, M. Uiberacker, V. Yakovlev, A. Scrinzi, Th. Westerwalbesloh, U. Kleineberg, U. Heinzmann & F. Krausz, Time-resolved atomic inner-shell spectroscopy, Nature, Vol. 419, October 2002, pp. 803-807.

[134] John B. Asbury, Randy J. Ellingson, Hirendra N. Ghosh, Suzanne Ferrere, Arthur J. Nozik, and Tianquan Lian, Femtosecond IR Study of Excited-State Relaxation and Electron-Injection Dynamics of Ru(dcbpy)$_2$(NCS)$_2$ in Solution and on Nanocrystalline TiO2 and Al2O3 Thin Films, Journal of Physical Chemistry B, Vol. 103, No. 16, February 1999, pp. 3110-3119.

[135] M. G. Bawendi, W. L. Wilson, L. Rothberg, P. J. Carroll, T. M. Jedju, M. L. Steigerwald, and L. E. Brus, Electronic structure and photoexcited-carrier dynamics in nanometer-size CdSe clusters, Physical Review Letters, Vol. 65, No. 13, September 1990, pp. 1623-1626.

[136] L. Casperson and A. Yariv, Pulse Propagation in a High- Gain Medium, Physical Review Letters, Vol. 26, No. 6, February 1971, pp 293-295.

[137] L. I. Gudzenko, L. A. Shelepin, and S. I. Yakovlenko, Amplification in recombining plasmas (plasma lasers), Soviet Physics-Uspekhi, Vol. 17, No.6, June 1975, pp 848-863.

[138] R. D. Milroy, C. E. Capjack, and C. R. James, Plasma laser pulse amplifier using induced Raman or Brillouin processes, Physics of Plasmas, Vol. 22, No.10, October 1979, pp 1922-1931.

[139] C. Chenais-Popovics, R. Corbett, C. J. Hooker, M. H. Key, G. P. Kiehn, C. L. S. Lewis, G. J. Pert, C. Regan, S. J. Rose, S. Sadaat, R. Smith, T. Tomie, and O. Willi, Laser Amplification at 18.2 nm in Recombining Plasma from a Laser- Irradiated Carbon Fiber, Physical Review Letters, Vol. 59, No. 19, November 1987, pp 2161-2164.

[140] A. A. Andreev, C. Riconda, and V. T. Tikhonchuk and S. Weber, Short light pulse amplification and compression by stimulated Brillouin scattering in plasmas in the strong coupling regime, Physics of Plasmas, Vol. 13, May 2006, pp 053110-1-5.

[141] A. V. Ponomarev, P. S. Strelkov, and A. G. Shkvarunets, Tunable Plasma Relativistic Microwave Amplifier, Plasma Physics Reports, Vol. 26, No. 7, 2007, pp 592-597.

**Smijesh N.**
*Senior Research Fellow*
*Laser and Nonlinear Optics lab*
*Department of Physics*
*National Institute of Technology Calicut*
*Kerala, India*
*Phone: +91-9946326883/7829528635*
Contact: *smiju5247@gmail.com*

## ACADEMIC QUALIFICATIONS

- Submitted **Ph. D. Thesis** titled "**Spectral and temporal characterization of nanosecond and femtosecond laser produced plasma from metallic targets**" to the National Institute of Technology Calicut (*http://nitc.ac.in*), Kerala, India [Supervisors: Dr. K. Chandrasekharan (*csk@nitc.ac.in*) and Dr. Reji Philip (*reji@rri.res.in*)].
- **M. Tech. in Electronics and Communication (Optoelectronics and Optical Communication)** in November 2009 from the University of Kerala, Thiruvananthapuram, Kerala, India.
- **M. Sc. (Physics)** in 2006 from Mahatma Gandhi University Kottayam, Kerala, India.

## RESEARCH INTERESTS

Intense light-matter interaction, Laser Produced Plasmas, Laser Induced Breakdown Spectroscopy, Nonlinear Optics.

## JOURNAL PUBLICATIONS

1. Time of flight emission spectroscopy of a laser produced expanding nickel plasma: short-pulse and ultrafast excitations
   N. Smijesh, K. Chandrasekharan, J. C. Joshi, R. Philip, *J. Appl. Phys.* **116**, 013301 (2014)
2. Emission dynamics of an expanding ultrafast-laser produced Zn plasma under different ambient pressures
   N. Smijesh and R. Philip, *J. Appl. Phys.* **114**, 093301 (2013)
3. Organic dye impregnated poly(vinyl alcohol) nanocomposite as an efficient optical limiter: structure, morphology and photophysical properties
   S. Sreeja, S. Sreedhanya, N. Smijesh, R. Philip, C. I. Muneera, *J. Mater. Chem. C*, **1**, 3851(2013)
4. Size-dependent optical properties of Au nanorods
   S.L Smitha, K.G. Gopchandran, N. Smijesh, R. Philip, *Progress in Natural Science-Material International* **23**(1), 36 (2013)

5. Nonlinear optical properties of composite napthalocyanine thin films with nanocrystalline morphology
   N. S. Panicker, N. Smijesh, R. Philip, C.S. Menon, *Materials Letters* **89**, 188 (2012).

6. Electrochemical and Nonlinear Optical Studies of New DA Type pi-Conjugated Polymers Carrying 3, 4-Benzyloxythiophene, Oxadiazole, and 3, 4-Alkoxythiophene Systems
   M. S. Sunitha, A. V. Adhikari, K. A. Vishnumurthy, N. Smijesh, R. Philip, *Chemistry Letters* **41**, 234 (2012).

7. Two-Photon Absorption and optical limiting in Tristhiourea Cadmium Sulphate
   S. Dhanuskodi, T. C. Sabari Girisun, N. Smijesh, R. Philip, *Chemical Physics Letters* **486**, 8083 (2010).

**PUBLICATIONS IN CONFERENCE PROCEEDINGS**

1. Short and ultrafast laser produced Aluminum plasma: A fluence dependent study
   Pranitha Sankar, Jijil J. J. Nivas, N. Smijesh, Reji Philip (To appear in the conference proceedings of 29th National Symposium on plasma science and Technology- Plasma 2014, India)

2. Expansion dynamics of laser produced Zn plasma: Short pulse and ultrafast excitation
   N. Smijesh, Anitta R. Thomas, Kavya H. Rao and Reji Philip (To appear in the conference proceedings of 29th National Symposium on plasma science and Technology- Plasma 2014, India)

3. Optical Time of flight measurements of laser produced metal plasmas: Short pulse and ultrafast excitations
   N. Smijesh, K. Chandrasekharan, Reji Philip (To appear in the conference proceedings of 24th Swadeshi Science Congress, National Seminar, Kerala, India)

4. Acceleration of neutrals in an expanding laser produced Zn plasma
   N. Smijesh, Kavya H. Rao and Reji Philip (To appear in the conference proceedings of the Theme meeting on ultrafast science (UFS)- 2014, DAE- BRNS, India)

5. Influence of laser pulse width on the emission dynamics of laser produced Zn plasma in nitrogen ambient
   N. Smijesh, K. Chandrasekharan, Reji Philip (Appeared in the conference proceedings of National Laser Symposium- 22, DAE- BRNS, India)

6. Time-resolved spectroscopy of CI and CII line emissions from an ultrafast laser produced solid graphite plasma
   N. Smijesh, K. Chandrasekharan, R. Philip, AIP Conf. Proc.1620, 517 (2014))

7. Time-resolved studies of the 723 nm CII transition in a Q-switched Nd: YAG laser-induced carbon plasma
   N. Smijesh, K. Chandrasekharan, R. Philip (*Accepted for publication in an IOP conference proceedings- Material Science and Engineering*)

8. Indigo carmine dye polymer nanocomposite films for optical limiting applications
   S. Sreeja, S. Mayadevi, S. R. Suresh, P. G. L. Frobel, N. Smijesh, R. Philip, C. I. Muneera, AIP Conf. Proc. 1391, 618 (2011).
9. Thermal diffusivity measurements of dental resin using photoacoustic effect
   N. Smijesh, L. K. Joseph, A. Kurien, V. P. N. Nampoori, IEEE-ICIAS-2007 (DOI:10.1109/ICIAS.2007.4658397).

**MANUSCRIPTS TO BE COMMUNICATED**

1. Acceleration of neutrals in an expanding nanosecond laser produced nickel plasma
   N. Smijesh, K. Chandrasekharan, R. Philip
2. Time of flight dynamics of neutrals in an expanding femtosecond laser produced zinc plasma
   N. Smijesh, Kavya H. Rao, R. Philip
3. Time-resolved emission spectroscopy of an expanding laser produced Al-Plasma: effect of pulse width and ambient pressure
   J. J. J. Nivas, Pranitha Sankar, N. Smijesh, R. Philip
4. Stark broadening of Zn I transitions in ultrafast laser produced Zinc plasma: Recombination Effects
   Kavya H. Rao, N. Smijesh, R. Philip
5. Influence of laser pulse width on the ablation of Zinc in nitrogen ambient
   N. Smijesh, K. Chandrasekharan, R. Philip
6. Emission dynamics of ions and neutrals in an ultrafast laser produced expanding Carbon plasma
   N. Smijesh, K. Chandrasekharan, R. Philip

**INVITED TALKS**

1. "Spectroscopy and optical time of flight studies of laser produced metal plasmas: short pulse and ultrafast excitations" at the Raman Research Instutute, Bangalore, India – 560 080 on 10 September 2014.

**PROJECTS UNDERGONE**

1. **M.Tech. Major project:** "**Investigation of Ultrafast laser induced plasma in condensed matter**" at the *Raman Research Institute, Bangalore, India* (December 2008–December 2009). The femtosecond intense short laser pulse causes the ionization of the matter and emits radiations like x-rays, UV and VUV etc. This experiment describes the comparison of the amount of x-rays produced by water-soluble dyes and water upon the irradiation with the femtosecond laser pulses.

2. **M.Sc. Project:** "**Thermal diffusivity measurements of dental resin using photoacoustic effect**" at *catholicate college Pathanamthitta, Kerala, India* (Jan-May 2006). The laser induced non-destructive photoacoustic technique has been employed to measure the thermal diffusivity of dental resin. The thermal diffusivity value was evaluated by knowing the transition frequency between the thermally thin to thermally thick region from the log-log plot of photoacoustic amplitude versus chopping frequency. Analysis of the data was carried out on the basis of the one-dimensional model of Rosencwaig and Gersho.

**ACADEMIC ACHIEVEMENTS**

- Best poster award for the poster titled "Laser produced plasma emission studies in solid targets" by N. Smijesh, M. Shafi Ollakkan, and R. Philip, in the Fifth SERB (http://www.serb.gov.in) school on "Tokamaks & magnetized Plasma Fusion," held during 25 Feb to 15 Mar 2013 at the Institute for Plasma Research, Gandhinagar, India.

- Qualified **GATE** (Graduate Aptitude Test in Engineering) **in 2007**.

**RESEARCH/ TEACHING EXPERIENCE**

- Worked as a Project Assistant in the Ultrafast and Nonlinear Optics lab of the Light and Matter Physics (LAMP) Group, Raman Research Institute (http://www.rri.res.in/), Bangalore, India, from May 2010 to July 2011.
- Worked as a Guest Lecturer in the Post Graduate and Research Department of Physics, Catholicate College (http://www.catholicatecollege.co.in/), Pathanamthitta, India, from January 2007 to November 2007.
- Worked as a Post Graduate Teacher in Shalom Public School (ICSE syllabus) Chenneerkara, Pathanamthitta, India.

**PERSONAL DETAILS**

| | |
|---|---|
| Name | : Smijesh N. |
| Address | : Kilithattil Veedu, |
| |   Anakottoor (P O), |
| |   Kottarakara, Kollam (*Dist*) |
| |   Kerala, India- 691505. |
| Date of Birth | : 04th April 1984 |
| Citizenship | : Indian |
| Language proficiency | : English, Malayalam, Kannada, Tamil. |

**REFERENCES**

**Dr. K. Chandrasekharan**
Professor
Department of Physics
National institute of Technology Calicut
Kerala-673 601, India
E-mail: csk@nitc.ac.in

**Dr. Reji Philip**
Associate Professor and Coordinator
Light and Matter Physics
Raman Research Institute
Bangalore-560 080, India
E-mail: reji@rri.res.in

**Dr. S. S. Harilal**
Scientist
Pacific Northwest National Laboratory
Richland, WA 99352
Phone: 509.372.4926
Fax: 509.375.6497
hari@pnnl.gov